\documentclass[prb,twocolumn,showpacs,amssymb,floatfix,10pt,aps]{revtex4-1}

\usepackage{graphicx}
\usepackage{dcolumn}
\usepackage{bm}
\usepackage{amssymb}

\begin{document}


\title{The NIST compilation of ionization potentials revisited (I): From He-like to Xe-like ions}

\author{Gabriel Gil$^{1,2}$ and Augusto Gonzalez$^1$}
\affiliation{$^1$Institute of Cybernetics, Mathematics and Physics, Havana, Cuba}
\affiliation{$^2$CNR Institute of Nanosciences S3, Modena, Italy}

\begin{abstract}

The National Institute of Standards and Technology (NIST) database on ionization potentials for neutral atoms and ions is examined. For each isoelectronic sequence, we construct a regularized perturbative series that exactly matches the large-$Z$ and $Z\approx N-1$ regions. Comparison of the NIST data with this series allows the identification of problematic values in the reported data.

\end{abstract}


\maketitle

Physics is indeed a mature science. Large amounts of data, of very different kinds, have been accumulated along decades. Sometimes, a fresh look at these compilations, on the basis of simple -physically meaningful- models, leads to a qualitative understanding of the data. 

In previous papers \cite{PRA1,PRA2}, on the basis of the scaling characteristic of Thomas-Fermi theory \cite{TF}, we have shown that both the NIST data on ionization potentials of atomic ions$~$\cite{NIST}, and the correlation energies of atomic ions can be accomodated along single universal curves. 

In the present work, we go a step further and develop a simple model allowing the identification of problematic isolated reported values for the ionization potentials in a particular isoelectronic sequence. Because of the fact that the compilation is very often used for different purposes, indications of which data could be wrong and should be re-examined is of great importance. For instance, as commented in Refs. [\onlinecite{Sansonetti}] and [\onlinecite{Nandy}], accurate description of the spectra is useful in order to model some dynamical features of stellar sources, and also to interpret astronomical data.

Our model for the ionization potential is a smooth interpolation that matches both the large-$Z$ (heavy ion) and the $Z\approx N-1$ (anion) regions.\cite{RPT} We call it the ``regularized perturbation theory'' (RPT). We have used similar approaches in order to compute the energy of relatively large quantum dots,\cite{qdot} atomic ions in a harmonic trap,\cite{htrap_a} neutral atoms in traps,\cite{htrap_b} and Rydberg-like impurity levels in a quantum well.\cite{qwell}

The RPT series for the ionization potential of an $N$-electron atomic ion with nuclear charge $Z$ is written as (atomic units are used everywhere):\cite{RPT,ourMPLB}

\begin{equation}
I_p=a_2 Z^2 + a_1 Z + a_0 + a_{-1}/Z.
\label{ip}
\end{equation}

\noindent Coefficients $a_2$ and $a_1$ are obtained from the large-$Z$ limit.\cite{RPT} $a_2$ comes from the leading term (free electrons in the nuclear field), whereas $a_1$ is computed in the next-to-leading approximation, where Coulomb repulsion between electrons is perturbatively treated. Relativistic corrections should be included because the NIST data span the range of nuclear charges up to very heavy ions, i.e. $N \leq Z \leq 110$. Detailed expressions for $a_2$ and $a_1$, with explicit relativistic corrections, can be found in Ref. [\onlinecite{RPT}].

The next two terms of the series have the functional form suggested by higher-order perturbation theory on the Coulomb repulsion. However, in order to determine them, we shall follow a different strategy. We force expression (\ref{ip}) to match the expected value and the slope at $Z=N-1$.\cite{RPT} In this sense, it is a ``regularized perturbative series''. These conditions are written as follows:

\begin{eqnarray}
\nonumber \left. I_p\right|_{Z=N-1}&=&E_a, \\
\left. \frac{d I_p}{dZ}\right|_{Z=N-1}=&s&=\frac{\int_{R}^{\infty}{dr} e^{-2 \kappa r}/r}{\int_{R}^{\infty}{dr} e^{-2 \kappa r}}.
\label{restric}
\end{eqnarray}

\noindent $E_a$ is the electron affinity of the neutral atom with $N-1$ electrons.\cite{RSC} The slope $s$, on the other hand, is computed in terms of $E_a$ and $R$, the latter is a characteristic radius of the $(N-1)$-electron system, which we estimate as the covalent radius.\cite{ourMPLB} Note that $\kappa = \sqrt{2 E_a}$. The expression of the slope makes use of the fact that, at $Z=N-1$, the outer electron interacts with a neutral core. Thus, the interaction is short-ranged and the wave function at large distances is solely determined by the binding energy, i.e. $E_a$. We use this function in order to perturbatively compute the residual Coulomb interaction of the outer electron with the core, when $Z$ is slightly displaced from $N-1$. The explicit expression for $s$ was derived earlier in Ref. [\onlinecite{ourMPLB}].

\begin{center}
\begin{figure}[!ht]
\includegraphics[width=0.9\linewidth,angle=0]{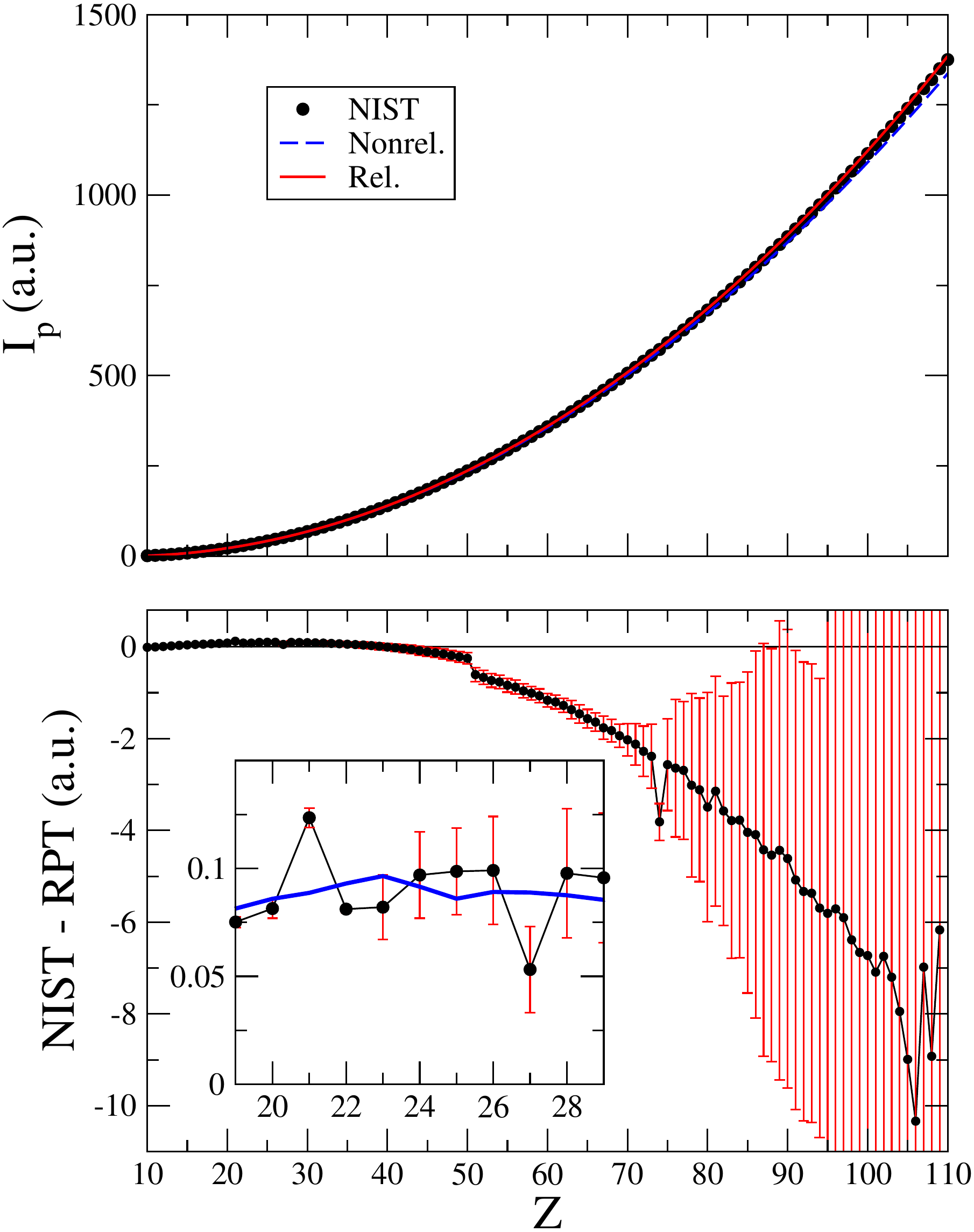}
\caption{\label{fig10} (Color online) Ne-like ions ($N=10$). Upper panel: Ionization potentials taken from the NIST compilation along with our nonrelativistic (discontinuous, blue) and relativistically corrected (continuous, red) RPT predictions versus atomic number. Curves are smooth at any scale. Lower panel: The difference between NIST reported values and the relativistic RPT series versus atomic number. Reported error bars are shown. Inconsistencies  at $Z = 21$, 22, 27, 74, and the abrupt jump at $Z=50-51$ are noticed. A 5-points running average curve (blue line) is used to identify problematic points, as shown in the inset. We point out also a great dispersion of the data for $Z \ge 102$.}
\end{figure}
\end{center}

Once the coefficients in Eq. (\ref{ip}) are determined, our RPT series provides an interpolation for intermediate values of $Z$. We show in Fig. \ref{fig10}, in quality of example, the series for Ne-like ions ($N=10$) along with the corresponding NIST data.

Some comments should be added to this figure. First, NIST reported values are experimental points only when $Z$ is very close to $N$. For larger $Z$, data come from calculations or interpolations. Thus, we expect relatively high errors in this regime. Relative deviations of the RPT from NIST data are only a few percents in the intermediate-$Z$ region, which is a common feature of interpolants.\cite{qdot,htrap_a,htrap_b,qwell}

Second, the dependence of $I_p$ on $Z$ is smooth. In the upper panel, it is difficult to distinguish a problematic point, even if we change the scale. An abrupt change in $I_p$ could only be related to a rearrange of the occupied electronic levels. This may take place at specific values of $Z$, also reported in the NIST compilation. Thus, in spite of the fact that our RPT series lacks of spectroscopic precision, every jump or deviation in NIST-RPT, not coinciding with a spectrum rearrangement point, can be taken as a signature of possible errors. Problematic points are identified by comparison with a smoothed curve, which we construct by means of a 5-points running average. 

The remarkable deviations at $Z=74~$,\cite{W-ions1} and the abrupt jump at $Z= 50-51$,\cite{Dirac-Fock1} just to mention a few examples common to many of the isoelectronic sequences, are clear indications of deficiencies in the computed data reported by NIST. The dispersion of the data for very heavy ions should also motivate a reconsideration of the Dirac-Fock calculations.\cite{Dirac-Fock1}  

In the appendices, we examine $42$ isoelectronic sequences (five rows of the Mendeleev Table), from He-like to Xe-like ions. In this range, only the sequences which do not have a stable singly charged anion are excluded. Heavier ions, from La-like to Ac-like, are to be analyzed in a subsequent paper \cite{NISTrev2}.

We hope that the present analysis will be helpful in order to improve the reference data.

\begin{acknowledgments}
The authors are grateful to the Caribbean Network for Quantum Mechanics, Particles and Fields (ICTP) for support. G.G. also acknowledge financial support from the European Community's FP7 through the Marie Curie ITN-INDEX.
\end{acknowledgments}

\appendix

\section{The He-like sequence ($N=2$)}
\label{He}

RPT coefficients:
\begin{eqnarray}
\nonumber a_2 &=& 0.5 + 0.125~(Z/137.036)^2,\\
\nonumber a_1 &=& -0.625 -0.0989536~(Z/137.036)^2,\\
\nonumber a_0 &=& 0.368987,\\
\nonumber a_{-1} &=& -0.216282.
\label{he_coeff}
\end{eqnarray}

Conditions at $Z=N-1$:
\begin{eqnarray}
\nonumber E_a({\rm H}) &=& 0.0277063,\\
s &=& 0.59129.
\label{rest_he}
\end{eqnarray}

The slope was computed from $E_a({\rm H})$ and $R_{cov}({\rm H})=0.604712$. Here, and in the analysis below, data for $E_a$ and $R_{cov}$ are taken from Ref. \onlinecite{RSC}.

\begin{center}
\begin{figure}[!ht]
\includegraphics[width=0.9\linewidth,angle=0]{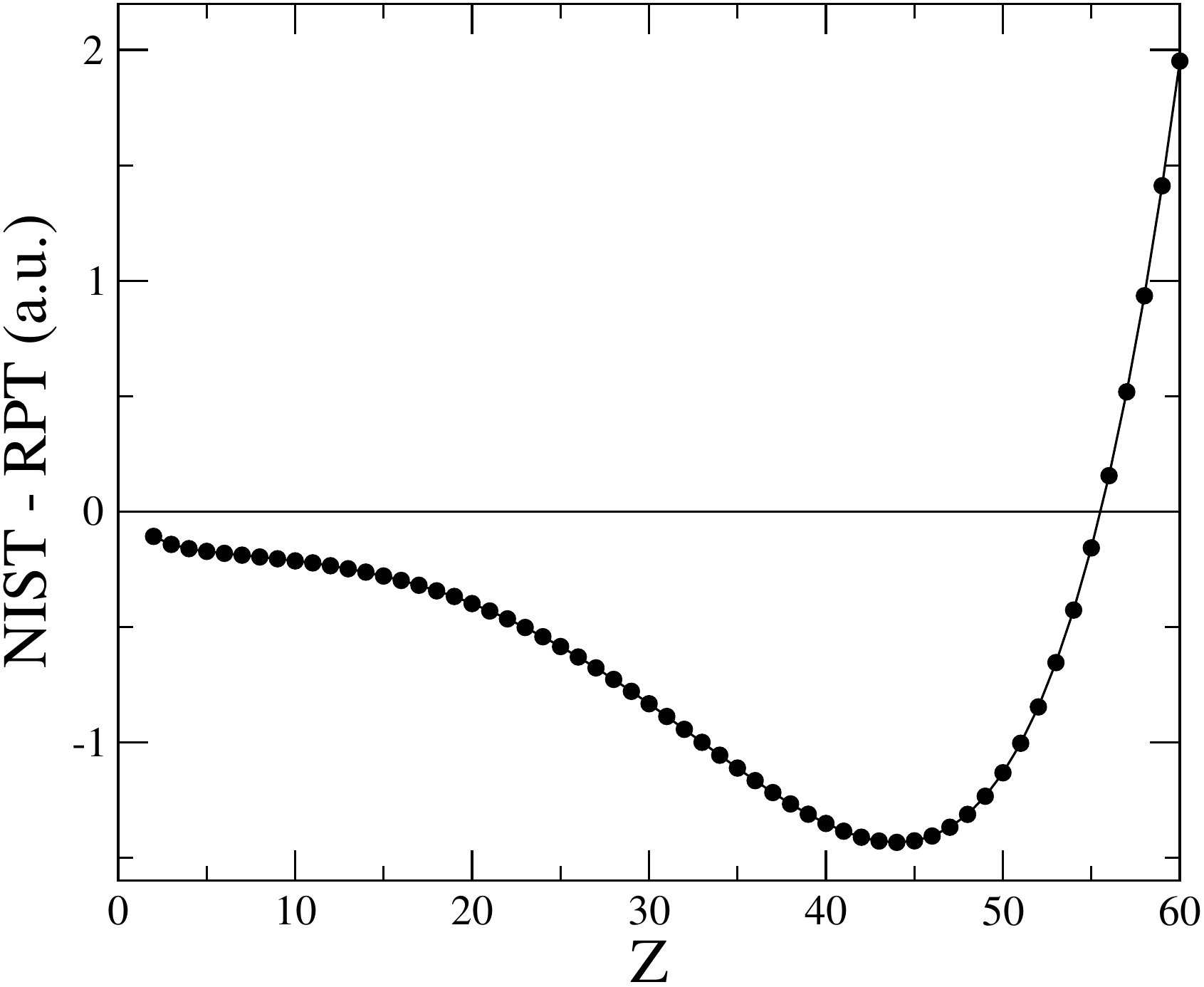}
\caption{\label{fig2} (Color online) The He-like systems ($N=2$). The difference between NIST values and the  relativistic RPT versus atomic number. Error bars are very small and can not be seen in the scale of the figure. No inconsistencies were detected.}
\end{figure}
\end{center}

\textbf{Discussion:}

Let us comment on some features of the He-like case. In the range $2 \le Z \le 60$, the difference between our RPT series and the reported values is very well behaved. In the intermediate region, the maximum relative error is around $0.15\%$. 

When $Z>56$,  NIST-RPT  rises.  NIST ionization potentials in the $12 \le Z \le 100$ range come mainly from \textit{ab intio} QED calculations by Artemyev et al., \cite{HeQED} which include finite nuclear-size effects. Our perturbative treatment of relativity cannot reproduce their results for very large $Z$.

\section{Second row elements}

\subsection{The Be-like sequence ($N=4$)}
\label{Be}

RPT coefficients:
\begin{eqnarray}
\nonumber a_2 &=& 0.125 + 0.0390625~(Z/137.036)^2,\\
\nonumber a_1 &=& -0.548196 -0.0952831~(Z/137.036)^2,\\
\nonumber a_0 &=& 0.675679,\\
\nonumber a_{-1} &=& -0.400253.
\label{be_coeff}
\end{eqnarray}

Conditions at $Z=N-1$:
\begin{eqnarray}
\nonumber E_a({\rm Li}) &=& 0.0227050,\\
s &=& 0.246343.
\label{rest_be}
\end{eqnarray}

The slope was computed from $E_a({\rm Li})$ and $R_{cov}({\rm Li})=2.456644$.

\begin{center}
\begin{figure}[!ht]
\includegraphics[width=0.9\linewidth,angle=0]{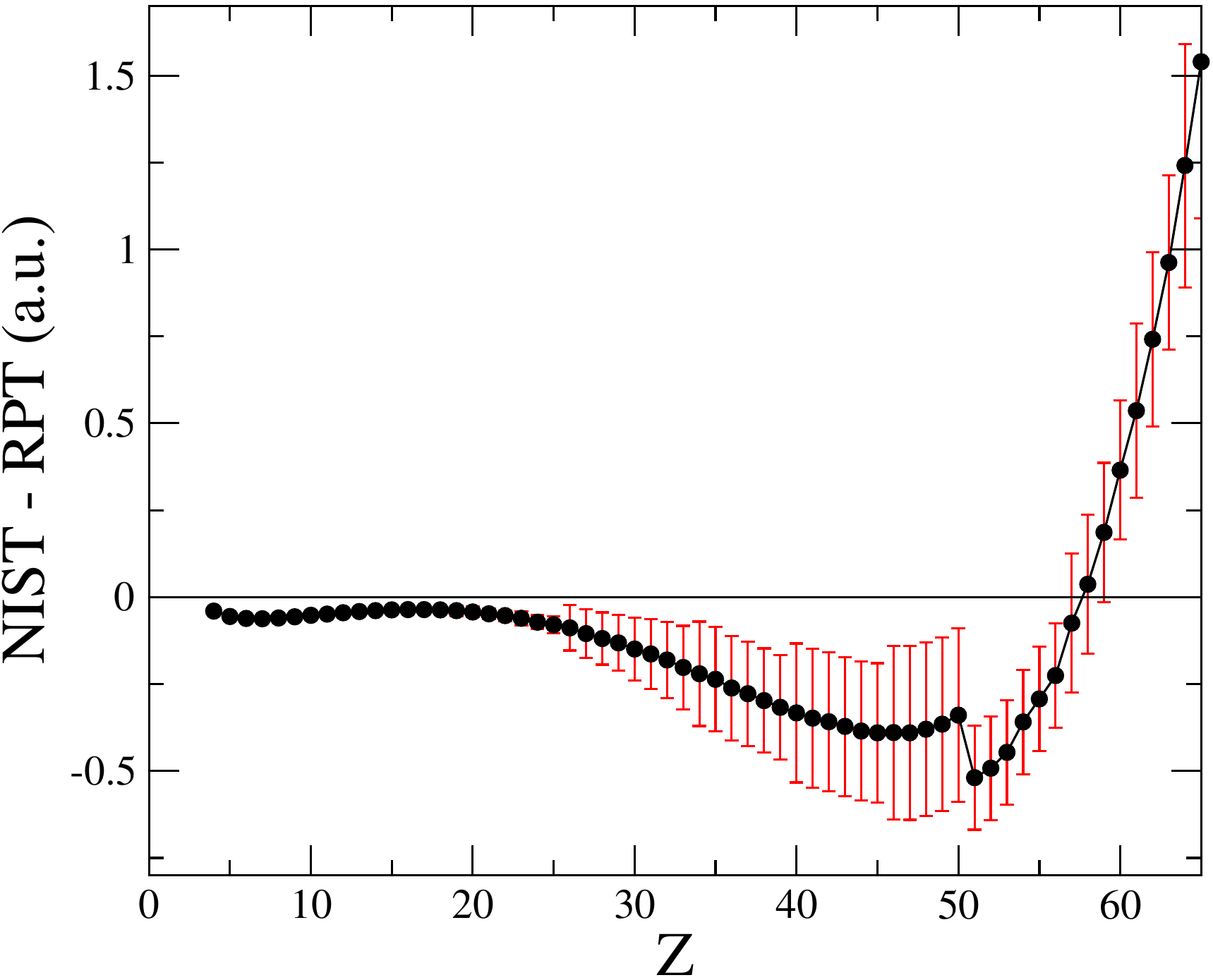}
\caption{\label{fig4} (Color online) The Be-like isoelectronic sequence ($N=4$). The abrupt jump at $Z =50-51$ can be, however, accomodated within error bars.}
\end{figure}
\end{center}

\textbf{Discussion:}

The comparison between NIST and RPT values for the Be-like sequence shows an abrupt jump of $0.28$ a.u. at $Z=50-51$ (see Fig. \ref{fig4}). These are numbers based on Dirac-Fock calculations of $I_p$ computed by different groups. When $Z\le 50$, numbers come from Ref. [\onlinecite{Interpol_N42-50}], whereas for $Z \ge 51$ almost all reported numbers come from Rodrigues et al. \cite{Dirac-Fock1}. We notice that, in Ref. [\onlinecite{Interpol_N42-50}], a formula like Eq. (\ref{ip}) is used as a fit to correct the computed values. 

The observed jump is consistent with the natural dispersion of points, as suggested by the reported error bars. Thus, no inconsistency is detected.

\subsection{The C-like sequence ($N=6$)}
\label{C}

RPT coefficients:
\begin{eqnarray}
\nonumber a_2 &=& 0.125 + 0.0390625~(Z/137.036)^2,\\
\nonumber a_1 &=& -0.945089 -0.241928~(Z/137.036)^2,\\
\nonumber a_0 &=& 0.0390625,\\
\nonumber a_{-1} &=& 0.643271.
\label{c_coeff}
\end{eqnarray}

Conditions at $Z=N-1$:
\begin{eqnarray}
\nonumber E_a({\rm B}) &=& 0.0102761,\\
s &=& 0.279379.
\label{rest_c}
\end{eqnarray}

The slope was computed from $E_a({\rm B})$ and $R_{cov}({\rm B})=1.58737$.

\begin{center}
\begin{figure}[!ht]
\includegraphics[width=0.9\linewidth,angle=0]{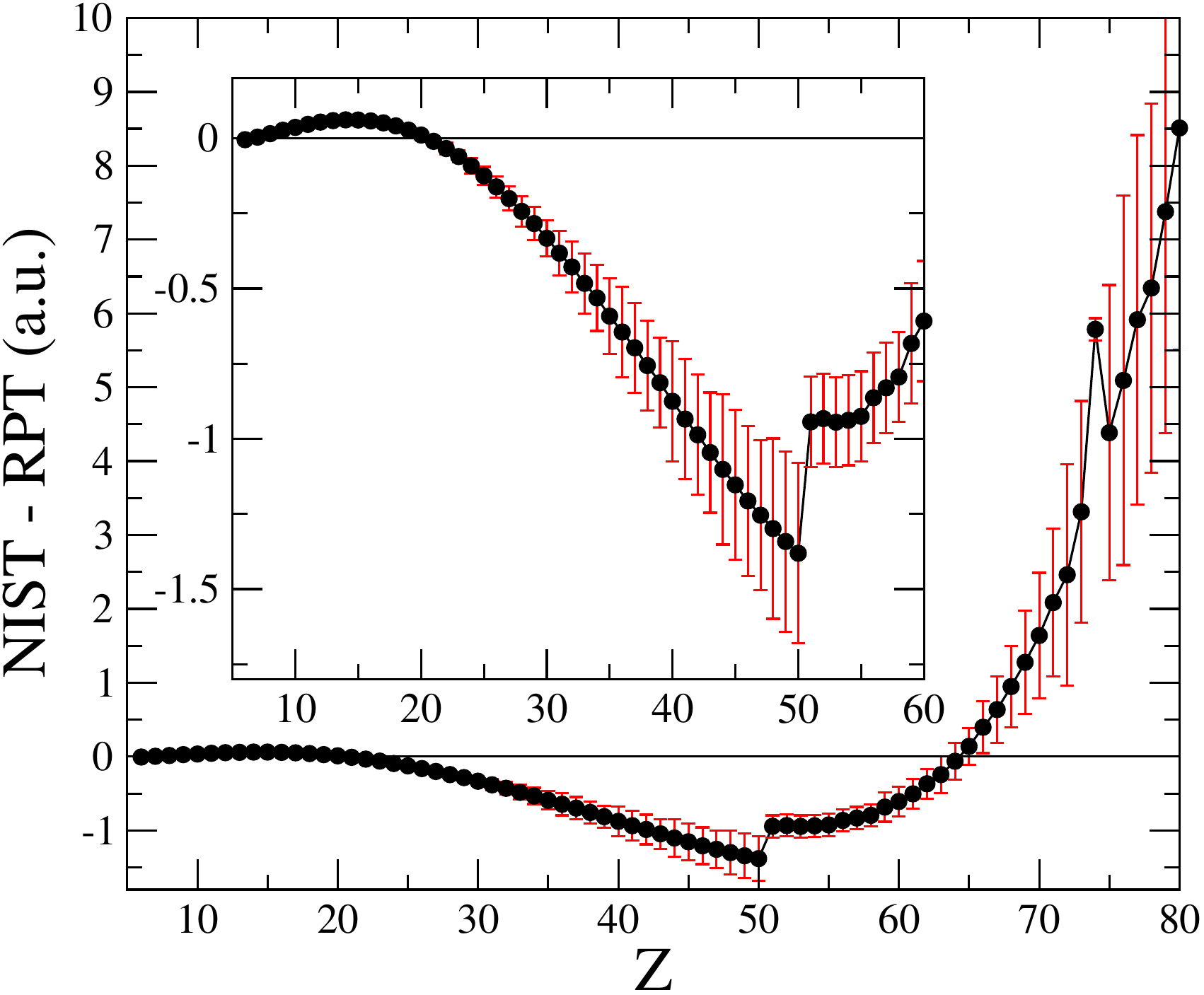}
\caption{\label{fig6} (Color online) The C-like systems ($N=6$). An inconsistency at $Z=74$ and an abrupt jump at $Z=50-51$) are noticed.}
\end{figure}
\end{center}

\textbf{Discussion:}

Analysis of Fig. \ref{fig6} (C-like sequence) also shows an abrupt jump of nearly $0.43$ a.u. at $Z=50- 51$. As in the Be-like case, this jump is associated to changes in the calculation methodology.
However, the jump exactly equals the reported error bars. Thus, we recommend revision of this data.

In addition, an apparent deviation in the ionization potential is noticed at $Z=74$. This number, corresponding to a tungsten heavy ion (W$^{+68}$), was collected by NIST compilers Kramida and Reader with the help of  a semi-empirical approach \cite{W-ions1}. Comparison with the average curve suggests that the reported value for W$^{+68}$ is overestimated in 1.577 a.u. 

\subsection{The N-like sequence ($N=7$)}
\label{N}

RPT coefficients:
\begin{eqnarray}
\nonumber a_2 &=& 0.125 + 0.0078125~(Z/137.036)^2,\\
\nonumber a_1 &=& -1.10915 -0.152346~(Z/137.036)^2,\\
\nonumber a_0 &=& 2.26192,\\
\nonumber a_{-1} &=& -0.356612.
\label{n_coeff}
\end{eqnarray}

Conditions at $Z=N-1$:
\begin{eqnarray}
\nonumber E_a({\rm C}) &=& 0.0463657,\\
s &=& 0.400642.
\label{rest_n}
\end{eqnarray}

The slope was computed from $E_a({\rm C})$ and $R_{cov}({\rm C})=1.417295$.

\begin{center}
\begin{figure}[!ht]
\includegraphics[width=0.9\linewidth,angle=0]{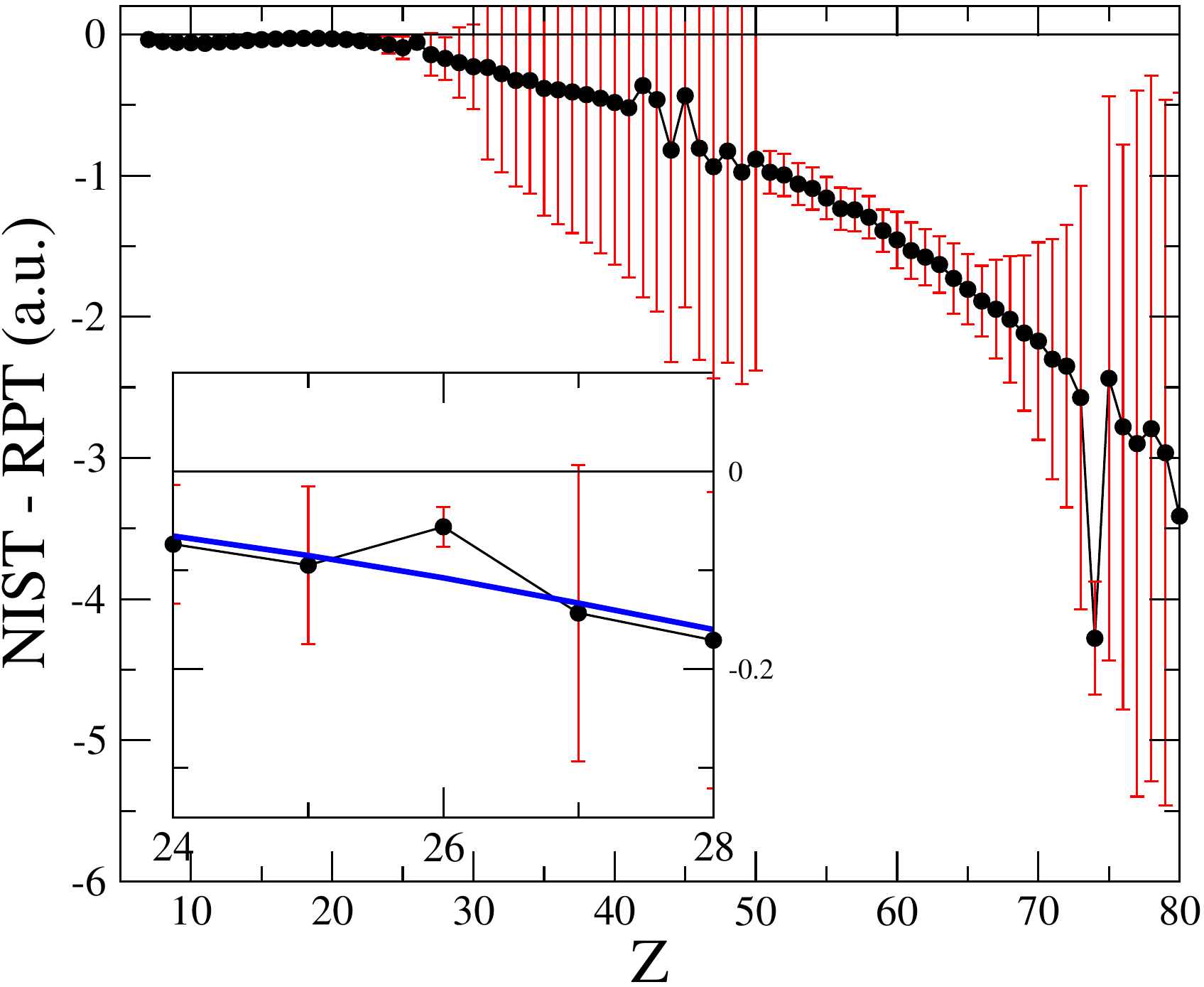}
\caption{\label{fig7} (Color online) The N-like ions ($N=7$). Deviations at $Z=26$ and 74, and a strong dispersion of points in the region $Z=42-50$ are noticed. However, the latter is consistent with the relatively high error bars.}
\end{figure}
\end{center}

\textbf{Discussion:}

The N-like sequence (Fig. \ref{fig7}) presents noticeable jumps at $Z=26$ and, once more, at $Z=74$. 

The ionization potential for $Z=26$ was taken from the paper by Sugar and Corliss.\cite{Sugar} According to our procedure, an overestimation of 0.051 a.u. is noticed.

The $Z=74$ case is again taken from Ref. [\onlinecite{W-ions1}]. It seems to be an inconsistent point, underestimated by $1.395$ a.u.  

Data for $Z=42-50$ come from the Dirac-Fock calculations by Bi\'emont et al. \cite{Interpol_N42-50} The reported large uncertainties in the data, of around $1.5$ a.u., are consistent with the observed deviations. 

Besides, the great dispersion for $Z \ge 96$, questions the consistency of Dirac-Fock calculations by Rodrigues et al.\cite{Dirac-Fock1} for highly charged ions. Notice, however, that deviations are within error bars, which are remarkably high (from $2.0$ to $20$ a.u.) for $Z > 74$.    

\subsection{The F-like sequence ($N=9$)}
\label{F}

RPT coefficients:
\begin{eqnarray}
\nonumber a_2 &=& 0.125 + 0.0078125~(Z/137.036)^2,\\
\nonumber a_1 &=& -1.4654 -0.181681~(Z/137.036)^2,\\
\nonumber a_0 &=& 2.79034,\\
\nonumber a_{-1} &=& 7.91836.
\label{f_coeff}
\end{eqnarray}

Conditions at $Z=N-1$:
\begin{eqnarray}
\nonumber E_a({\rm O}) &=& 0.0536759,\\
s &=& 0.410681.
\label{rest_f}
\end{eqnarray}

The slope was computed from $E_a({\rm O})$ and $R_{cov}({\rm O})=1.417295$.

\begin{center}
\begin{figure}[!ht]
\includegraphics[width=0.9\linewidth,angle=0]{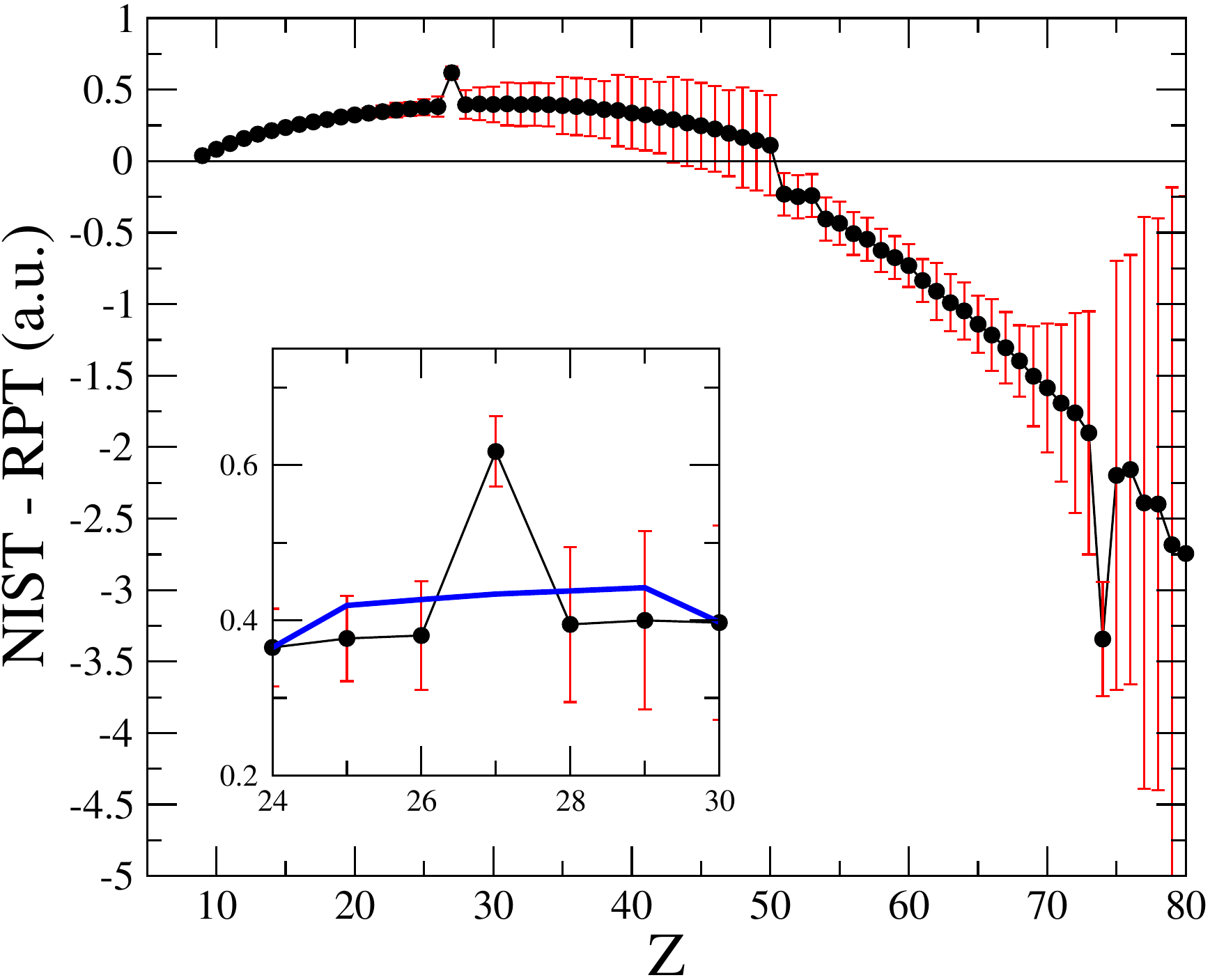}
\caption{\label{fig9} (Color online) The F-like sequence ($N=9$). There are remarkable inconsistencies  at $Z=27$ and 74. The jump at $Z=50-51$ is within error bars.}
\end{figure}
\end{center}

\textbf{Discussion:}

In the F-like sequence (Fig. \ref{fig9}), high deviations are noticed for $Z=27$ and 74. $I_p$ for $Z=27$ comes from Ref. [\onlinecite{Sugar}], the very same paper by Sugar and Corliss cited in the N-like case. According to our Fig., it seems to be overestimated in 0.183 a.u.

On the other hand, $Z=74$ belongs to the case of tungsten ions (W$^{+65}$), and the $I_p$ value is, once more, taken from Ref. [\onlinecite{W-ions1}]. In this case, an underestimation of 1.071 a.u. is apparent.

Finally, the observed jump at $Z=50-51$ is consistent with the reported error bars, and there is also a  great dispersion of the data for $Z > 100$, coming again from the Dirac-Fock calculations of Ref. [\onlinecite{Dirac-Fock1}]. It is also worth mentioning the large error bars accompanying the data for $Z \ge 80$.

\subsection{The Ne-like sequence ($N=10$)}
\label{Ne}

RPT coefficients:
\begin{eqnarray}
\nonumber a_2 &=& 0.125 + 0.0078125~(Z/137.036)^2,\\
\nonumber a_1 &=& -1.63649 -0.195621~(Z/137.036)^2,\\
\nonumber a_0 &=& 4.13824,\\
\nonumber a_{-1} &=& 5.3556.
\label{ne_coeff}
\end{eqnarray}

Conditions at $Z=N-1$:
\begin{eqnarray}
\nonumber E_a({\rm F}) &=& 0.124985,\\
s &=& 0.547149.
\label{rest_ne}
\end{eqnarray}

The slope was computed from $E_a({\rm F})$ and $R_{cov}({\rm F})=1.13384$.

\textbf{Discussion:}

In the Ne-like sequence (Fig. \ref{fig10}), deviations at $Z = 21$, 22, and 27 are noticed. Points are related to Ref. [\onlinecite{Sugar}]. We suggest correcting these values in -0.034, +0.012 and +0.036 a.u., respectively.

The $Z = 74$ case, \cite{W-ions1} is also clearly inconsistent. The ionization potential seems to be underestimated in 1.077 a.u.

The previously discussed abrupt jump at $Z =50- 51$, should be revised by the NIST team because the jump is greater than the data uncertainty. 

Finally, the great dispersion of points for $Z \ge 101$, related to the Dirac-Fock calculations in Ref. \onlinecite{Dirac-Fock1}, are within uncertainty bars. 

\section{Third row elements}

\subsection{The Mg-like sequence ($N=12$)}
\label{Mg}

RPT coefficients:
\begin{eqnarray}
\nonumber a_2 &=& 0.0555556 + 0.0138889~(Z/137.036)^2,\\
\nonumber a_1 &=& -0.931477 -0.0906944~(Z/137.036)^2,\\
\nonumber a_0 &=& 2.62996,\\
\nonumber a_{-1} &=& 10.0077.
\label{mg_coeff}
\end{eqnarray}

Conditions at $Z=N-1$:
\begin{eqnarray}
\nonumber E_a({\rm Na}) &=& 0.0201287,\\
s &=& 0.209421.
\label{rest_mg}
\end{eqnarray}

The slope was computed from $E_a({\rm Na})$ and $R_{cov}({\rm Na})=3.023562$.

\begin{center}
\begin{figure}[!ht]
\includegraphics[width=0.9\linewidth,angle=0]{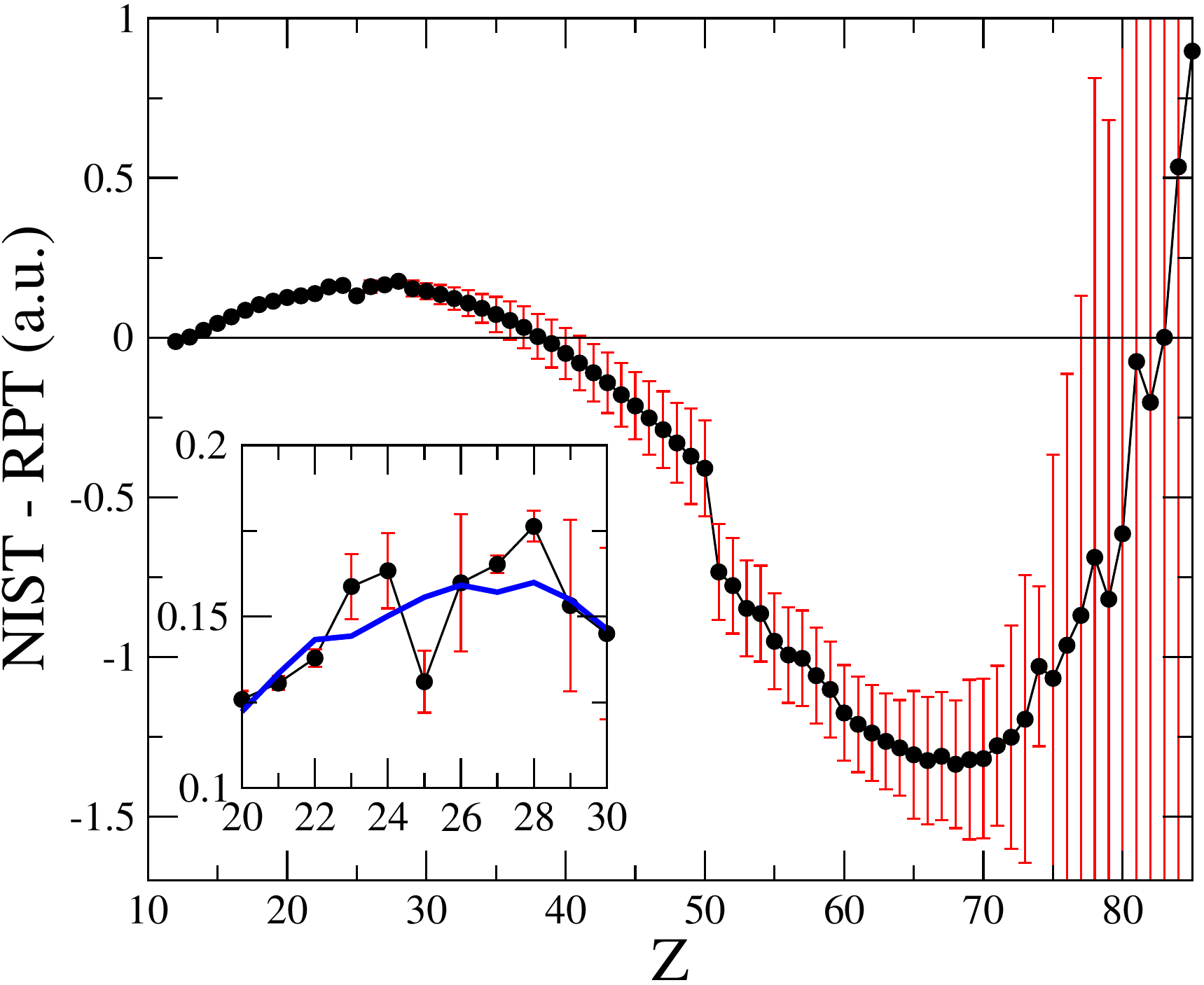}
\caption{\label{fig12} (Color online) Mg-like systems ($N=12$). Inconsistencies at $Z=23$, 24, 25, 27 and 28, and a jump at $Z =50- 51$ are apparent.}
\end{figure}
\end{center}

\textbf{Discussion:}

The Mg-like sequence (Fig. \ref{fig12}) of ions shows clear deviations at $Z=23$, 24, 25, 27 and 28.\cite{Sugar} The figure suggests corrections of -0.015, -0.013, +0.026, -0.008 and -0.016 a.u. to these points, respectively. 

The jump at $Z=50- 51$ was discussed above. It is clearly not consistent with error bars, thus we suggest revision of these data.

\subsection{The Si-like sequence ($N=14$)}
\label{Si}

RPT coefficients:
\begin{eqnarray}
\nonumber a_2 &=& 0.0555556 + 0.0138889~(Z/137.036)^2,\\
\nonumber a_1 &=& -1.13489 -0.148767~(Z/137.036)^2,\\
\nonumber a_0 &=& 4.44303,\\
\nonumber a_{-1} &=& 12.1396.
\label{si_coeff}
\end{eqnarray}

\begin{center}
\begin{figure}[!ht]
\includegraphics[width=0.9\linewidth,angle=0]{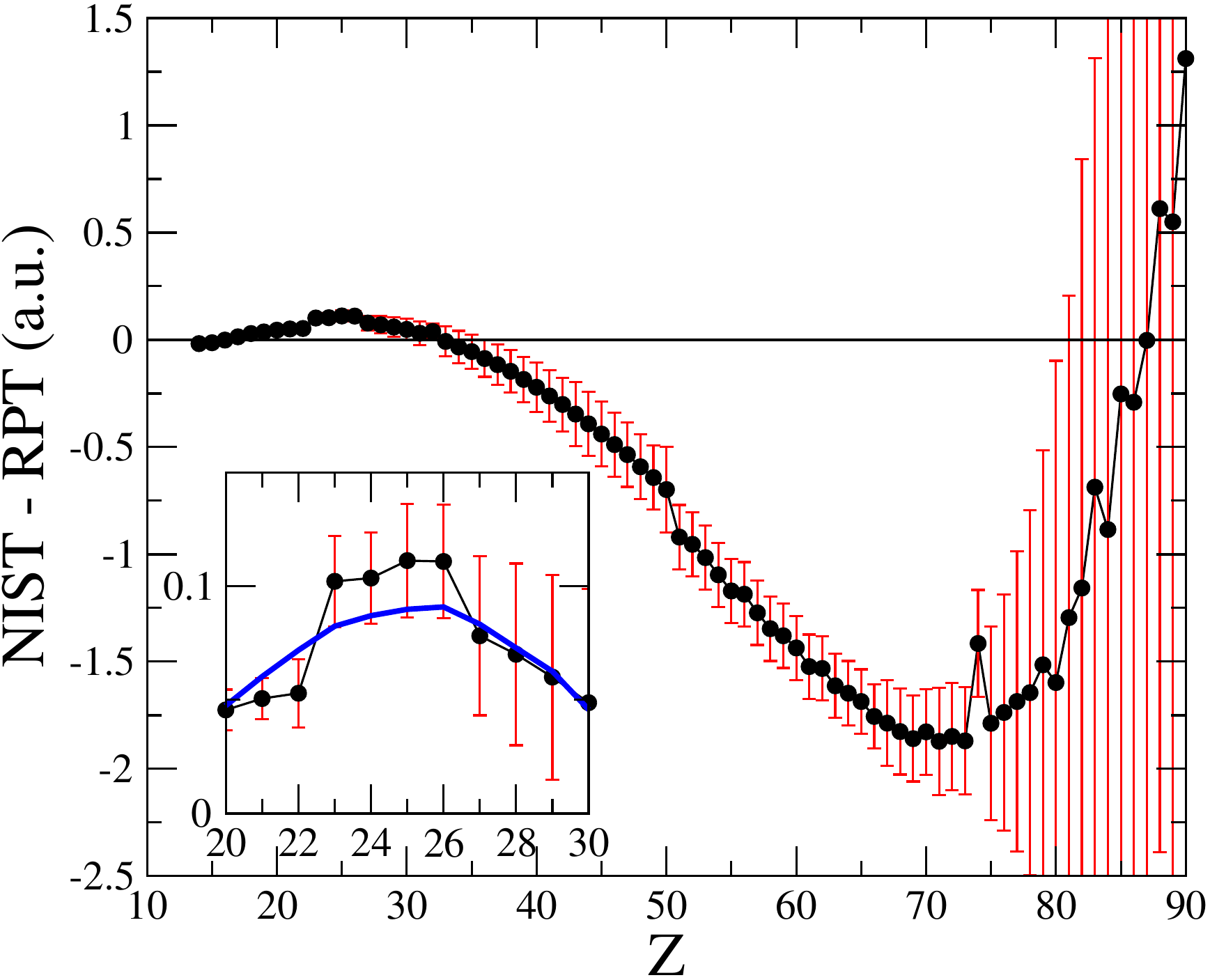}
\caption{\label{fig14} (Color online) The Si-like sequence ($N=14$). Deviations at $Z =21-26$ and Z=74 are remarkable.}
\end{figure}
\end{center}

Conditions at $Z=N-1$:
\begin{eqnarray}
\nonumber E_a({\rm Al}) &=& 0.0159006,\\
s &=& 0.239636.
\label{rest_si}
\end{eqnarray}

The slope was computed from $E_a({\rm Al})$ and $R_{cov}({\rm Al})=2.343260$.

\textbf{Discussion:}

It seems that there is a group of four points, $Z=23-26$, out of the general trend. Thus, we use a 7-points running average (instead of 5-points) in order to create a smooth curve. We suggest using the top of the error bars as values for $Z=21-22$, and the bottom of the error bars for $Z=23-26$. 

The $Z=74$ point, on the other hand, should be corrected in -0.329 a.u.

\subsection{The P-like sequence ($N=15$)}
\label{P}

RPT coefficients:
\begin{eqnarray}
\nonumber a_2 &=& 0.0555556 + 0.00462963~(Z/137.036)^2,\\
\nonumber a_1 &=& -1.19956 -0.109484~(Z/137.036)^2,\\
\nonumber a_0 &=& 5.24386,\\
\nonumber a_{-1} &=& 10.0615.
\label{p_coeff}
\end{eqnarray}

Conditions at $Z=N-1$:
\begin{eqnarray}
\nonumber E_a({\rm Si}) &=& 0.0510817,\\
s &=& 0.304873.
\label{rest_p}
\end{eqnarray}

The slope was computed from $E_a({\rm Si})$ and $R_{cov}({\rm Si})=2.154288$.

\begin{center}
\begin{figure}[!ht]
\includegraphics[width=0.9\linewidth,angle=0]{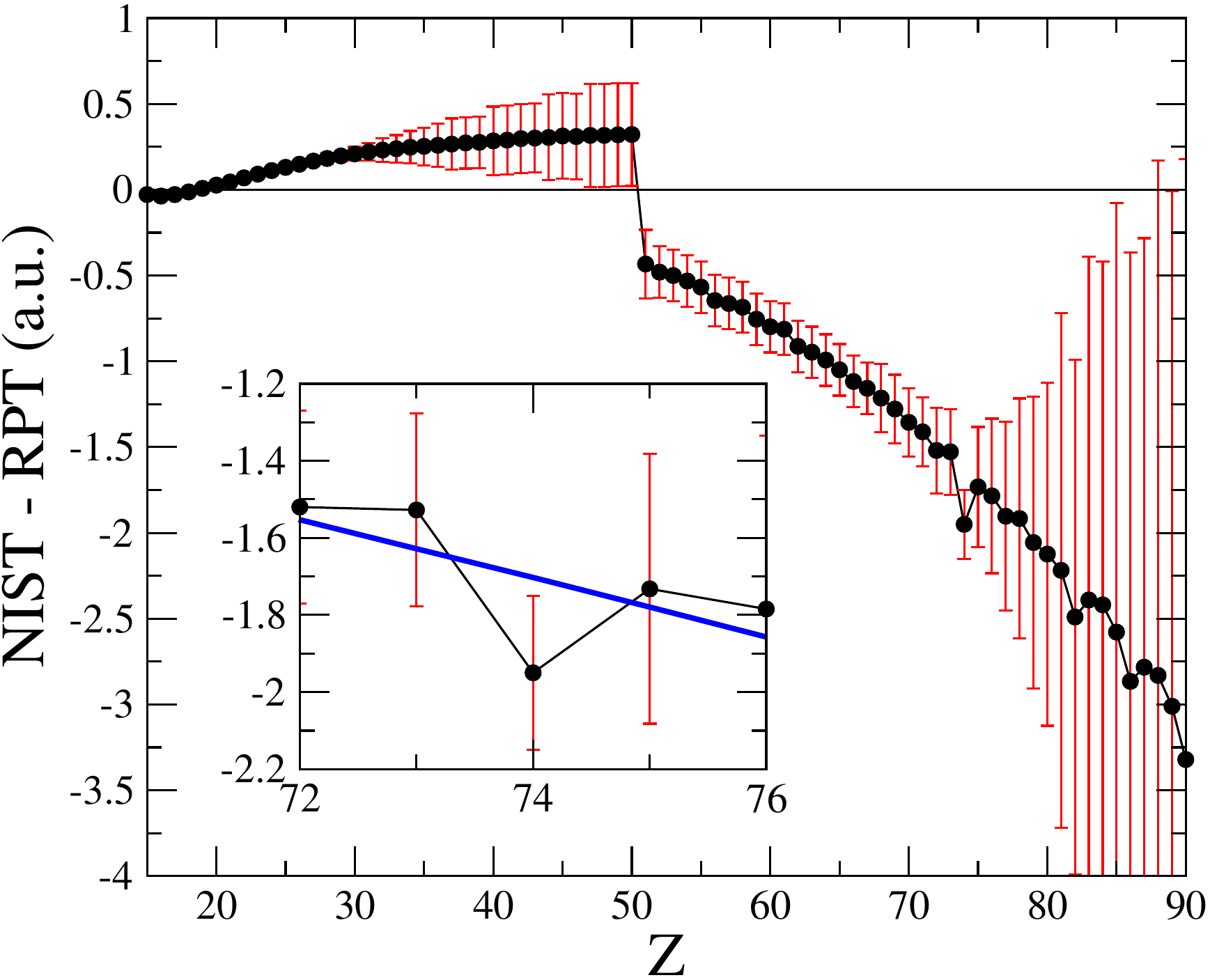}
\caption{\label{fig15} (Color online) The P-like systems ($N=15$). A big jump at $Z =50- 51$, and a  noticeable deviation at $Z=74$ are apparent.}
\end{figure}
\end{center}

\textbf{Discussion:}

We suggest a revision of the data near $Z=50$, and correcting the $Z=74$ value in + 0.247 a.u.

\subsection{The S-like sequence ($N=16$)}
\label{S}

RPT coefficients:
\begin{eqnarray}
\nonumber a_2 &=& 0.0555556 + 0.00462963~(Z/137.036)^2,\\
\nonumber a_1 &=& -1.26711 -0.114919~(Z/137.036)^2,\\
\nonumber a_0 &=& 4.83765,\\
\nonumber a_{-1} &=& 25.569.
\label{s_coeff}
\end{eqnarray}

\begin{center}
\begin{figure}[!ht]
\includegraphics[width=0.9\linewidth,angle=0]{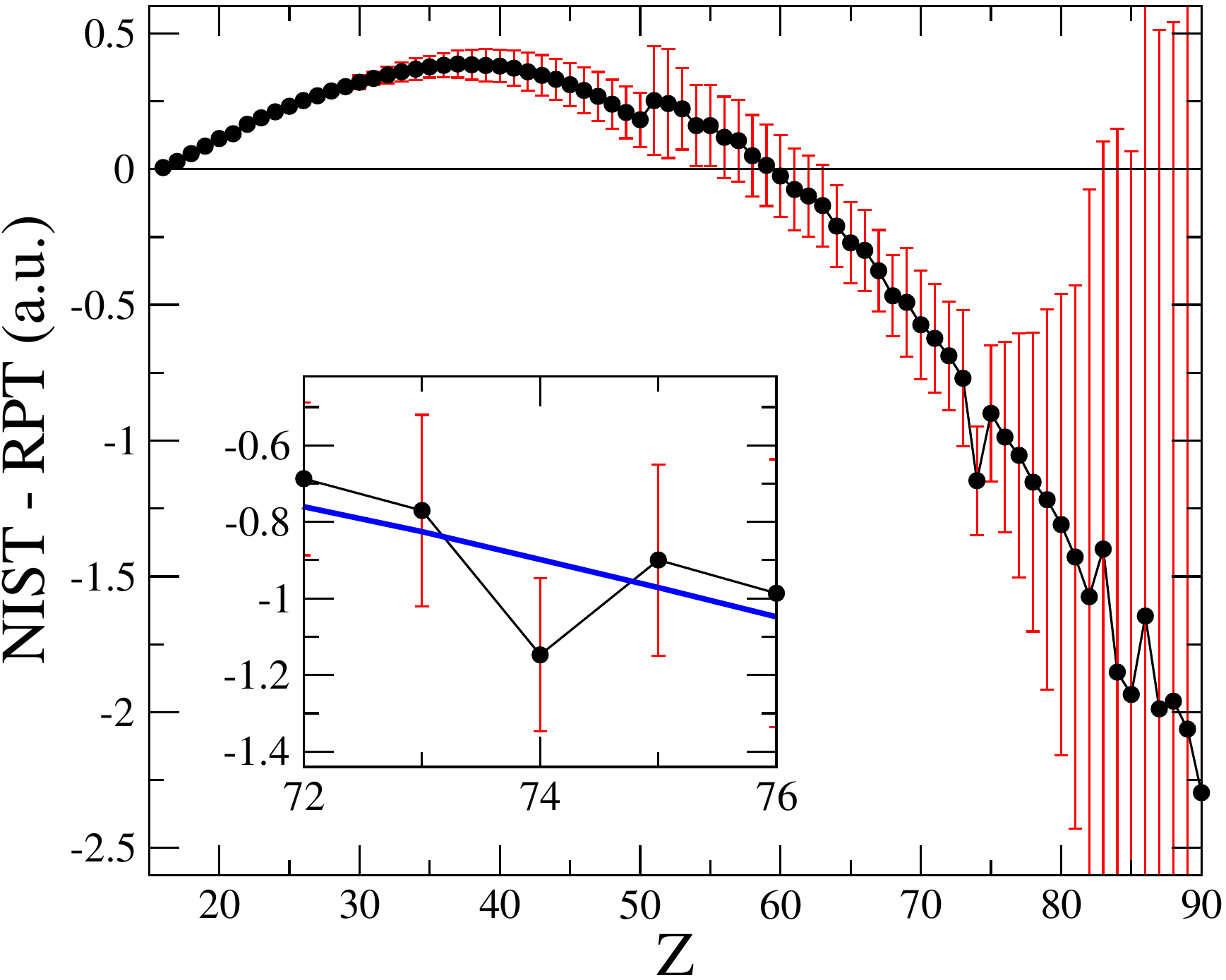}
\caption{\label{fig16} (Color online) The S-like ions ($N=16$). The only detected inconsistency is the deviation at $Z=74$.}
\end{figure}
\end{center}

Conditions at $Z=N-1$:
\begin{eqnarray}
\nonumber E_a({\rm P}) &=& 0.0274519,\\
s &=& 0.286207.
\label{rest_s}
\end{eqnarray}

The slope was computed from $E_a({\rm P})$ and $R_{cov}({\rm P})=2.0598015$.

\textbf{Discussion:}

Although error bars are relatively high in this case, we suggest a correction of + 0.249 a.u. to the $Z=74$ value of $I_p$.

\subsection{The Cl-like sequence ($N=17$)}
\label{Cl}

RPT coefficients:
\begin{eqnarray}
\nonumber a_2 &=& 0.0555556 + 0.00462963~(Z/137.036)^2,\\
\nonumber a_1 &=& -1.34042 -0.121136~(Z/137.036)^2,\\
\nonumber a_0 &=& 5.81594,\\
\nonumber a_{-1} &=& 23.9212.
\label{cl_coeff}
\end{eqnarray}

Conditions at $Z=N-1$:
\begin{eqnarray}
\nonumber E_a({\rm S}) &=& 0.0763283, \\
s &=& 0.344289.
\label{rest_cl}
\end{eqnarray}

The slope was computed from $E_a({\rm S})$ and $R_{cov}({\rm S})=1.965315$.

\begin{center}
\begin{figure}[!ht]
\includegraphics[width=0.9\linewidth,angle=0]{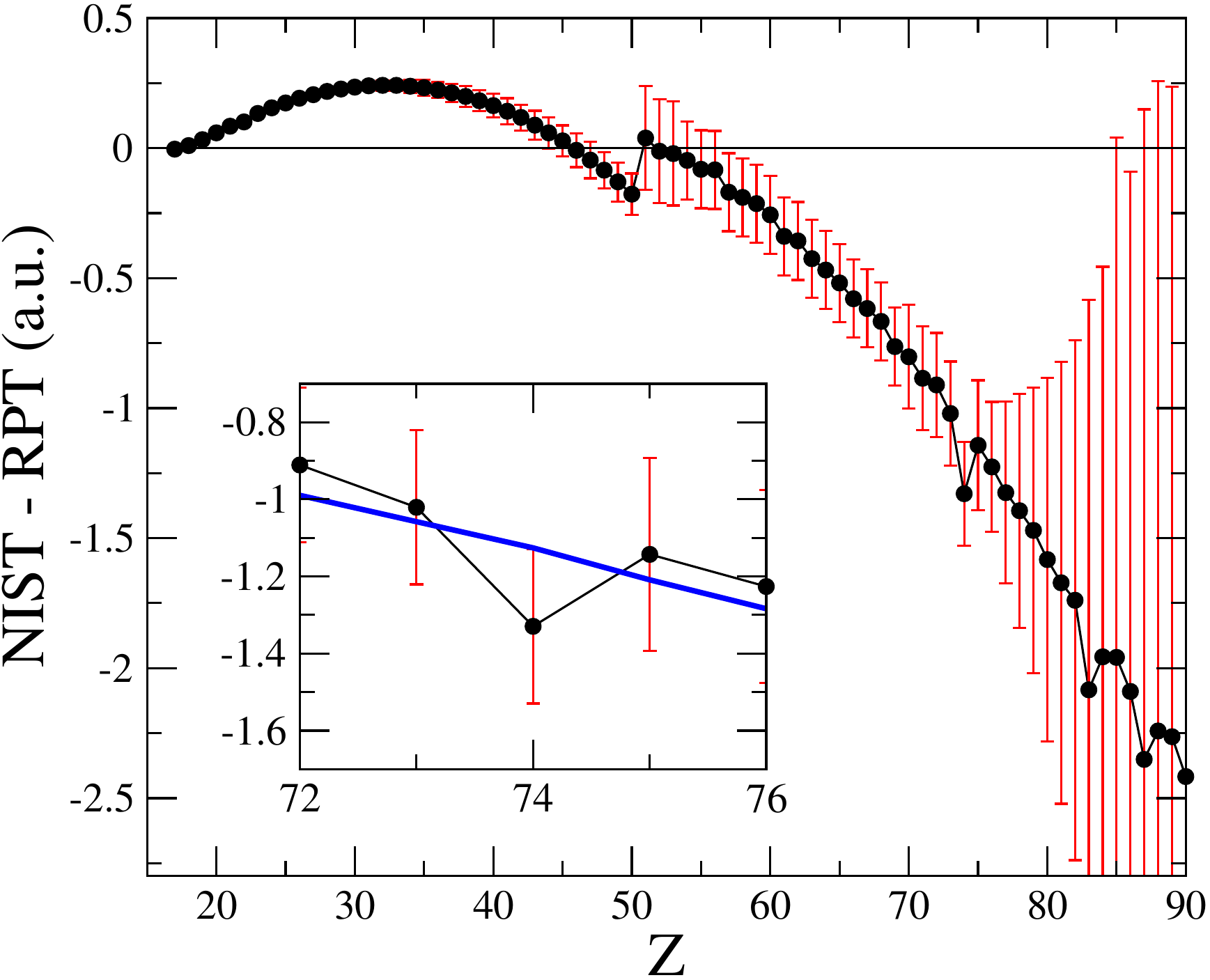}
\caption{\label{fig17} (Color online) Cl-like systems ($N=17$). The jump at $Z=50- 51$, and the deviation at $Z=74$ are noticeable.}
\end{figure}
\end{center}

\textbf{Discussion:}

Although error bars are relatively high in this case, we suggest a revision of the data near $Z=50$, which show a jump of 0.216 a.u., and using in the $Z=74$ case the top value of the error bar.

\subsection{The Ar-like sequence ($N=18$)}
\label{Ar}

\begin{center}
\begin{figure}[!ht]
\includegraphics[width=0.9\linewidth,angle=0]{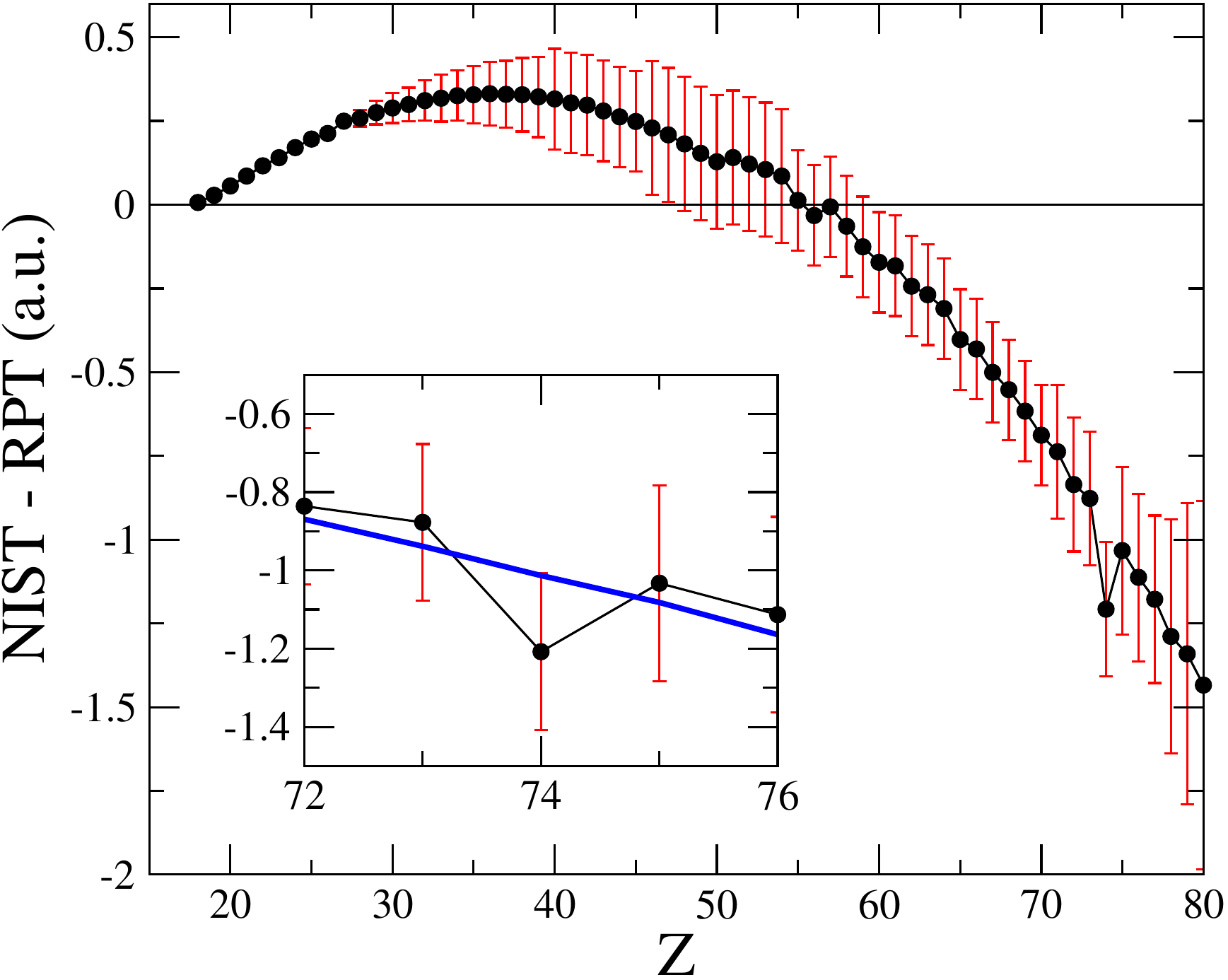}
\caption{\label{fig18} (Color online) The Ar-like systems ($N=18$). In this case, we distinguish only the deviation at $Z=74$.\cite{W-ions1}}
\end{figure}
\end{center}

RPT coefficients:
\begin{eqnarray}
\nonumber a_2 &=& 0.0555556 + 0.00462963~ (Z/137.036)^2,\\
\nonumber a_1 &=& -1.40796 -0.12657~ (Z/137.036)^2,\\
\nonumber a_0 &=& 6.30212,\\
\nonumber a_{-1} &=& 29.2914.
\label{ar_coeff}
\end{eqnarray}

Conditions at $Z=N-1$:
\begin{eqnarray}
\nonumber E_a({\rm Cl}) &=& 0.132775, \\
s &=& 0.380045.
\label{rest_ar}
\end{eqnarray}

The slope was computed from $E_a({\rm Cl})$ and $R_{cov}({\rm Cl})=1.889726$.

\textbf{Discussion:}

Although error bars are relatively high, we suggest using the top of the error bar as the $Z=74$ value of $I_p$.

\section{Four row elements}

\subsection{The Ca-like sequence ($N=20$)}
\label{Ca}

RPT coefficients:
\begin{eqnarray}
\nonumber a_2 &=& 0.0555556 + 0.00462963~ (Z/137.036)^2,\\
\nonumber a_1 &=& -1.70312 -0.200236~ (Z/137.036)^2,\\
\nonumber a_0 &=& 7.96996,\\
\nonumber a_{-1} &=& 83.4702.
\label{ca_coeff}
\end{eqnarray}

\begin{center}
\begin{figure}[!ht]
\includegraphics[width=0.9\linewidth,angle=0]{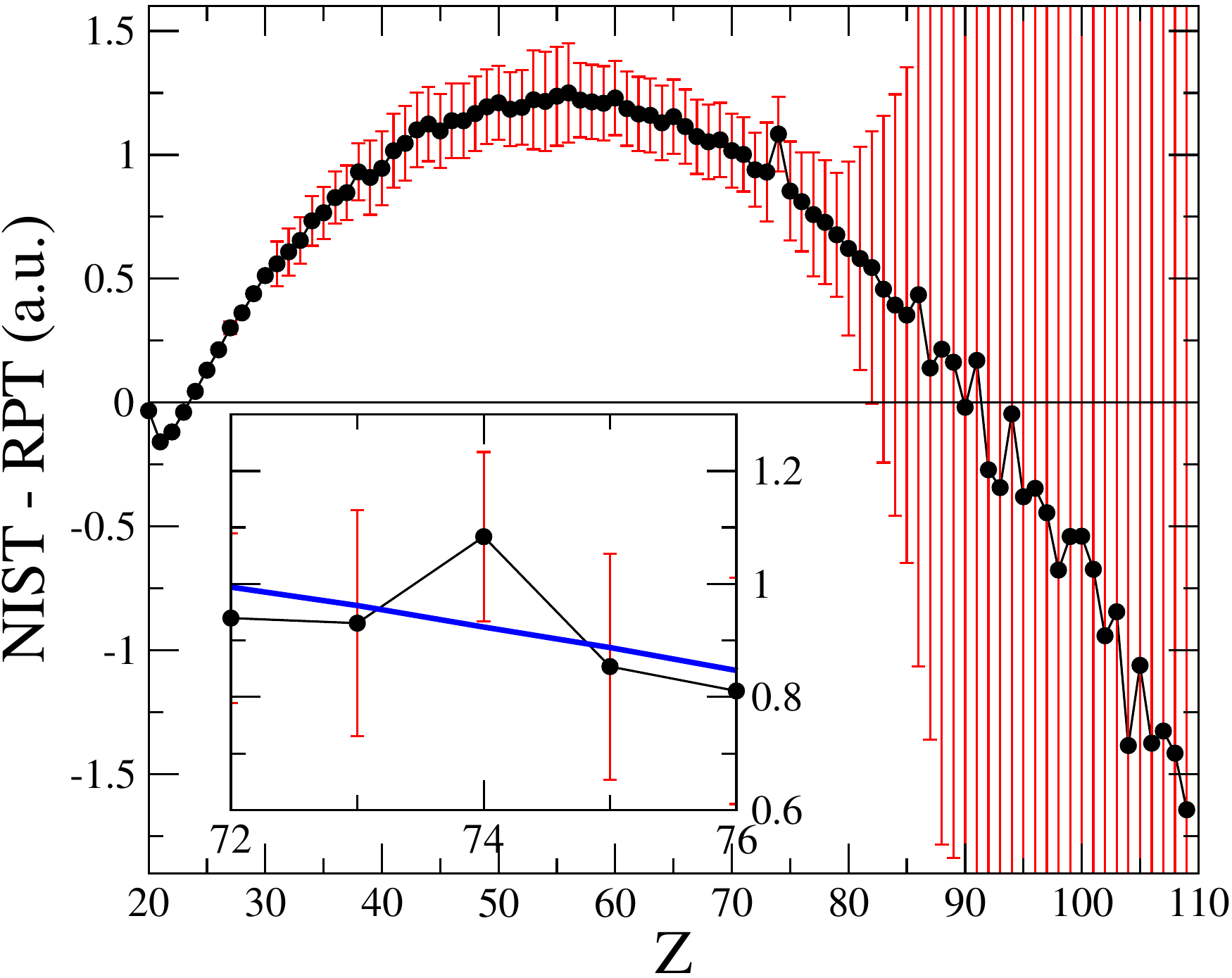}
\caption{\label{fig20} (Color online) The Ca-like systems ($N=20$). Only the $Z=74$ point is distinguished.}
\end{figure}
\end{center}

Conditions at $Z=N-1$:
\begin{eqnarray}
\nonumber E_a({\rm K}) &=& 0.0184220, \\
s &=& 0.176307.
\label{rest_ca}
\end{eqnarray}

The slope was computed from $E_a({\rm K})$ and $R_{cov}({\rm K})=3.779452$.

\textbf{Discussion:}

The Ca-like sequence is the first with a rearrangement of the electronic spectrum with the increase of $Z$. For $Z\approx N$ the last two electrons occupy the 4s subshell, whereas for larger $Z$ they move to the 3d orbital. The observed jump at $Z=20-21$ is surely related to this fact. 

On the other hand, we suggest using for $I_p$ at $Z=74$ the bottom of its error bar.

\subsection{The Sc-like sequence ($N=21$)}
\label{Sc}

RPT coefficients:
\begin{eqnarray}
\nonumber a_2 &=& 0.0555556 + 0.00462963~ (Z/137.036)^2,\\
\nonumber a_1 &=& -1.79098 -0.209669~ (Z/137.036)^2,\\
\nonumber a_0 &=& 7.18837,\\
\nonumber a_{-1} &=& 129.196.
\label{sc_coeff}
\end{eqnarray}

\begin{center}
\begin{figure}[!ht]
\includegraphics[width=0.9\linewidth,angle=0]{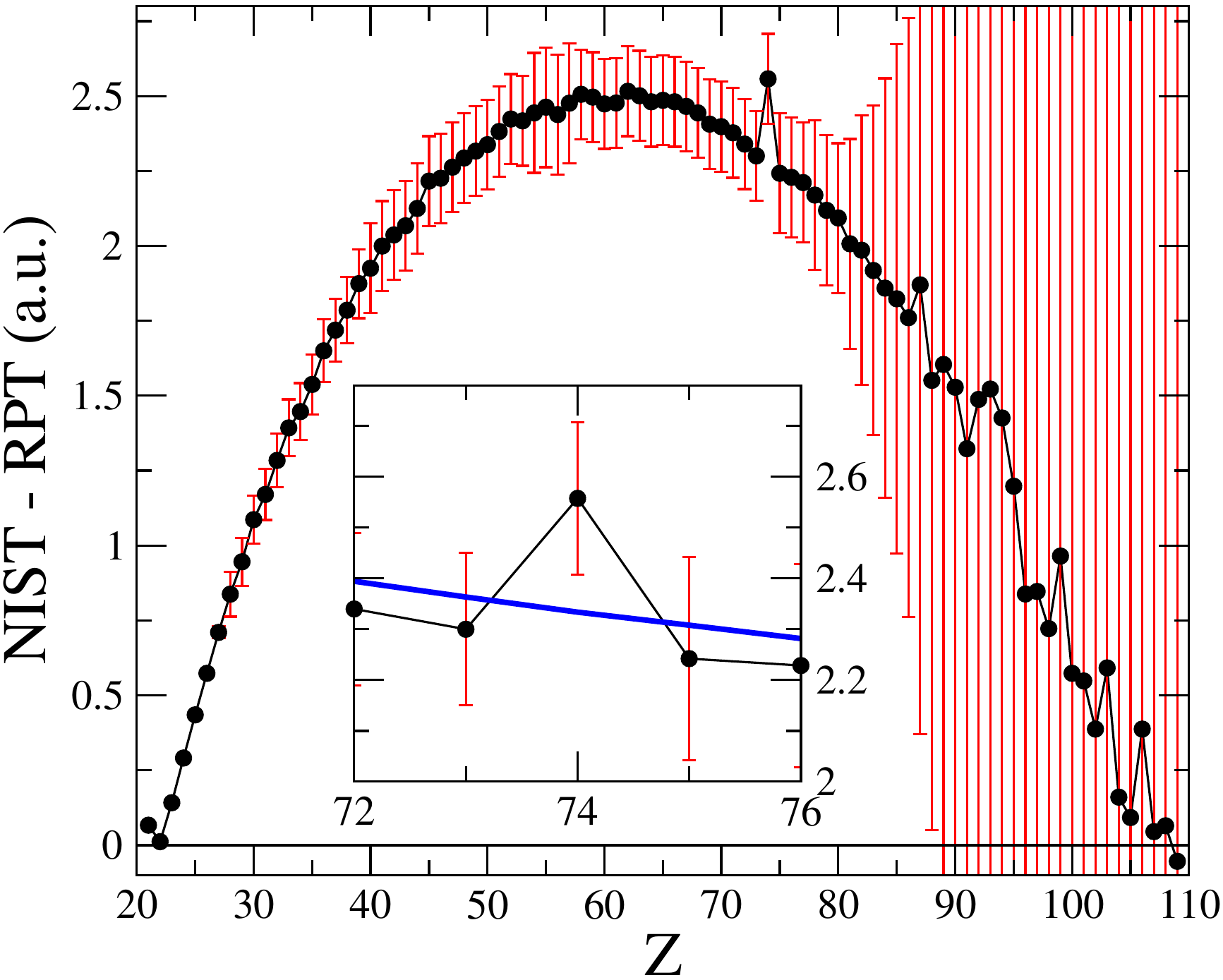}
\caption{\label{fig21} (Color online) The Sc-like systems ($N=21$). Only the $Z=74$ point is distinguished.}
\end{figure}
\end{center}

Conditions at $Z=N-1$:
\begin{eqnarray}
\nonumber E_a({\rm Ca}) &=& 0.000901924, \\
s &=& 0.107729.
\label{rest_sc}
\end{eqnarray}

The slope was computed from $E_a({\rm Ca})$ and $R_{cov}({\rm Ca})=3.288123$.

\textbf{Discussion:}

Very similar to the Ca sequence. We suggest correcting $I_p$ at $Z=74$ in -0.224 a.u.

\subsection{The Ti-like sequence ($N=22$)}
\label{Ti}

RPT coefficients:
\begin{eqnarray}
\nonumber a_2 &=& 0.0555556 + 0.00462963~ (Z/137.036)^2,\\
\nonumber a_1 &=& -1.87158 -0.218161~ (Z/137.036)^2,\\
\nonumber a_0 &=& 8.87906,\\
\nonumber a_{-1} &=& 125.803.
\label{ti_coeff}
\end{eqnarray}

Conditions at $Z=N-1$:
\begin{eqnarray}
\nonumber E_a({\rm Sc}) &=& 0.00690650, \\
s &=& 0.175933.
\label{rest_ti}
\end{eqnarray}

The slope was computed from $E_a({\rm Sc})$ and $R_{cov}({\rm Sc})=3.004665$.

\begin{center}
\begin{figure}[!ht]
\includegraphics[width=0.9\linewidth,angle=0]{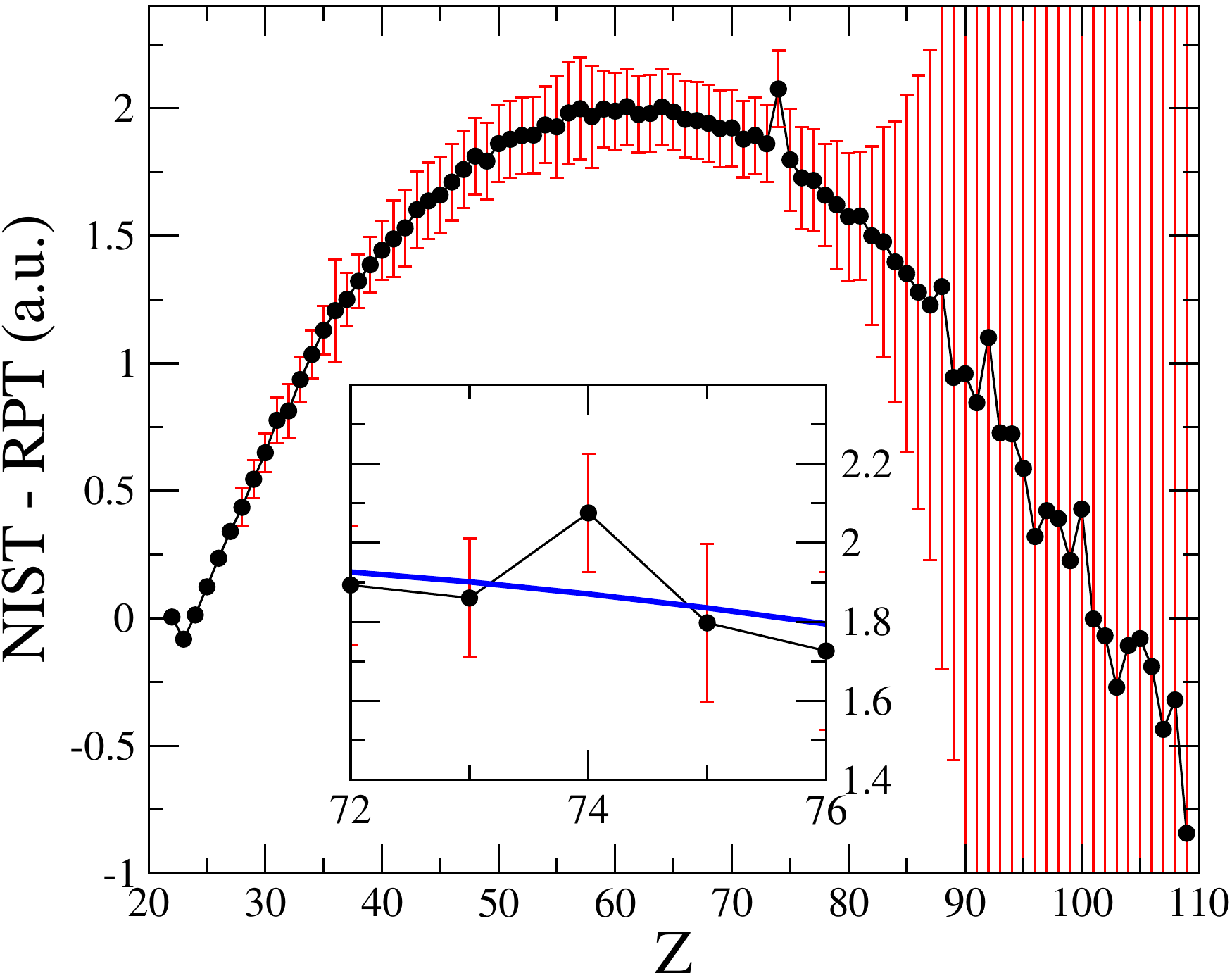}
\caption{\label{fig22} (Color online) Ti-like systems ($N=22$). Only the $Z=74$ point is distinguished.}
\end{figure}
\end{center}

\textbf{Discussion:}

Very similar to the previous sequences. We suggest correcting $I_p$ at $Z=74$ in -0.205 a.u.

\subsection{The V-like sequence ($N=23$)}
\label{V}

RPT coefficients:
\begin{eqnarray}
\nonumber a_2 &=& 0.0555556 + 0.00154321 (Z/137.036)^2,\\
\nonumber a_1 &=& -1.95847 -0.154589~ (Z/137.036)^2,\\
\nonumber a_0 &=& 8.95387,\\
\nonumber a_{-1} &=& 160.08.
\label{v_coeff}
\end{eqnarray}

\begin{center}
\begin{figure}[!ht]
\includegraphics[width=0.9\linewidth,angle=0]{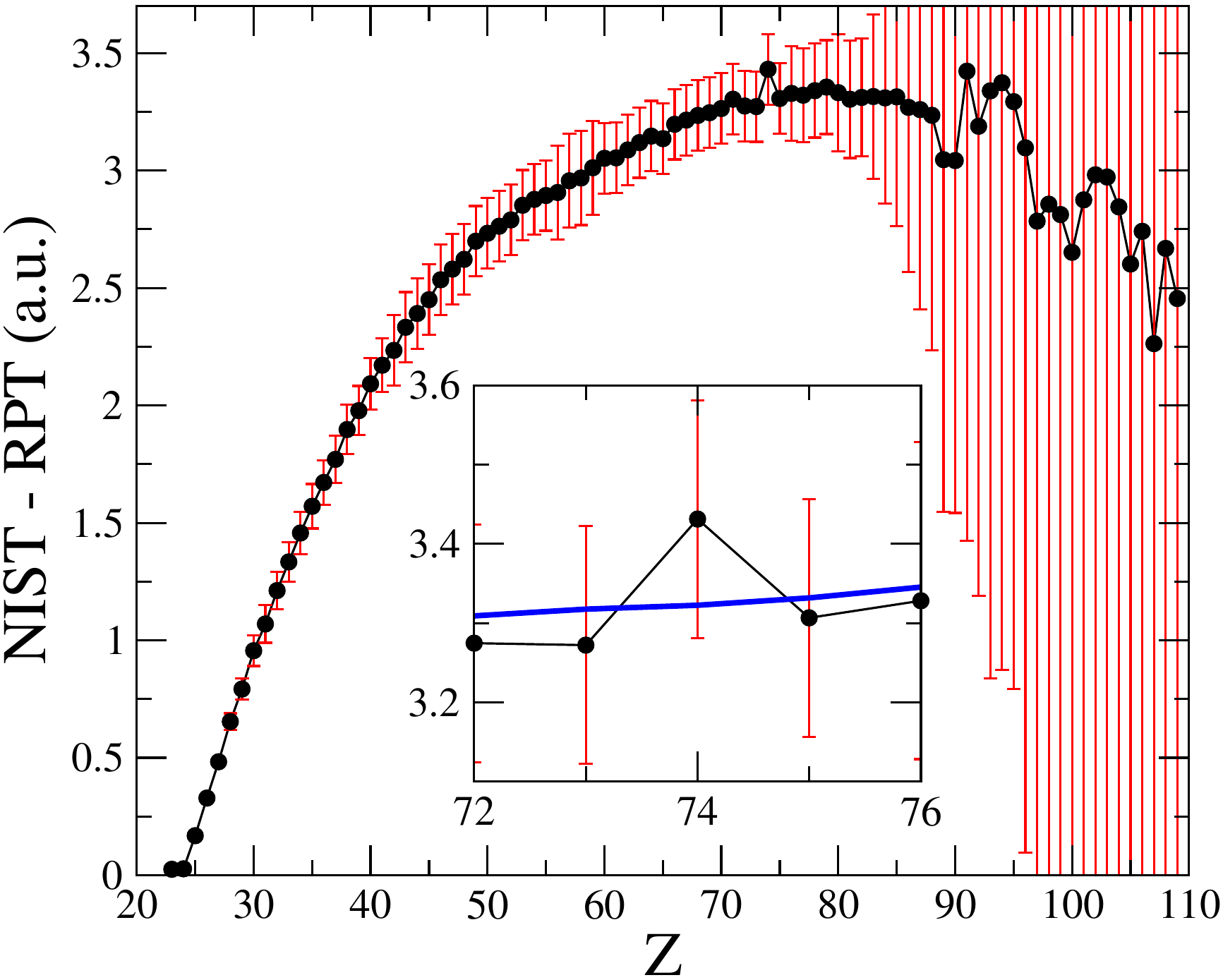}
\caption{\label{fig23} (Color online) The V-like ions ($N=23$). No inconsistencies were detected.}
\end{figure}
\end{center}

Conditions at $Z=N-1$:
\begin{eqnarray}
\nonumber E_a({\rm Ti}) &=& 0.00290230, \\
s &=& 0.156499.
\label{rest_v}
\end{eqnarray}

The slope was computed from $E_a({\rm Ti})$ and $R_{cov}({\rm Ti})=2.796794$.

\subsection{The Cr-like sequence ($N=24$)}
\label{Cr}

RPT coefficients:
\begin{eqnarray}
\nonumber a_2 &=& 0.0555556 + 0.00154321~ (Z/137.036)^2,\\
\nonumber a_1 &=& -2.03885 -0.15826~ (Z/137.036)^2,\\
\nonumber a_0 &=& 10.9246,\\
\nonumber a_{-1} &=& 153.618.
\label{cr_coeff}
\end{eqnarray}

\begin{center}
\begin{figure}[!ht]
\includegraphics[width=0.9\linewidth,angle=0]{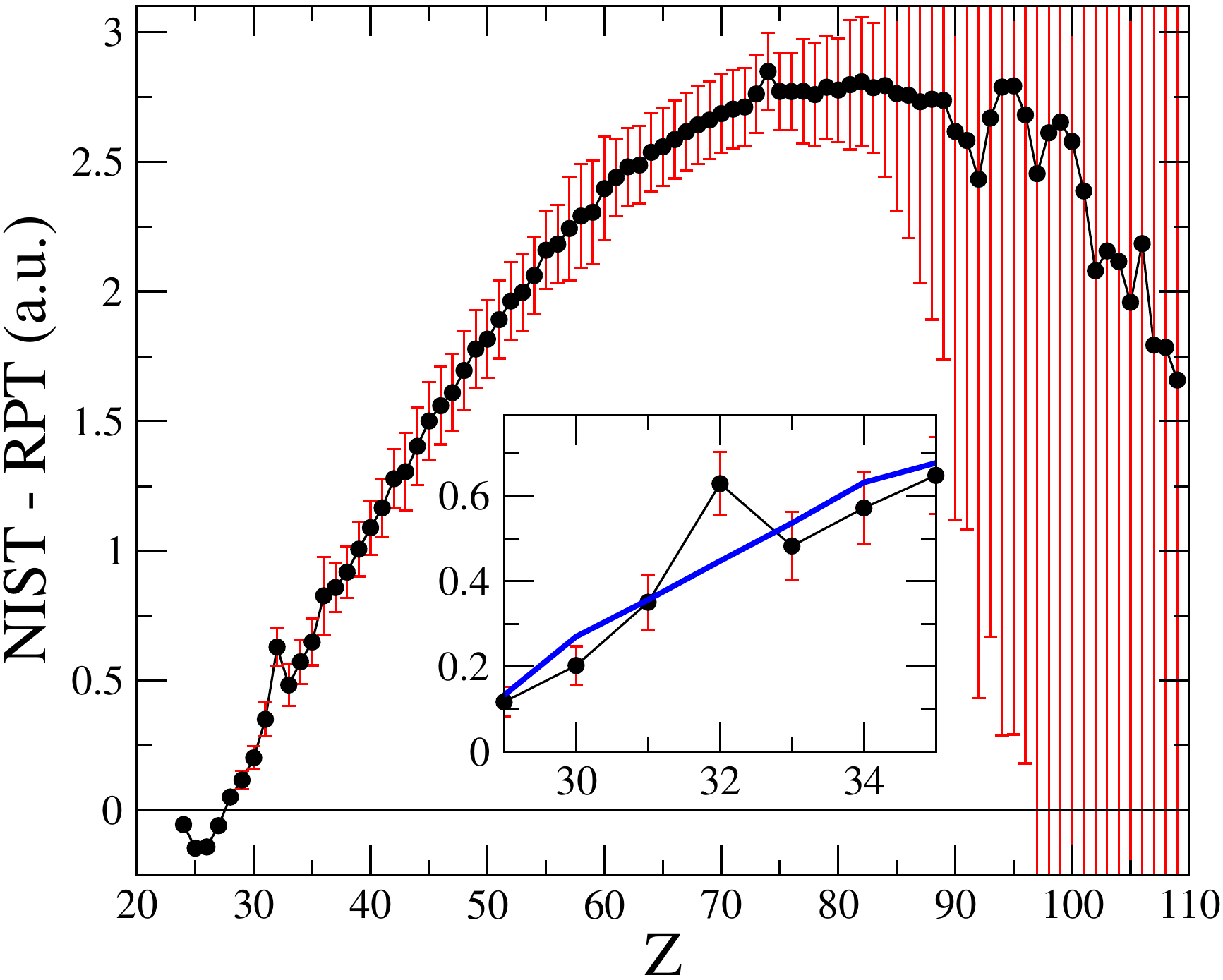}
\caption{\label{fig24} (Color online) The Cr-like sequence ($N=24$). An apparent deviation at $Z=32$
is detected.}
\end{figure}
\end{center}

Conditions at $Z=N-1$:
\begin{eqnarray}
\nonumber E_a({\rm V}) &=& 0.0192866, \\
s &=& 0.223850.
\label{rest_cr}
\end{eqnarray}

The slope was computed from $E_a({\rm V})$ and $R_{cov}({\rm V})=2.721206$.

\textbf{Discussion:}

In the Cr-like sequence, we shall distinguish the problematic point at $Z=32$, coming from the paper by Sugar and Musgrove \cite{Sugar1}. $I_p$ is overestimated in 0.181 a.u.

\subsection{The Mn-like sequence ($N=25$)}
\label{Mn}

RPT coefficients:
\begin{eqnarray}
\nonumber a_2 &=& 0.0555556 + 0.00154321~ (Z/137.036)^2,\\
\nonumber a_1 &=& -2.12264 -0.162119~ (Z/137.036)^2,\\
\nonumber a_0 &=& 12.0502,\\
\nonumber a_{-1} &=& 168.234.
\label{mn_coeff}
\end{eqnarray}

\begin{center}
\begin{figure}[!ht]
\includegraphics[width=0.9\linewidth,angle=0]{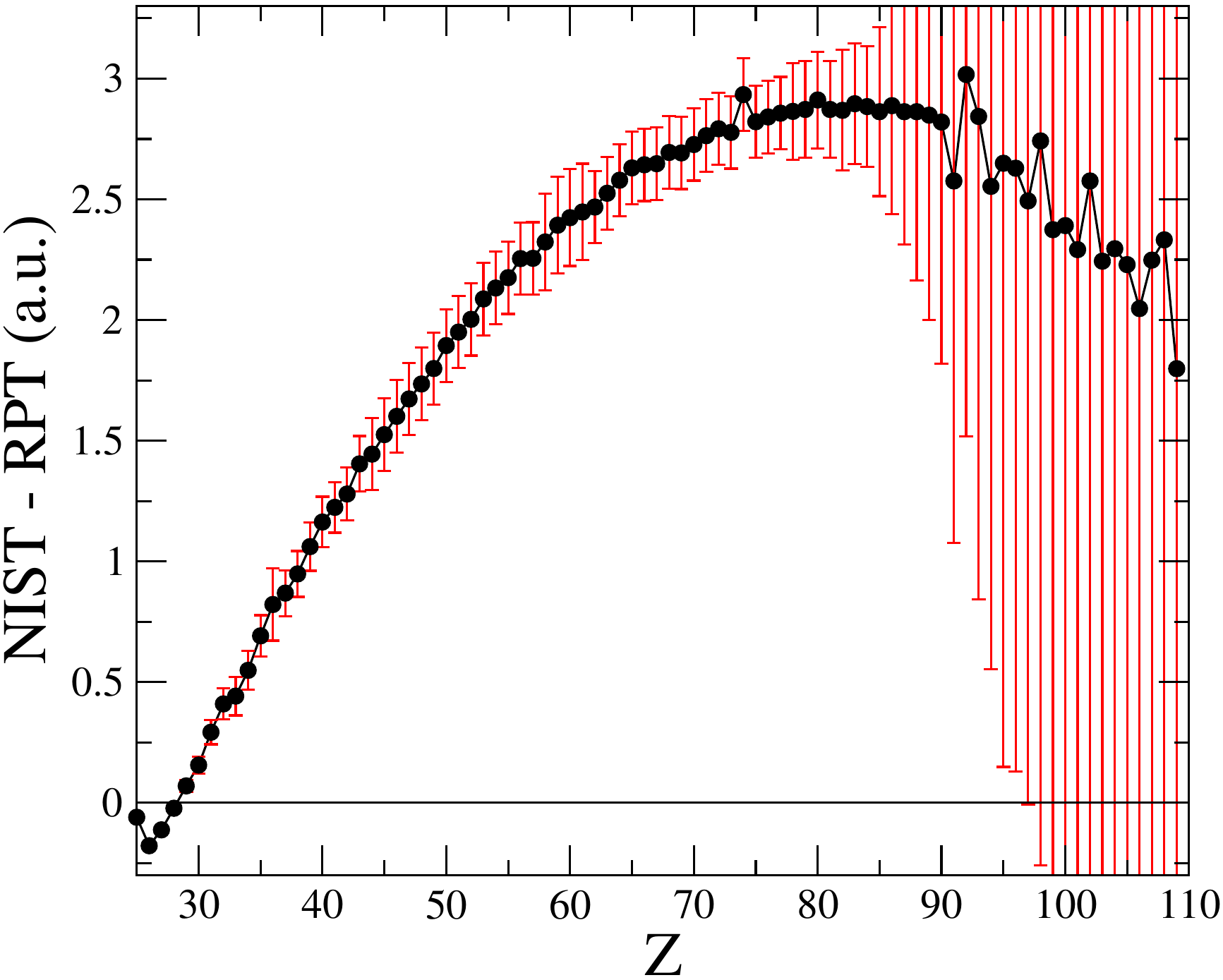}
\caption{\label{fig25} (Color online) The Mn-like systems ($N=25$). No inconsistency is detected.}
\end{figure}
\end{center}

Conditions at $Z=N-1$:
\begin{eqnarray}
\nonumber E_a({\rm Cr}) &=& 0.0244666, \\
s &=& 0.249254.
\label{rest_mn}
\end{eqnarray}

The slope was computed from $E_a({\rm Cr})$ and $R_{cov}({\rm Cr})=2.456644$.

\subsection{The Co-like sequence ($N=27$)}
\label{Co}

\begin{center}
\begin{figure}[!ht]
\includegraphics[width=0.9\linewidth,angle=0]{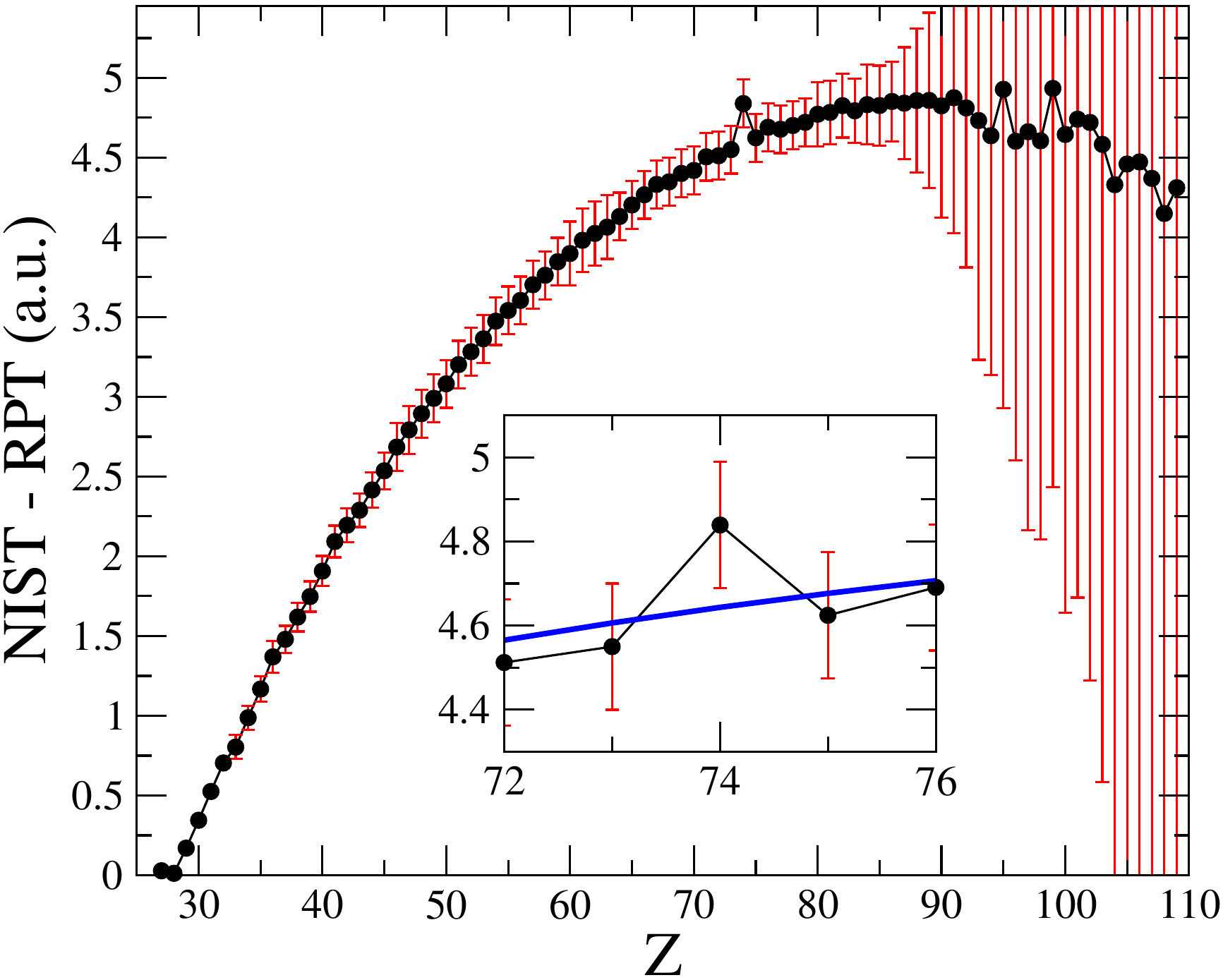}
\caption{\label{fig27} (Color online) The Co-like sequence ($N=27$). The only detected inconsistency is at $Z=74$.}
\end{figure}
\end{center}

RPT coefficients:
\begin{eqnarray}
\nonumber a_2 &=& 0.0555556 + 0.00154321~ (Z/137.036)^2,\\
\nonumber a_1 &=& -2.29511 -0.170117~ (Z/137.036)^2,\\
\nonumber a_0 &=& 11.9833,\\
\nonumber a_{-1} &=& 266.793.
\label{co_coeff}
\end{eqnarray}

Conditions at $Z=N-1$:
\begin{eqnarray}
\nonumber E_a({\rm Fe}) &=& 0.00554713, \\
s &=& 0.195878.
\label{rest_co}
\end{eqnarray}

The slope was computed from $E_a({\rm Fe})$ and $R_{cov}({\rm Fe})=2.343260$.

\textbf{Discussion:}

$I_p$ at $Z=74$ is overestimated in 0.196 a.u.

\subsection{The Ni-like sequence ($N=28$)}
\label{Ni}

RPT coefficients:
\begin{eqnarray}
\nonumber a_2 &=& 0.0555556 + 0.00154321~ (Z/137.036)^2,\\
\nonumber a_1 &=& -2.3755 -0.173789~ (Z/137.036)^2,\\
\nonumber a_0 &=& 14.2165,\\
\nonumber a_{-1} &=& 258.787.
\label{ni_coeff}
\end{eqnarray}

Conditions at $Z=N-1$:
\begin{eqnarray}
\nonumber E_a({\rm Co}) &=& 0.0243196, \\
s &=& 0.266002.
\label{rest_ni}
\end{eqnarray}

The slope was computed from $E_a({\rm Co})$ and $R_{cov}({\rm Co})=2.229877$.

\begin{center}
\begin{figure}[!ht]
\includegraphics[width=0.9\linewidth,angle=0]{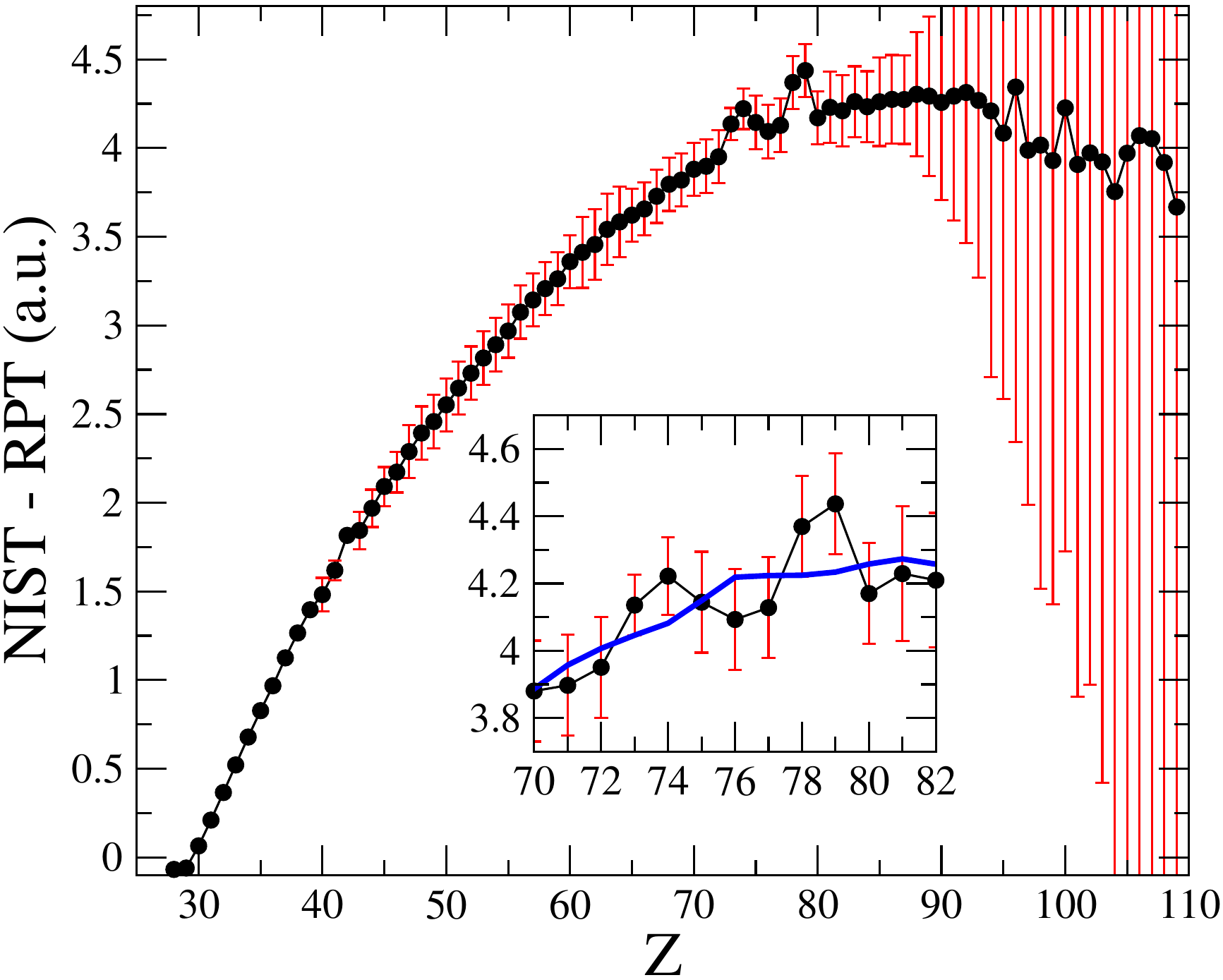}
\caption{\label{fig28} (Color online) The Ni-like systems ($N=28$). Inconsistencies are detected at $Z=42$, 74, and 79.}
\end{figure}
\end{center}

\textbf{Discussion:}

The Ni-like ions (Fig. \ref{fig28}) shows deviations at $Z=42$, 74 and 79, the latter coming from the paper by Tragin et al.\cite{Tragin}. 

As suggested by the figure, $I_p$ at $Z=42$ is overestimated in 0.071 a.u. Corrections of -0.141 and -0.203 a.u. should be added to the points at $Z=74$ and 79, respectively. 

\subsection{The Cu-like sequence ($N=29$)}
\label{Cu}

RPT coefficients:
\begin{eqnarray}
\nonumber a_2 &=& 0.03125 + 0.00634766~ (Z/137.036)^2,\\
\nonumber a_1 &=& -1.34713 -0.0620302~ (Z/137.036)^2,\\
\nonumber a_0 &=& 9.66505,\\
\nonumber a_{-1} &=& 96.9281.
\label{cu_coeff}
\end{eqnarray}

Conditions at $Z=N-1$:
\begin{eqnarray}
\nonumber E_a({\rm Ni}) &=& 0.0424673, \\
s &=& 0.291495.
\label{rest_cu}
\end{eqnarray}

The slope was computed from $E_a({\rm Ni})$ and $R_{cov}({\rm Ni})=2.210980$.

\begin{center}
\begin{figure}[!ht]
\includegraphics[width=0.9\linewidth,angle=0]{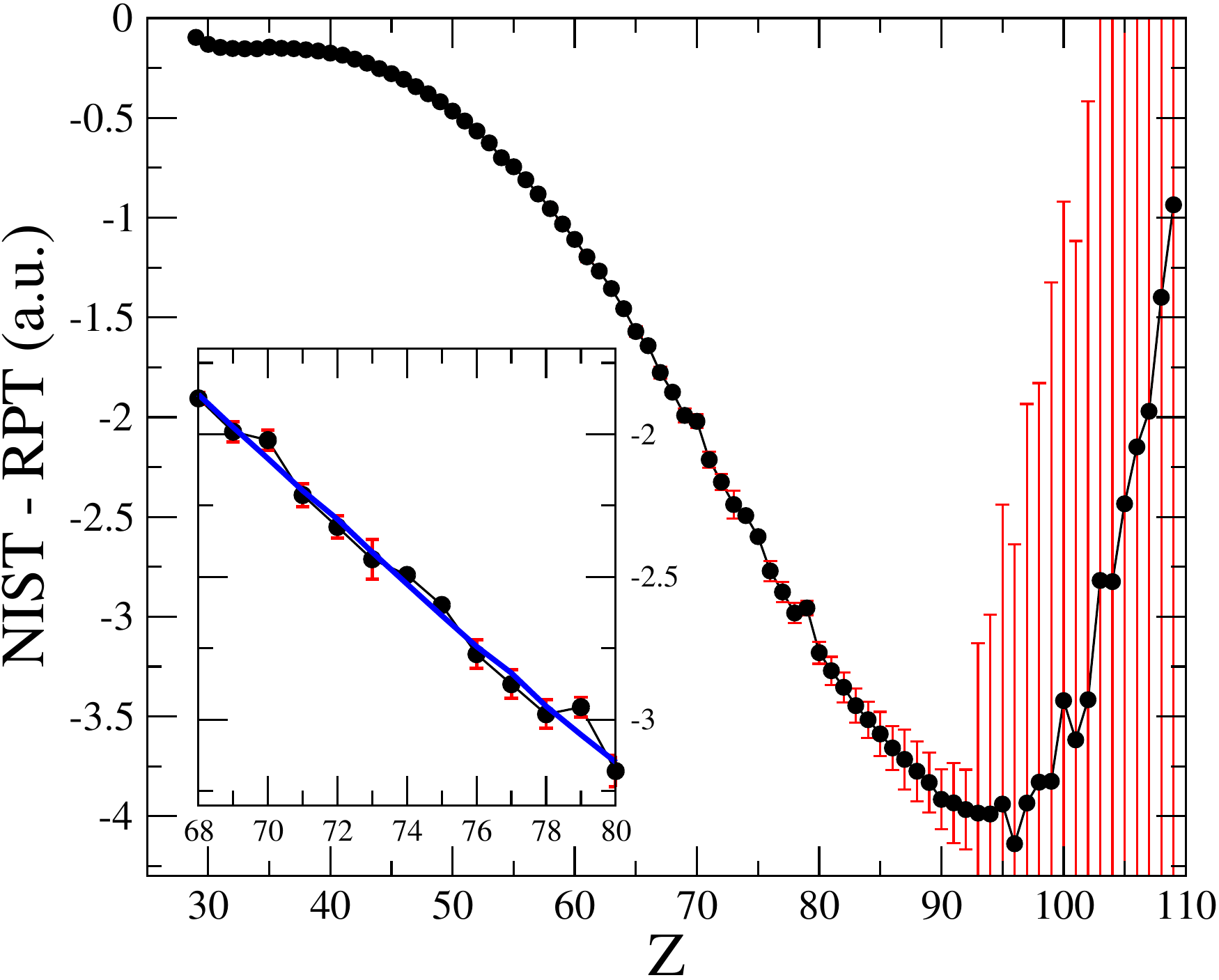}
\caption{\label{fig29} (Color online) Cu-like systems ($N=29$). Small deviations at $Z=70$ and 79 are noticed.}
\end{figure}
\end{center}

\textbf{Discussion:}

In the Cu-like sequence (Fig. \ref{fig29}), slight deviations at $Z=70$ and 79 are apparent. The data come from Ref. [\onlinecite{Tragin}]. Points seem to be overestimated in 0.064 and 0.097 a.u., respectively.

\subsection{The Zn-like sequence ($N=30$)}
\label{Zn}

RPT coefficients:
\begin{eqnarray}
\nonumber a_2 &=& 0.03125 + 0.00634766~ (Z/137.036)^2,\\
\nonumber a_1 &=& -1.3844 -0.0677735~ (Z/137.036)^2,\\
\nonumber a_0 &=& 9.2367,\\
\nonumber a_{-1} &=& 131.194.
\label{zn_coeff}
\end{eqnarray}

\begin{center}
\begin{figure}[!ht]
\includegraphics[width=0.9\linewidth,angle=0]{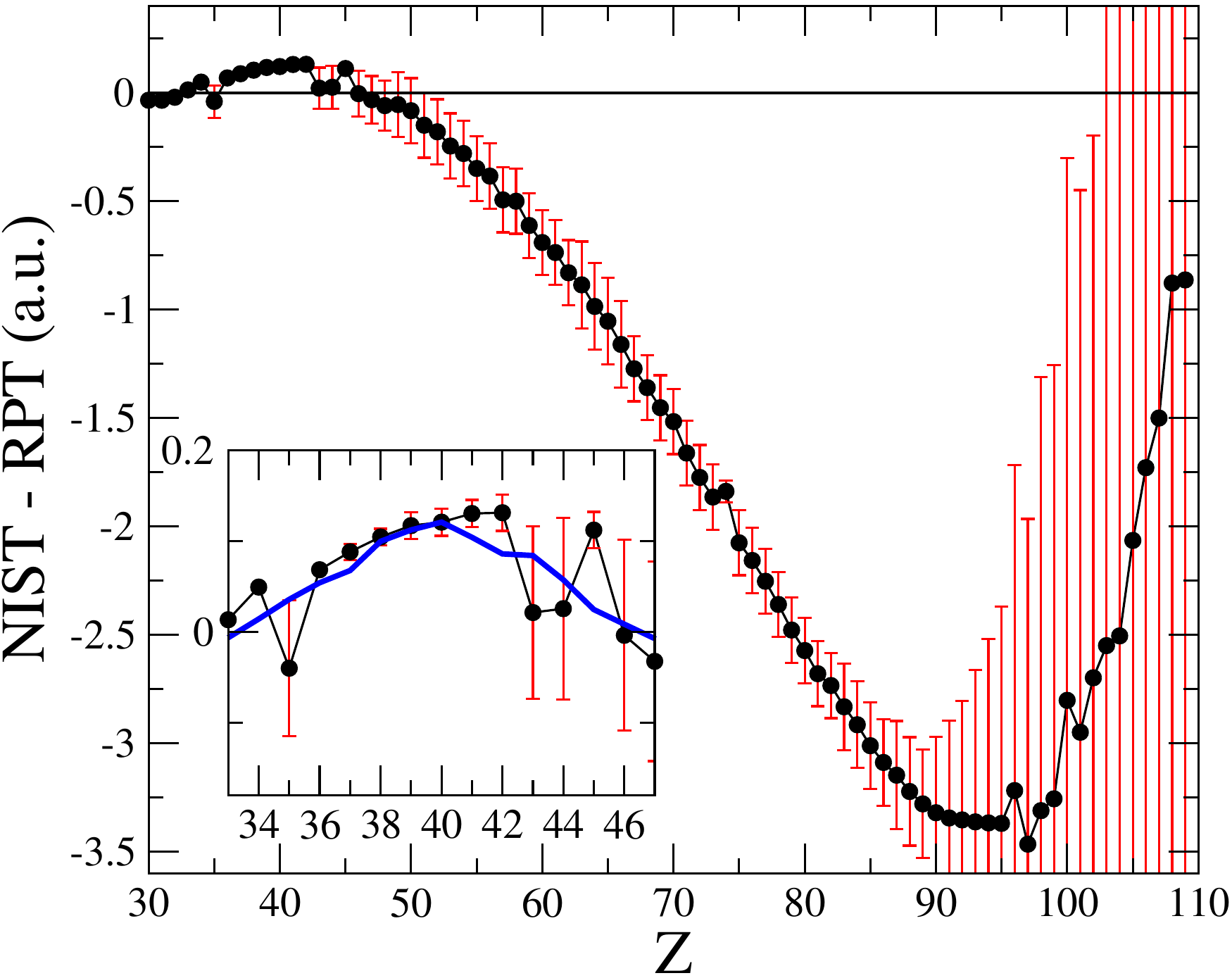}
\caption{\label{fig30} (Color online) The Zn-like systems ($N=30$). Inconsistencies are noticed at $Z=35$, 42, 45 and 74.}
\end{figure}
\end{center}

Conditions at $Z=N-1$:
\begin{eqnarray}
\nonumber E_a({\rm Cu}) &=& 0.0453693, \\
s &=& 0.285562.
\label{rest_zn}
\end{eqnarray}

The slope was computed from $E_a({\rm Cu})$ and $R_{cov}({\rm Cu})=2.305466$.

\textbf{Discussion:}

In Fig. \ref{fig30} (the Zn-like sequence), the data for $Z=35$ is taken from the Dirac-Fock calculation of Ref. [\onlinecite{Dirac-Fock1}], while the value for $Z=42$ is due to Refs. [\onlinecite{Sugar2}] and [\onlinecite{Litzen}]. These potentials should be corrected in +0.076 and -0.045 a.u., respectively.

On the other hand, $Z=45$ ionization potential was collected from the relativistic multireference many-body perturbation theory calculations of Vilkas et al. \cite{Vilkas}. It seems to be overestimated in 0.087 a.u.

The $Z=74$ ionization potential comes from Ref. [\onlinecite{W-ions1}], as before. It is 0.103 a.u. higher than the average curve.  

\subsection{The Ge-like sequence ($N=32$)}
\label{Ge}

RPT coefficients:
\begin{eqnarray}
\nonumber a_2 &=& 0.03125 + 0.00634766~ (Z/137.036)^2,\\
\nonumber a_1 &=& -1.51951 -0.0952627~ (Z/137.036)^2,\\
\nonumber a_0 &=& 10.9568,\\
\nonumber a_{-1} &=& 185.117.
\label{ge_coeff}
\end{eqnarray}

Conditions at $Z=N-1$:
\begin{eqnarray}
\nonumber E_a({\rm Ga}) &=& 0.0157966, \\
s &=& 0.240629.
\label{rest_ge}
\end{eqnarray}

\begin{center}
\begin{figure}[!ht]
\includegraphics[width=0.9\linewidth,angle=0]{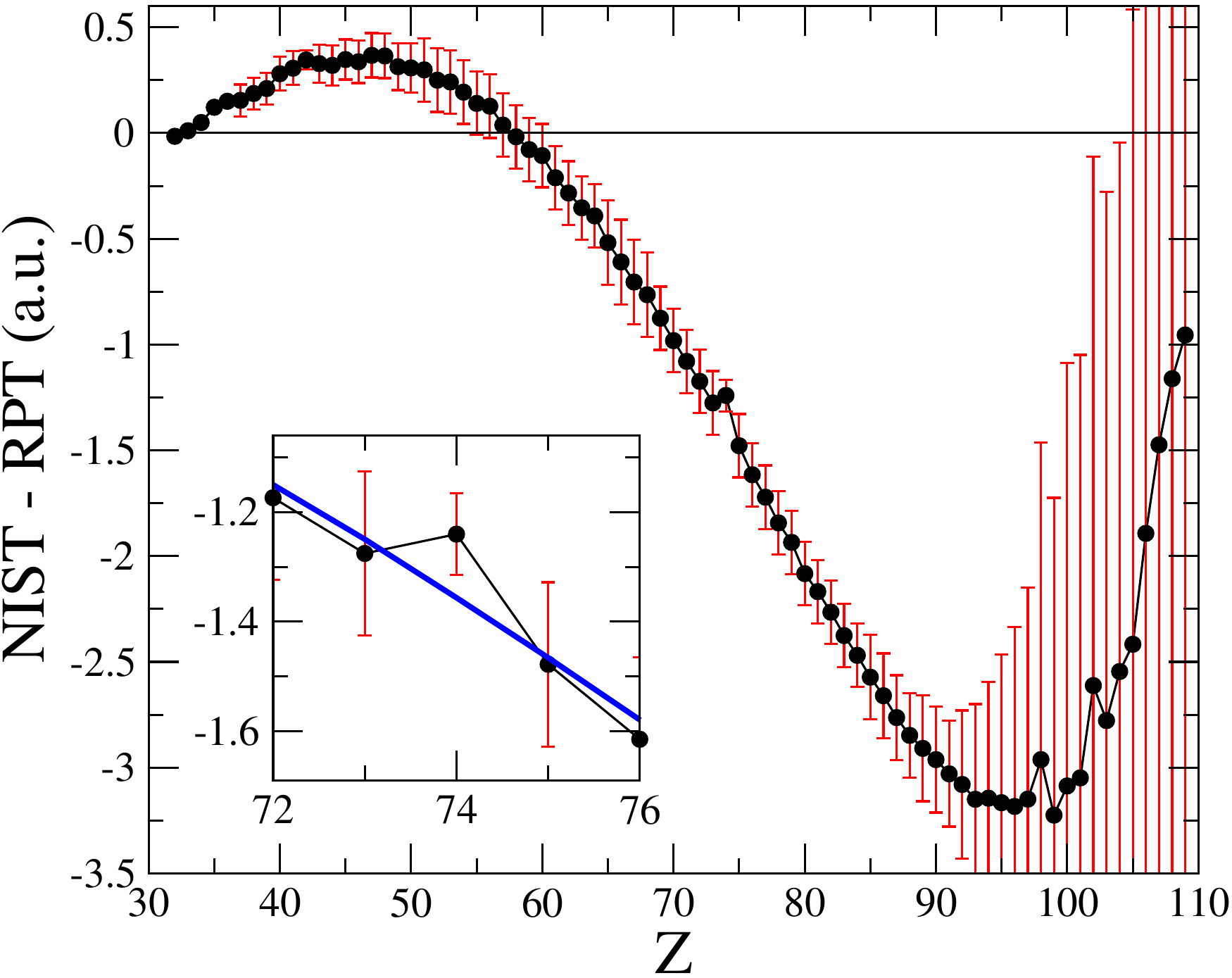}
\caption{\label{fig32} (Color online) Ge-like ions ($N=32$). Only the $Z=74$ point is distinguished.}
\end{figure}
\end{center}

The slope was computed from $E_a({\rm Ga})$ and $R_{cov}({\rm Ga})=2.324363$.

\textbf{Discussion:}

$I_p$ at $Z=74$ is overestimated in 0.116 a.u.

\subsection{The As-like sequence ($N=33$)}
\label{As}

RPT coefficients:
\begin{eqnarray}
\nonumber a_2 &=& 0.03125 + 0.00244141~ (Z/137.036)^2,\\
\nonumber a_1 &=& -1.55446 -0.0829917~ (Z/137.036)^2,\\
\nonumber a_0 &=& 12.6573,\\
\nonumber a_{-1} &=& 164.459.
\label{as_coeff}
\end{eqnarray}

Conditions at $Z=N-1$:
\begin{eqnarray}
\nonumber E_a({\rm Ge}) &=& 0.0453118, \\
s &=& 0.288928.
\label{rest_as}
\end{eqnarray}

The slope was computed from $E_a({\rm Ge})$ and $R_{cov}({\rm Ge})=2.267671$.

\begin{center}
\begin{figure}[!ht]
\includegraphics[width=0.9\linewidth,angle=0]{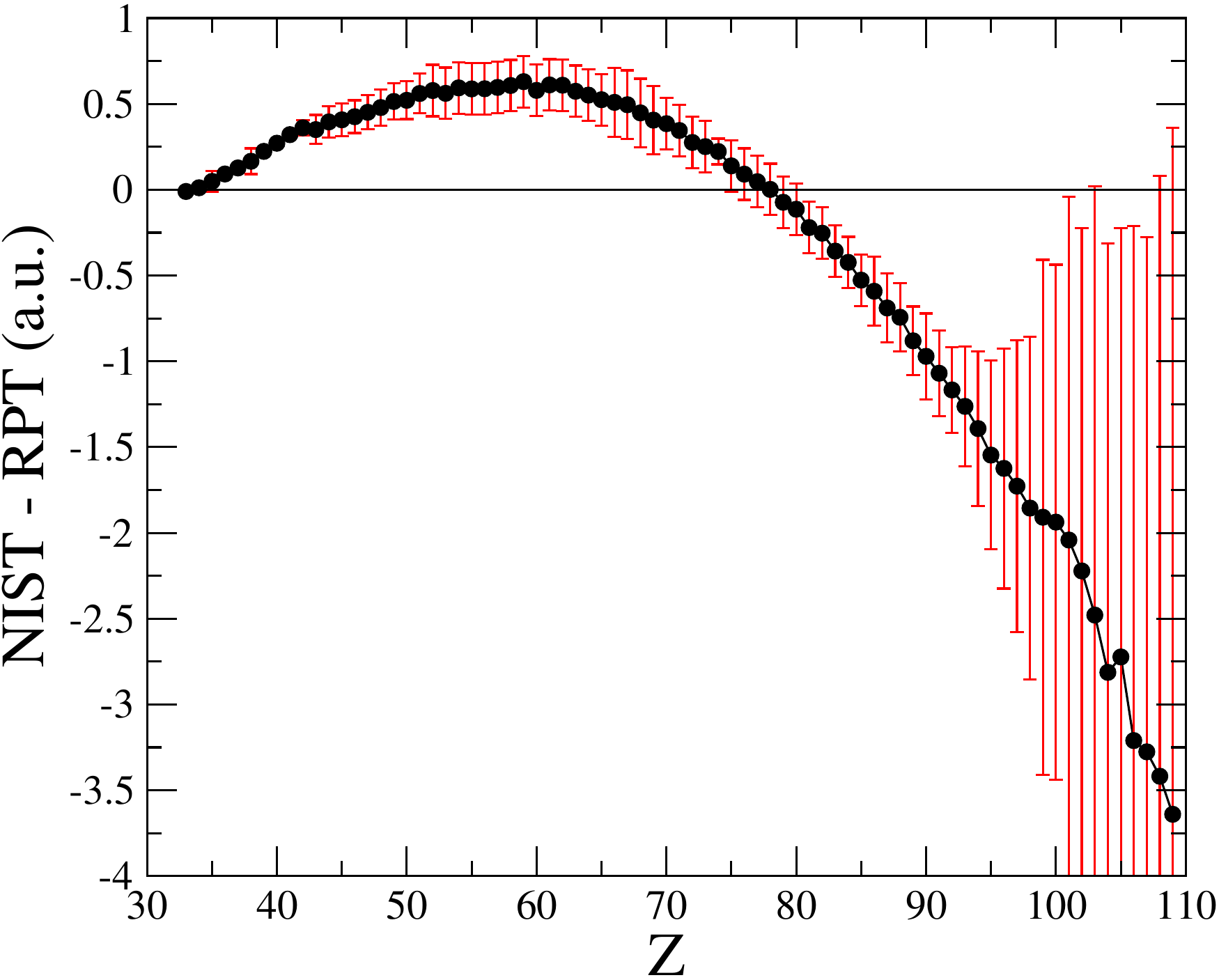}
\caption{\label{fig33} (Color online) The As-like systems ($N=33$). No inconsistencies are detected.}
\end{figure}
\end{center}

\subsection{The Se-like sequence ($N=34$)}
\label{Se}

RPT coefficients:
\begin{eqnarray}
\nonumber a_2 &=& 0.03125 + 0.00244141~ (Z/137.036)^2,\\
\nonumber a_1 &=& -1.59101 -0.0853527~ (Z/137.036)^2,\\
\nonumber a_0 &=& 11.7569,\\
\nonumber a_{-1} &=& 222.876.
\label{se_coeff}
\end{eqnarray}

Conditions at $Z=N-1$:
\begin{eqnarray}
\nonumber E_a({\rm As}) &=& 0.0295464, \\
s &=& 0.271227.
\label{rest_se}
\end{eqnarray}

The slope was computed from $E_a({\rm As})$ and $R_{cov}({\rm As})=2.267671$.

\begin{center}
\begin{figure}[!ht]
\includegraphics[width=0.9\linewidth,angle=0]{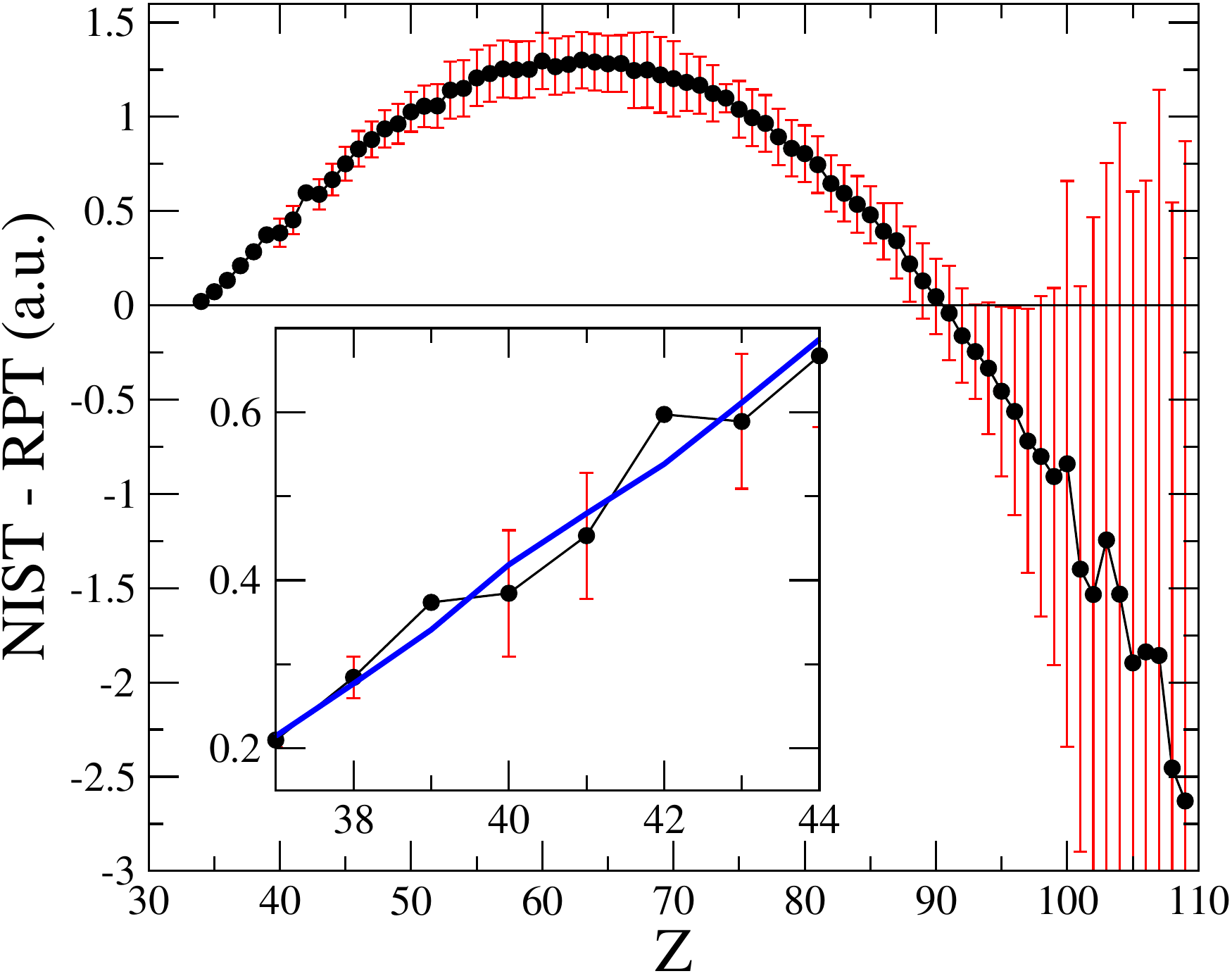}
\caption{\label{fig34} (Color online) The Se-like sequence ($N=34$). Slight deviations at $Z =39$ and 42 are apparent.}
\end{figure}
\end{center}

\textbf{Discussion:}

In Fig. \ref{fig34} (the Se-like sequence), we detect inconsistencies at $Z=39$ and 42. $I_p$ at $Z=39$ seems to be overestimated in 0.033 a.u.

The number for $I_p$ in Mo$^{+18}$ (i.e. $Z=42$, $N=34$) comes from Refs. [\onlinecite{Khatoon}] and [\onlinecite{Sugar2}]. It seems to be 0.059 a.u. higher than the average curve. 

\subsection{The Br-like sequence ($N=35$)}
\label{Br}

RPT coefficients:
\begin{eqnarray}
\nonumber a_2 &=& 0.03125 + 0.00244141~ (Z/137.036)^2,\\
\nonumber a_1 &=& -1.63074 -0.0881566~ (Z/137.036)^2,\\
\nonumber a_0 &=& 13.0595,\\
\nonumber a_{-1} &=& 215.756.
\label{br_coeff}
\end{eqnarray}

\begin{center}
\begin{figure}[!ht]
\includegraphics[width=0.9\linewidth,angle=0]{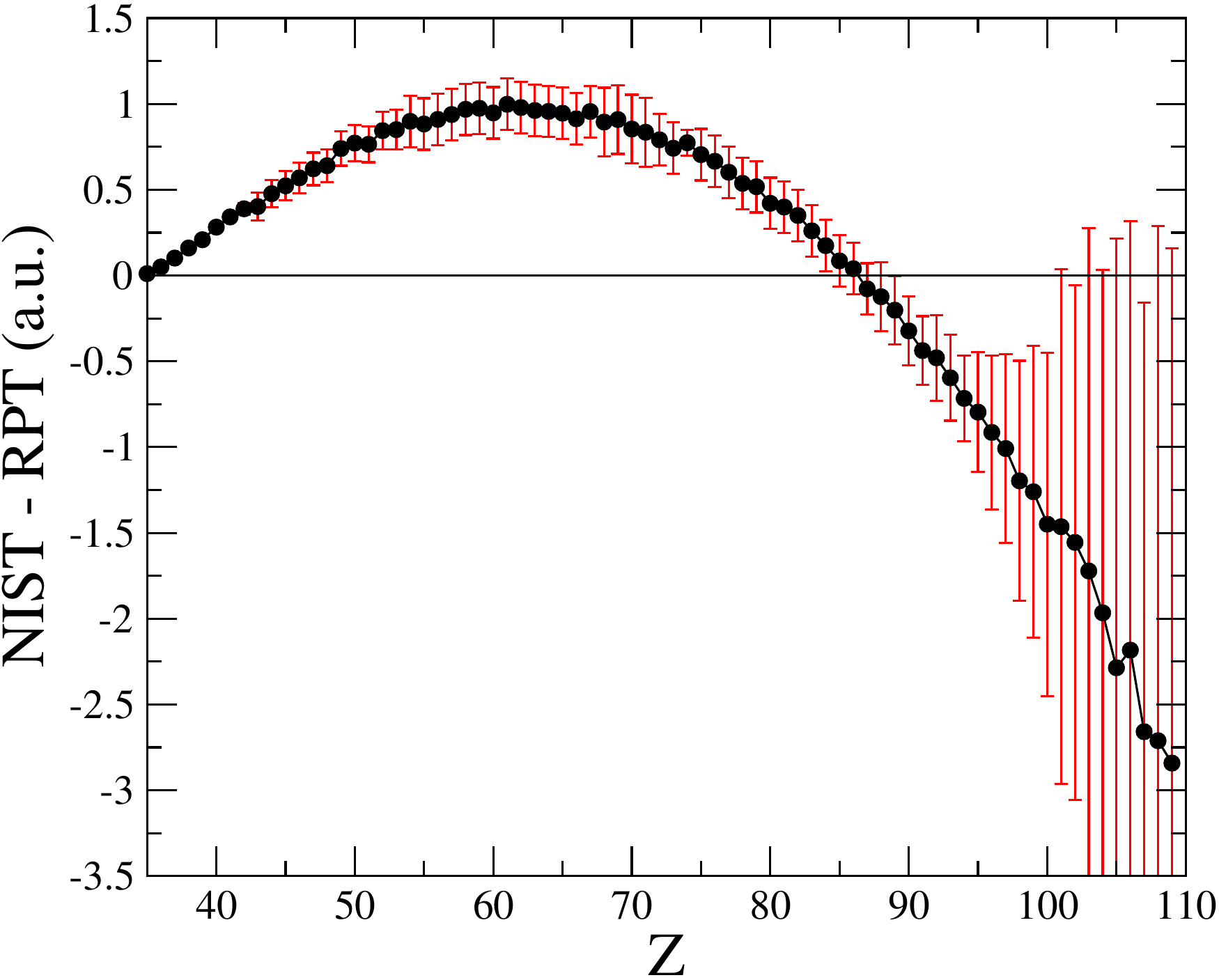}
\caption{\label{fig35} (Color online) Br-like systems ($N=35$). A smooth curve. No inconsistencies are detected.}
\end{figure}
\end{center}

Conditions at $Z=N-1$:
\begin{eqnarray}
\nonumber E_a({\rm Se}) &=& 0.0742704, \\
s &=& 0.312415.
\label{rest_br}
\end{eqnarray}

The slope was computed from $E_a({\rm Se})$ and $R_{cov}({\rm Se})=2.229877$.

\subsection{The Kr-like sequence ($N=36$)}
\label{Kr}

\begin{center}
\begin{figure}[!ht]
\includegraphics[width=0.9\linewidth,angle=0]{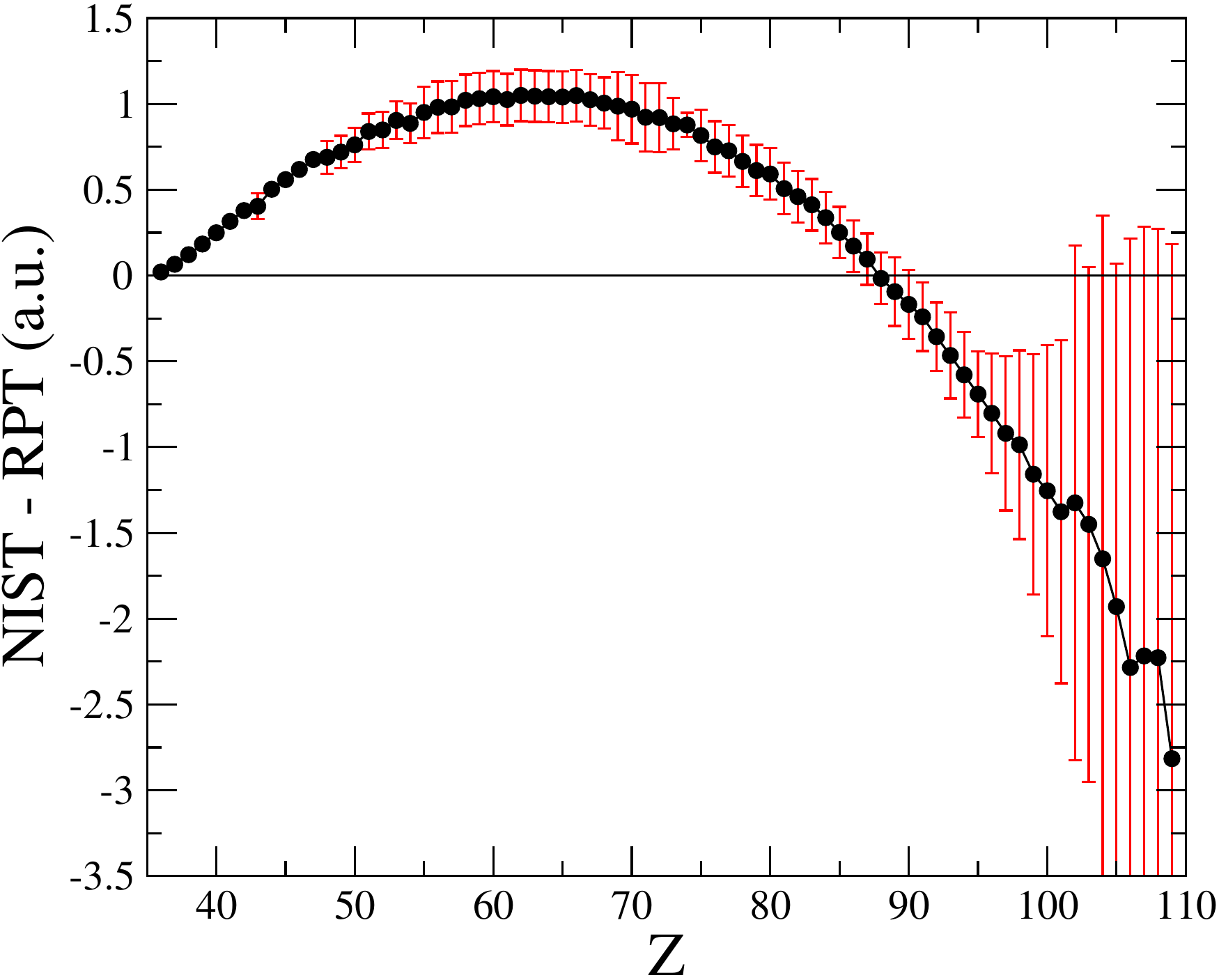}
\caption{\label{fig36} (Color online) The Kr-like systems ($N=36$). No inconsistencies are detected.}
\end{figure}
\end{center}

RPT coefficients:
\begin{eqnarray}
\nonumber a_2 &=& 0.03125 + 0.00244141~ (Z/137.036)^2,\\
\nonumber a_1 &=& -1.66728 -0.0905176~ (Z/137.036)^2,\\
\nonumber a_0 &=& 13.5057,\\
\nonumber a_{-1} &=& 234.611.
\label{kr_coeff}
\end{eqnarray}

Conditions at $Z=N-1$:
\begin{eqnarray}
\nonumber E_a({\rm Br}) &=& 0.123625, \\
s &=& 0.333943.
\label{rest_kr}
\end{eqnarray}

The slope was computed from $E_a({\rm Br})$ and $R_{cov}({\rm Br})=2.210980$.

\section{Fifth row elements}

\subsection{The Sr-like sequence ($N=38$)}
\label{Sr}

\begin{center}
\begin{figure}[!ht]
\includegraphics[width=0.9\linewidth,angle=0]{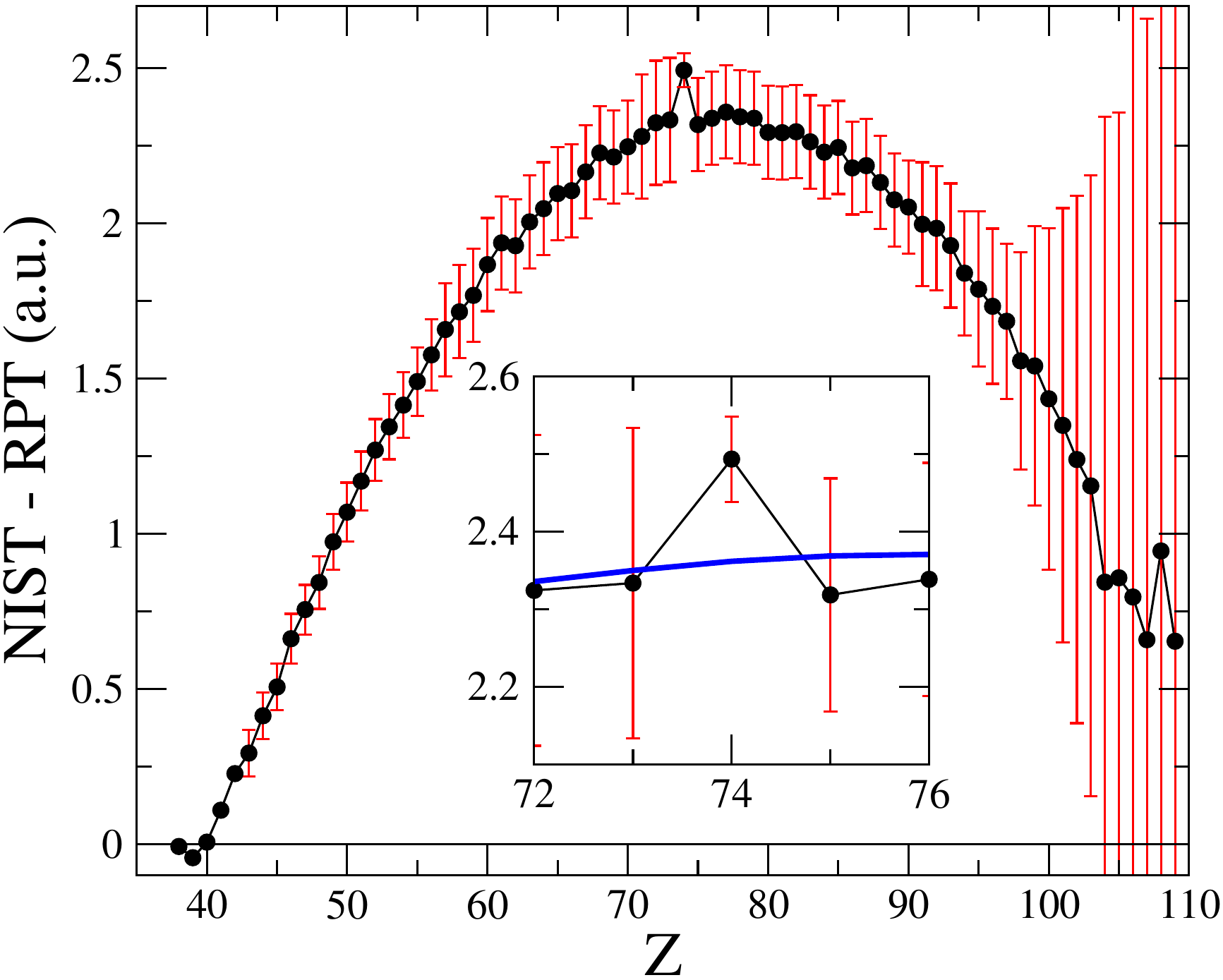}
\caption{\label{fig38} (Color online) The Sr-like systems ($N=38$). An apparent deviation at $Z=74$\cite{W-ions1} is shown.}
\end{figure}
\end{center}

RPT coefficients:
\begin{eqnarray}
\nonumber a_2 &=& 0.03125 + 0.00244141~ (Z/137.036)^2,\\
\nonumber a_1 &=& -1.86543 -0.136324~ (Z/137.036)^2,\\
\nonumber a_0 &=& 15.882,\\
\nonumber a_{-1} &=& 388.488.
\label{sr_coeff}
\end{eqnarray}

Conditions at $Z=N-1$:
\begin{eqnarray}
\nonumber E_a({\rm Rb}) &=& 0.0178511, \\
s &=& 0.166529.
\label{rest_sr}
\end{eqnarray}

The slope was computed from $E_a({\rm Rb})$ and $R_{cov}({\rm Rb})=4.062911$.

\textbf{Discussion:}

$I_p$ at $Z=74$ is overestimated in 0.132 a.u.

\subsection{The Y-like sequence ($N=39$)}
\label{Y}

RPT coefficients:
\begin{eqnarray}
\nonumber a_2 &=& 0.03125 + 0.00244141~ (Z/137.036)^2,\\
\nonumber a_1 &=& -1.90897 -0.139844~ (Z/137.036)^2,\\
\nonumber a_0 &=& 14.2754,\\
\nonumber a_{-1} &=& 504.63.
\label{y_coeff}
\end{eqnarray}

Conditions at $Z=N-1$:
\begin{eqnarray}
\nonumber E_a({\rm Sr}) &=& 0.00176347, \\
s &=& 0.120083.
\label{rest_y}
\end{eqnarray}

The slope was computed from $E_a({\rm Sr})$ and $R_{cov}({\rm Sr})=3.590480$.

\begin{center}
\begin{figure}[!ht]
\includegraphics[width=0.9\linewidth,angle=0]{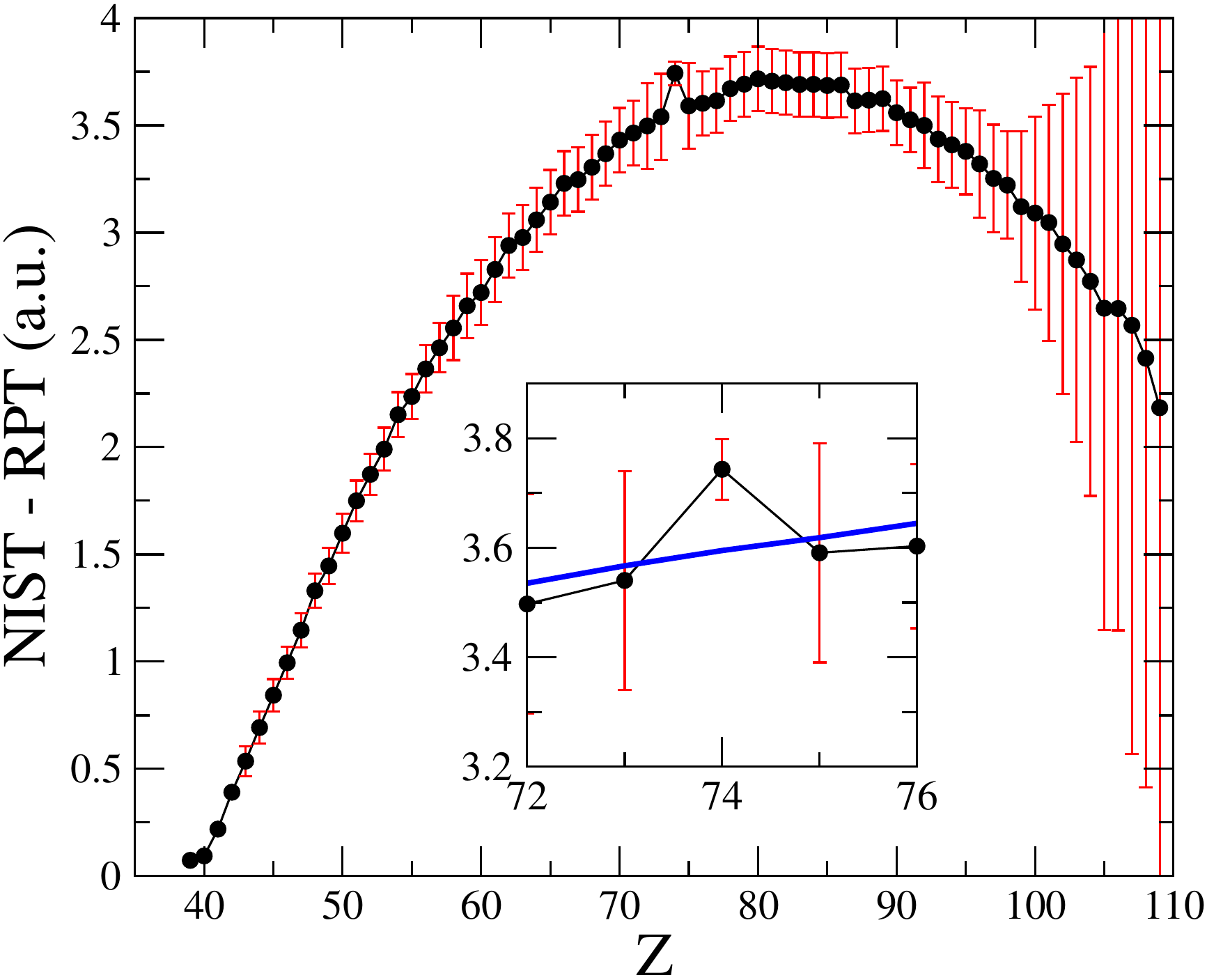}
\caption{\label{fig39} (Color online) The Y-like systems ($N=39$). The only detected inconsistency is  at $Z=74~$\cite{W-ions1}.}
\end{figure}
\end{center}

\textbf{Discussion:}

$I_p$ at $Z=74$ should be corrected in -0.148 a.u.

\subsection{The Zr-like sequence ($N=40$)}
\label{Zr}

RPT coefficients:
\begin{eqnarray}
\nonumber a_2 &=& 0.03125 + 0.00244141~ (Z/137.036)^2,\\
\nonumber a_1 &=& -1.94905 -0.142908~ (Z/137.036)^2,\\
\nonumber a_0 &=& 16.4347,\\
\nonumber a_{-1} &=& 476.155.
\label{zr_coeff}
\end{eqnarray}

\begin{center}
\begin{figure}[!ht]
\includegraphics[width=0.9\linewidth,angle=0]{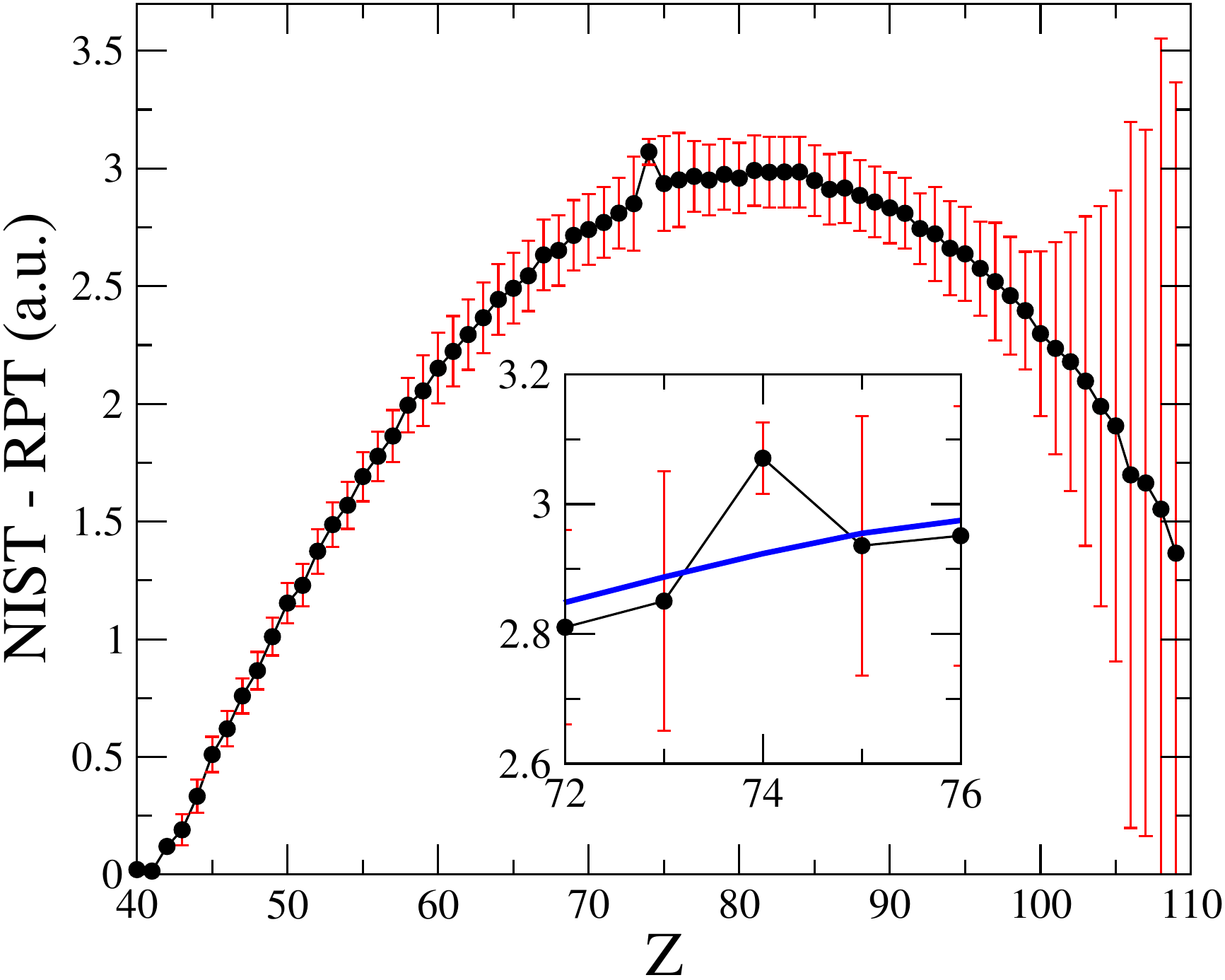}
\caption{\label{fig40} (Color online) The Zr-like sequence ($N=40$). The $Z=74$ point is distinguished.}
\end{figure}
\end{center}

Conditions at $Z=N-1$:
\begin{eqnarray}
\nonumber E_a({\rm Y}) &=& 0.0112782, \\
s &=& 0.179247.
\label{rest_zr}
\end{eqnarray}

The slope was computed from $E_a({\rm Y})$ and $R_{cov}({\rm Y})=3.325918$.

\textbf{Discussion:}

$I_p$ at $Z=74$ shows an overestimation of 0.147 a.u.

\subsection{The Nb-like sequence ($N=41$)}
\label{Nb}

RPT coefficients:
\begin{eqnarray}
\nonumber a_2 &=& 0.03125 + 0.00113932~ (Z/137.036)^2,\\
\nonumber a_1 &=& -1.99199 -0.112623~ (Z/137.036)^2,\\
\nonumber a_0 &=& 17.6019,\\
\nonumber a_{-1} &=& 492.881.
\label{nb_coeff}
\end{eqnarray}

\begin{center}
\begin{figure}[!ht]
\includegraphics[width=0.9\linewidth,angle=0]{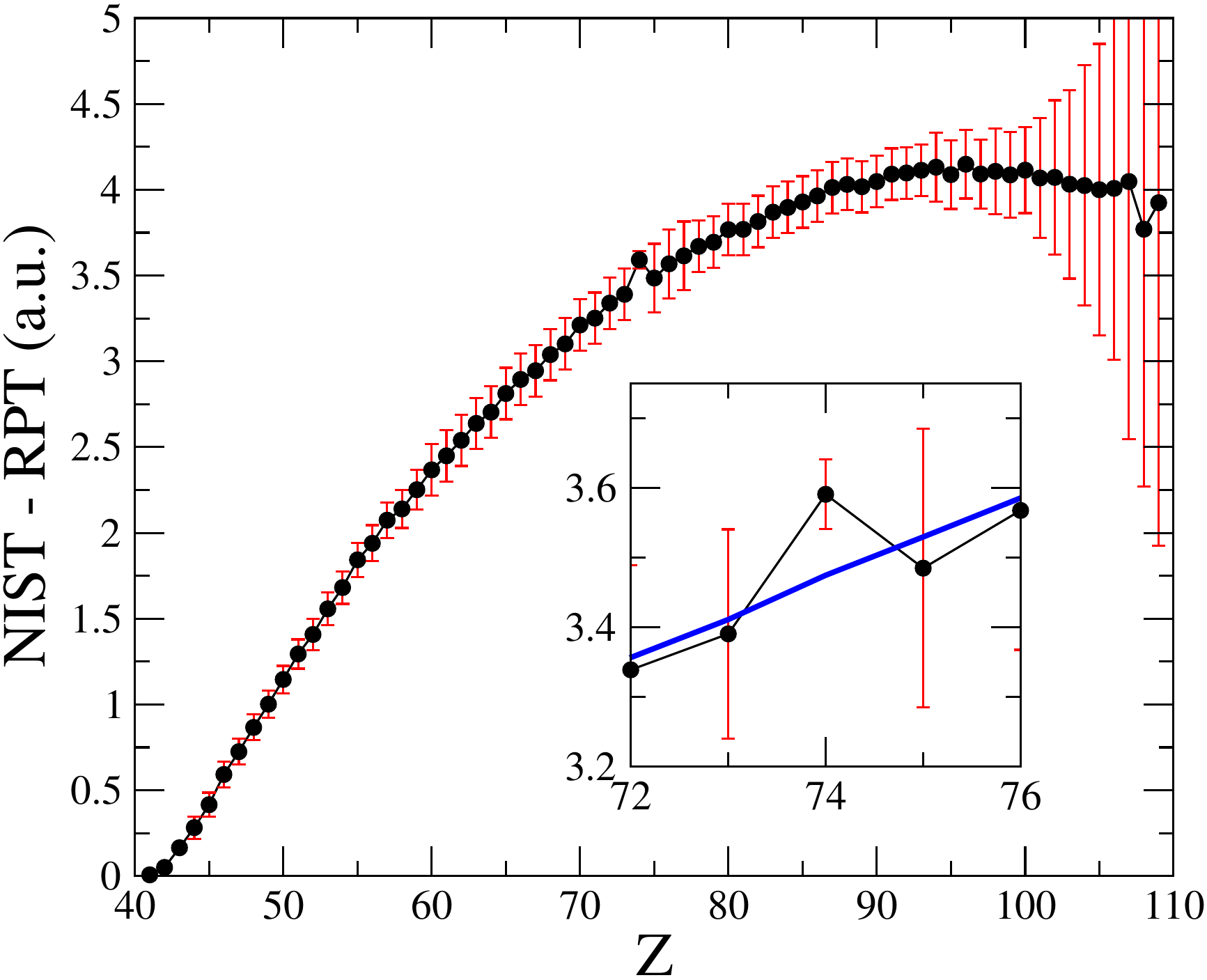}
\caption{\label{fig41} (Color online) The Nb-like systems ($N=41$). The only detected inconsistency is the $Z=74$ point.\cite{W-ions1}}
\end{figure}
\end{center}

Conditions at $Z=N-1$:
\begin{eqnarray}
\nonumber E_a({\rm Zr}) &=& 0.0156496, \\
s &=& 0.198128.
\label{rest_nb}
\end{eqnarray}

The slope was computed from $E_a({\rm Zr})$ and $R_{cov}({\rm Zr})=3.099151$.

\textbf{Discussion:}

$I_p$ at $Z=74$ is overestimated in 0.116 a.u.

\subsection{The Mo-like sequence ($N=42$)}
\label{Mo}

RPT coefficients:
\begin{eqnarray}
\nonumber a_2 &=& 0.03125 + 0.00113932~ (Z/137.036)^2,\\
\nonumber a_1 &=& -2.03196 -0.114317~ (Z/137.036)^2,\\
\nonumber a_0 &=& 18.7868,\\
\nonumber a_{-1} &=& 503.25.
\label{mo_coeff}
\end{eqnarray}

Conditions at $Z=N-1$:
\begin{eqnarray}
\nonumber E_a({\rm Nb}) &=& 0.0336625, \\
s &=& 0.229290.
\label{rest_mo}
\end{eqnarray}

The slope was computed from $E_a({\rm Nb})$ and $R_{cov}({\rm Nb})=2.947973$.

\begin{center}
\begin{figure}[!ht]
\includegraphics[width=0.9\linewidth,angle=0]{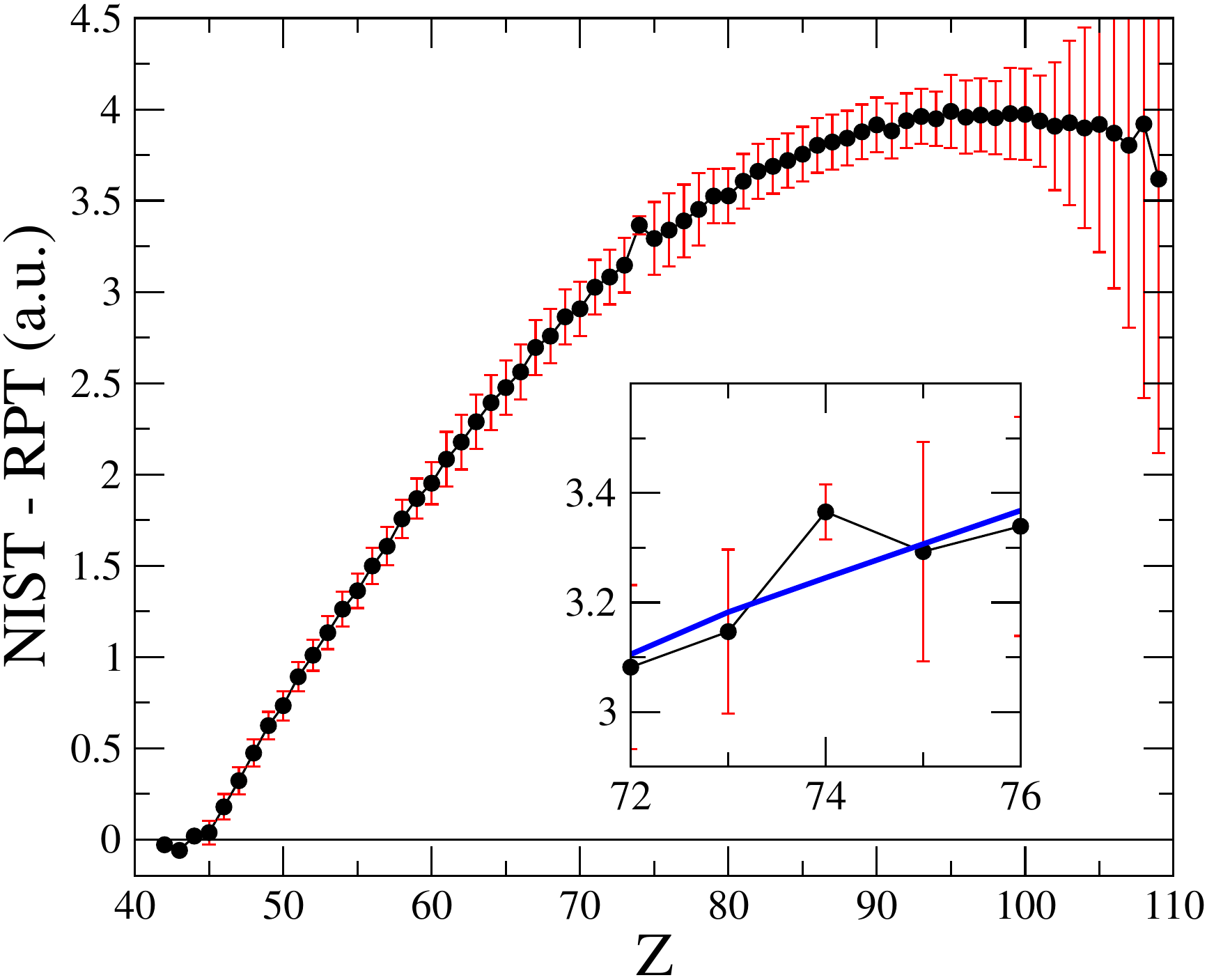}
\caption{\label{fig42} (Color online) The Mo-like ions ($N=42$). The only inconsistent point is $Z=74$\cite{W-ions1}.}
\end{figure}
\end{center}

\textbf{Discussion:}

$I_p$ at $Z=74$ should be corrected in -0.120 a.u.

\subsection{The Tc-like sequence ($N=43$)}
\label{Tc}

RPT coefficients:
\begin{eqnarray}
\nonumber a_2 &=& 0.03125 + 0.00113932~ (Z/137.036)^2,\\
\nonumber a_1 &=& -2.07354 -0.116109~ (Z/137.036)^2,\\
\nonumber a_0 &=& 19.0013,\\
\nonumber a_{-1} &=& 556.886.
\label{tc_coeff}
\end{eqnarray}

\begin{center}
\begin{figure}[!ht]
\includegraphics[width=0.9\linewidth,angle=0]{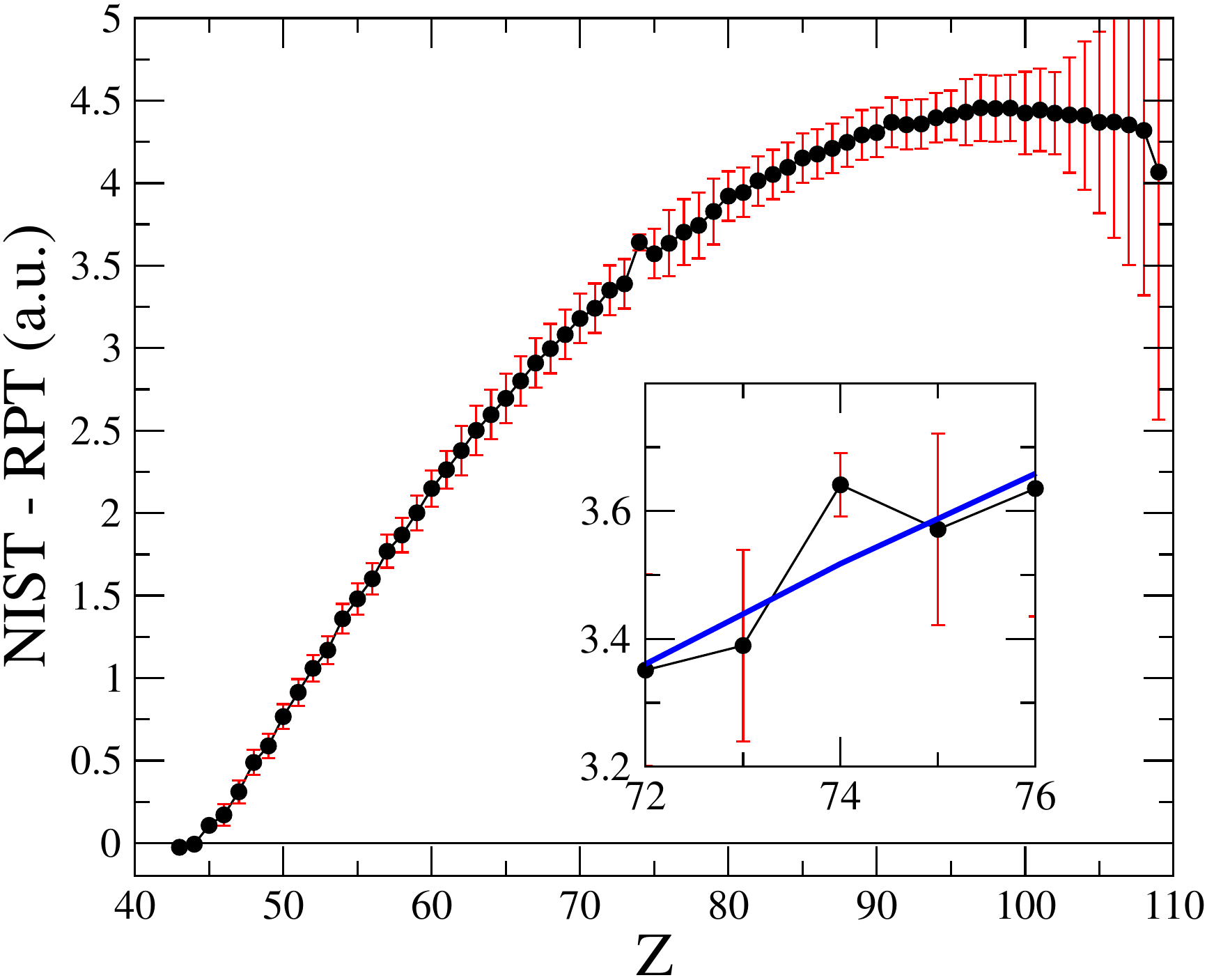}
\caption{\label{fig43} (Color online) The Tc-like ions ($N=43$). Only the $Z=74$ point is distinguished.\cite{W-ions1}.}
\end{figure}
\end{center}

Conditions at $Z=N-1$:
\begin{eqnarray}
\nonumber E_a({\rm Mo}) &=& 0.0274790, \\
s &=& 0.233850.
\label{rest_tc}
\end{eqnarray}

The slope was computed from $E_a({\rm Mo})$ and $R_{cov}({\rm Mo})=2.759000$.

\textbf{Discussion:}

Notice that $I_p$ at $Z=74$ is systematically overestimated (since the Ca sequence). This time in 0.123 a.u.

\subsection{The Ru-like sequence ($N=44$)}
\label{Ru}

\begin{center}
\begin{figure}[!ht]
\includegraphics[width=0.9\linewidth,angle=0]{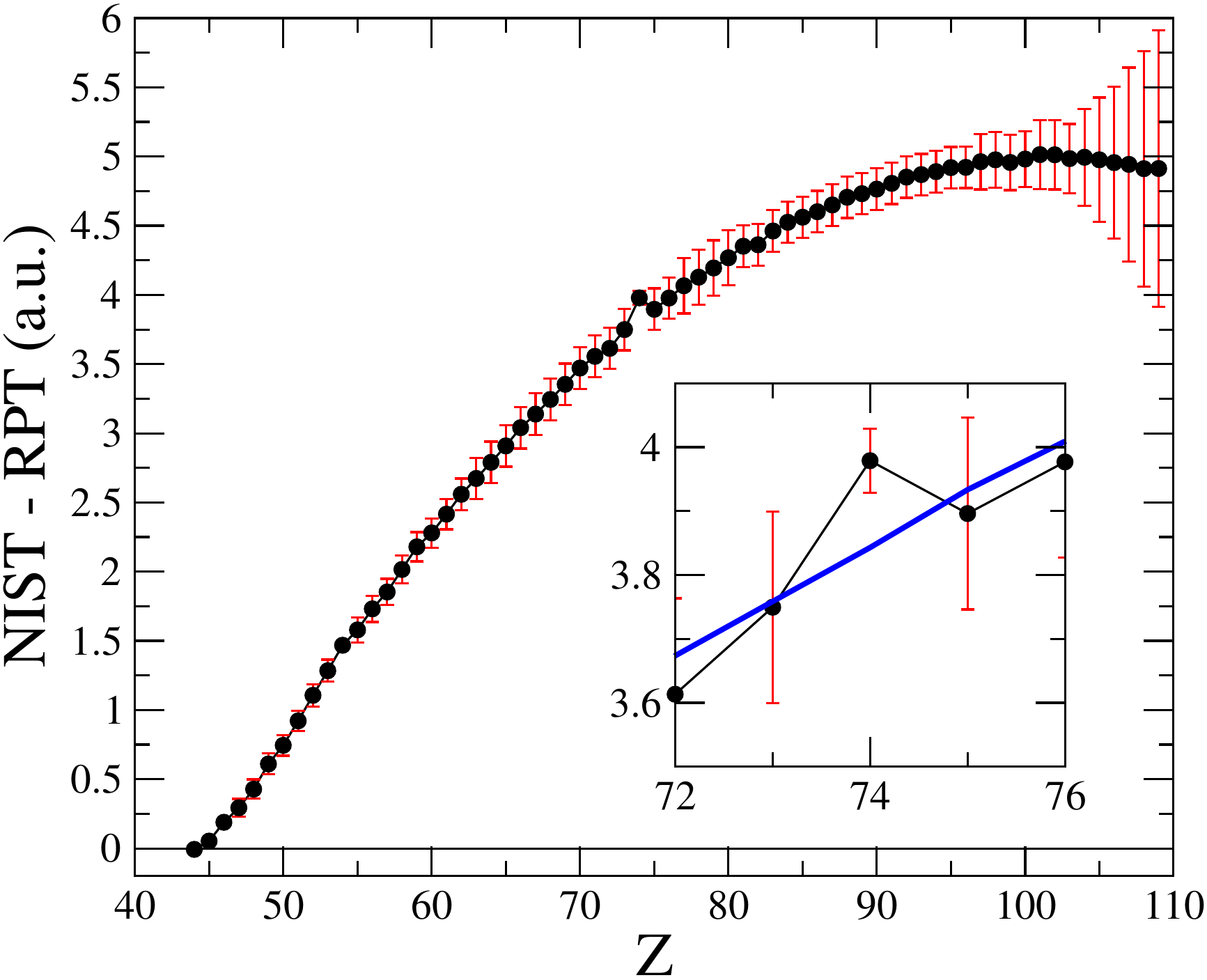}
\caption{\label{fig44} (Color online) The Ru-like systems ($N=44$). An inconsistency at $Z=74$ is apparent.}
\end{figure}
\end{center}

RPT coefficients:
\begin{eqnarray}
\nonumber a_2 &=& 0.03125 + 0.00113932~ (Z/137.036)^2,\\
\nonumber a_1 &=& -2.11747 -0.118053~ (Z/137.036)^2,\\
\nonumber a_0 &=& 19.1396,\\
\nonumber a_{-1} &=& 621.055.
\label{ru_coeff}
\end{eqnarray}

Conditions at $Z=N-1$:
\begin{eqnarray}
\nonumber E_a({\rm Tc}) &=& 0.0202049, \\
s &=& 0.232165.
\label{rest_ru}
\end{eqnarray}

The slope was computed from $E_a({\rm Tc})$ and $R_{cov}({\rm Tc})=2.607822$.

\textbf{Discussion:}

$I_p$ at $Z=74$ is overestimated in 0.136 a.u.

\subsection{The Rh-like sequence ($N=45$)}
\label{Rh}

\begin{center}
\begin{figure}[!ht]
\includegraphics[width=0.9\linewidth,angle=0]{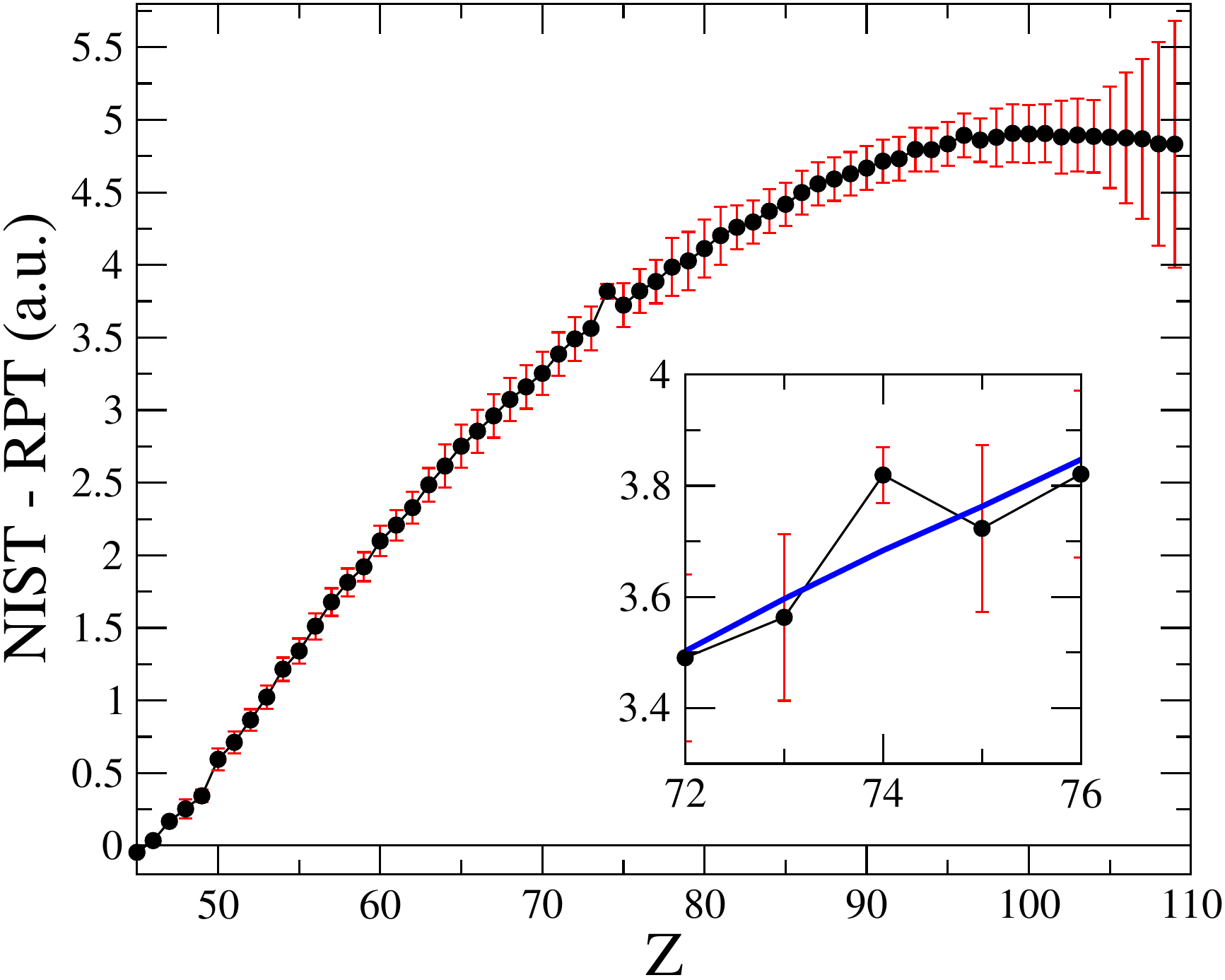}
\caption{\label{fig45} (Color online) Rh-like systems ($N=45$). There are apparent inconsistencies at $Z=49~$\cite{Dirac-Fock1} and  $Z=74~$\cite{W-ions1}.}
\end{figure}
\end{center}

RPT coefficients:
\begin{eqnarray}
\nonumber a_2 &=& 0.03125 + 0.00113932~ (Z/137.036)^2,\\
\nonumber a_1 &=& -2.15904 -0.119845~ (Z/137.036)^2,\\
\nonumber a_0 &=& 20.2974,\\
\nonumber a_{-1} &=& 640.445.
\label{rh_coeff}
\end{eqnarray}

Conditions at $Z=N-1$:
\begin{eqnarray}
\nonumber E_a({\rm Ru}) &=& 0.0385732, \\
s &=& 0.232165.
\label{rest_rh}
\end{eqnarray}

The slope was computed from $E_a({\rm Ru})$ and $R_{cov}({\rm Ru})=2.570028$.

\textbf{Discussion:}

$I_p$ at $Z=49$ is underestimated in 0.072 a.u., whereas the $Z=74$ point is 0.136 a.u. higher than the average curve. 

\subsection{The Pd-like sequence ($N=46$)}
\label{Pd}

RPT coefficients:
\begin{eqnarray}
\nonumber a_2 &=& 0.03125 + 0.00113932~ (Z/137.036)^2,\\
\nonumber a_1 &=& -2.19902 -0.121538~ (Z/137.036)^2,\\
\nonumber a_0 &=& 20.4131,\\
\nonumber a_{-1} &=& 703.998.
\label{pd_coeff}
\end{eqnarray}

\begin{center}
\begin{figure}[!ht]
\includegraphics[width=0.9\linewidth,angle=0]{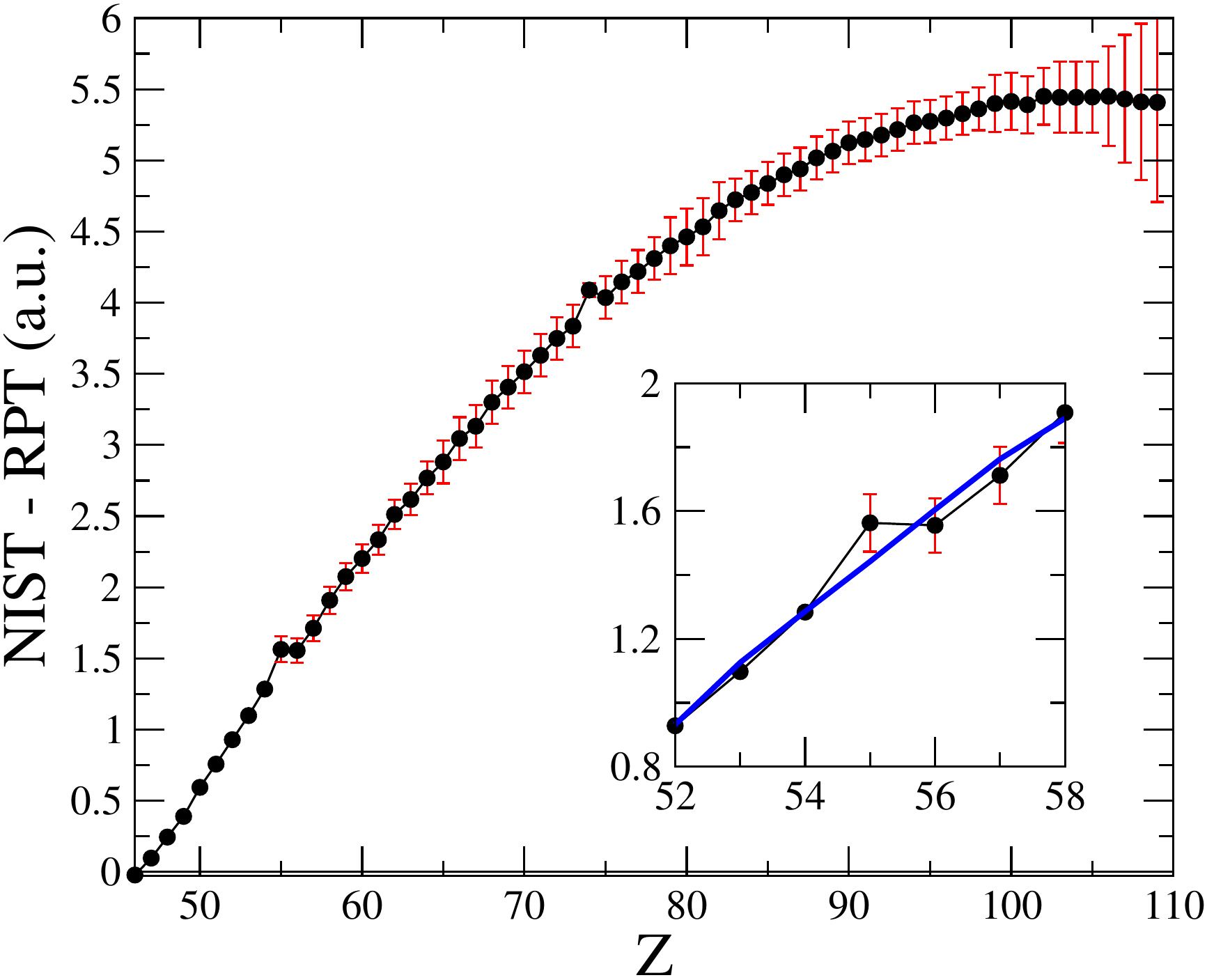}
\caption{\label{fig46} (Color online) The Pd-like sequence ($N=46$). Inconsistencies at $Z=55$ and 74 are noticed.}
\end{figure}
\end{center}

Conditions at $Z=N-1$:
\begin{eqnarray}
\nonumber E_a({\rm Rh}) &=& 0.0417692, \\
s &=& 0.258127.
\label{rest_pd}
\end{eqnarray}

The slope was computed from $E_a({\rm Rh})$ and $R_{cov}({\rm Rh})=2.532233$.

\textbf{Discussion:}

In the Pd-like sequence (Fig. \ref{fig46}), in addition to $Z=74$, an apparent inconsistency at $Z=55$ comes from the experimental work of Churilov et al.\cite{Churilov} The $Z=55$ and 74 points are, respectively,  0.120 and 0.118 a.u. above the average curve.

\subsection{The Ag-like sequence ($N=47$)}
\label{Ag}

RPT coefficients:
\begin{eqnarray}
\nonumber a_2 &=& 0.03125 + 0.00113932~ (Z/137.036)^2,\\
\nonumber a_1 &=& -2.41502 -0.161518~ (Z/137.036)^2,\\
\nonumber a_0 &=& 35.8484,\\
\nonumber a_{-1} &=& 446.393.
\label{ag_coeff}
\end{eqnarray}

\begin{center}
\begin{figure}[!ht]
\includegraphics[width=0.9\linewidth,angle=0]{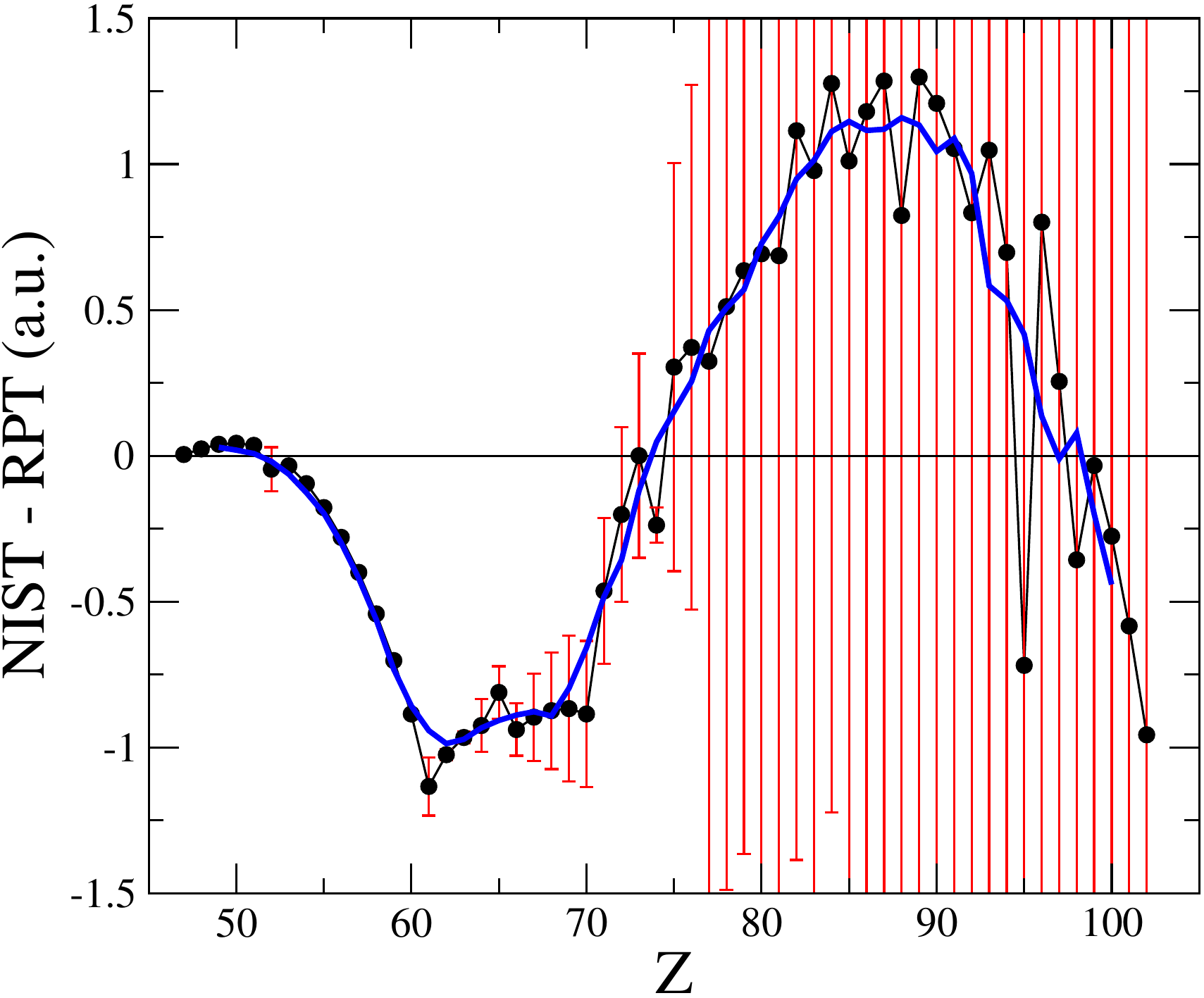}
\caption{\label{fig47} (Color online) The Ag-like systems ($N=47$). The average curve is drawn (blue) in order to identify problematic points. Inconsistencies at $Z=61$ and 74 are apparent. The $Z=95$) point also distinghuishes in spite of its huge error bar.}
\end{figure}
\end{center}

Conditions at $Z=N-1$:
\begin{eqnarray}
\nonumber E_a({\rm Pd}) &=& 0.0206460, \\
s &=& 0.263779.
\label{rest_ag}
\end{eqnarray}

The slope was computed from $E_a({\rm Pd})$ and $R_{cov}({\rm Pd})=2.456644$.

\textbf{Discussion:}

A qualitatively different picture appears in the Ag-like sequence, Fig. \ref{fig47}. The maximum of $|{\rm RPT-NIST}|$ diminished, as compared with the Pd-like sequence, but the dispersion of points has significantly increased. 

The problematic point at $Z=61$ comes from Ref. [\onlinecite{Dirac-Fock1}]. It seems to be 0.191 a.u. below the average curve. $I_p$ at $Z=74$, on the other hand, is underestimated in 0.275 a.u.

Data for $Z\ge 77$ is the entire responsibility of Carlson et al.\cite{Carlson}, who employ a simple spherical shell model in order to compute the ionization potentials. The indicated error bars are very high. The $Z=95$ point, with a deviation of -1.134 a.u. should, however, be noticed. 

\subsection{The Cd-like sequence ($N=48$)}
\label{Cd}

RPT coefficients:
\begin{eqnarray}
\nonumber a_2 &=& 0.03125 + 0.00113932~ (Z/137.036)^2,\\
\nonumber a_1 &=& -2.46174 -0.163993~ (Z/137.036)^2,\\
\nonumber a_0 &=& 37.7704,\\
\nonumber a_{-1} &=& 449.257.
\label{cd_coeff}
\end{eqnarray}

Conditions at $Z=N-1$:
\begin{eqnarray}
\nonumber E_a({\rm Ag}) &=& 0.0478309, \\
s &=& 0.265692.
\label{rest_cd}
\end{eqnarray}

The slope was computed from $E_a({\rm Ag})$ and $R_{cov}({\rm Ag})=2.570028$.

\begin{center}
\begin{figure}[!ht]
\includegraphics[width=0.9\linewidth,angle=0]{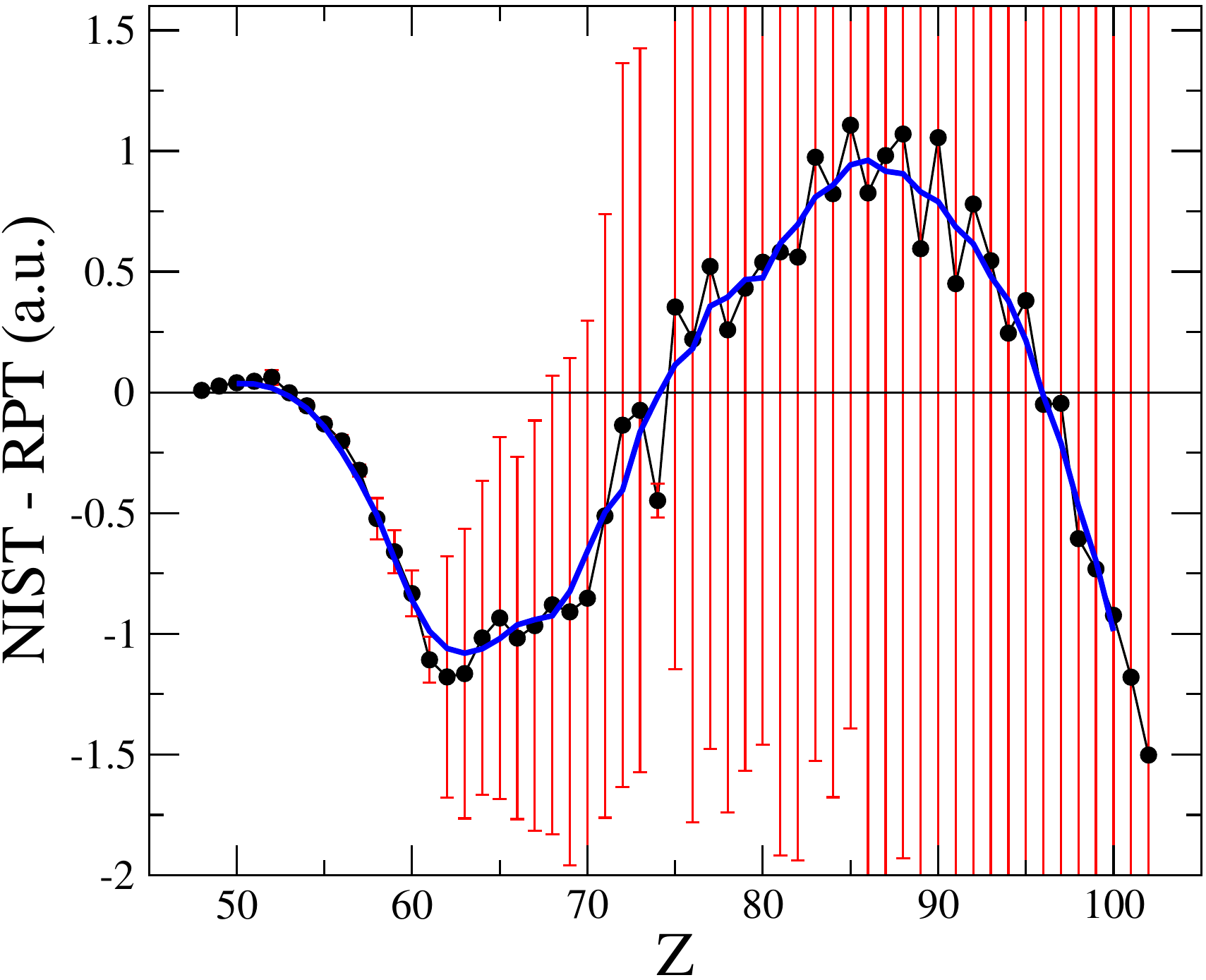}
\caption{\label{fig48} (Color online) The Cd-like systems ($N=48$). The only inconsistent point is $Z=74$.}
\end{figure}
\end{center}

\textbf{Discussion:}

$I_p$ at $Z=74$ is underestimated in 0.431 a.u. 

\subsection{The Sn-like sequence ($N=50$)}
\label{Sn}

\begin{center}
\begin{figure}[!ht]
\includegraphics[width=0.9\linewidth,angle=0]{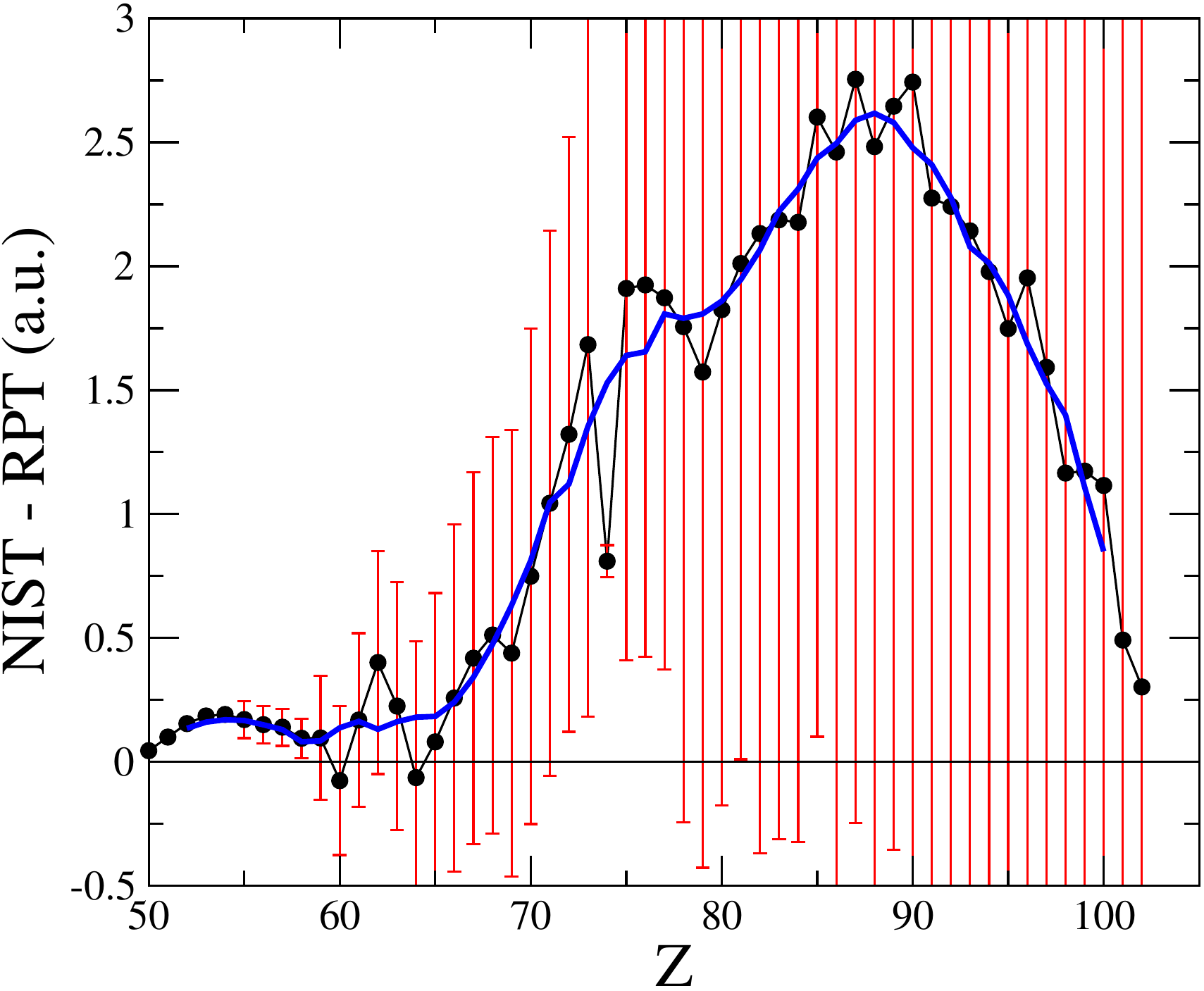}
\caption{\label{fig50} (Color online) The Sn-like systems ($N=50$). A deviation at $Z=74~$ \cite{W-ions1} is apparent.}
\end{figure}
\end{center}

RPT coefficients:
\begin{eqnarray}
\nonumber a_2 &=& 0.03125 + 0.00113932~ (Z/137.036)^2,\\
\nonumber a_1 &=& -2.56241 -0.169401~ (Z/137.036)^2,\\
\nonumber a_0 &=& 37.191,\\
\nonumber a_{-1} &=& 688.884.
\label{sn_coeff}
\end{eqnarray}

Conditions at $Z=N-1$:
\begin{eqnarray}
\nonumber E_a({\rm In}) &=& 0.0110208, \\
s &=& 0.205786.
\label{rest_sn}
\end{eqnarray}

The slope was computed from $E_a({\rm In})$ and $R_{cov}({\rm In})=2.683411$.

\textbf{Discussion:}

$I_p$ at $Z=74$ shows a huge deviation of -0.720 a.u. We may notice also that, apparently, there are crossing points at $Z=60$ and 64. Thus the group of points in $Z=61-63$ should be below the $x$ axis. However, we do not have a precise way of estimating these potentials.

\subsection{The Sb-like sequence ($N=51$)}
\label{Sb}

\begin{center}
\begin{figure}[!ht]
\includegraphics[width=0.9\linewidth,angle=0]{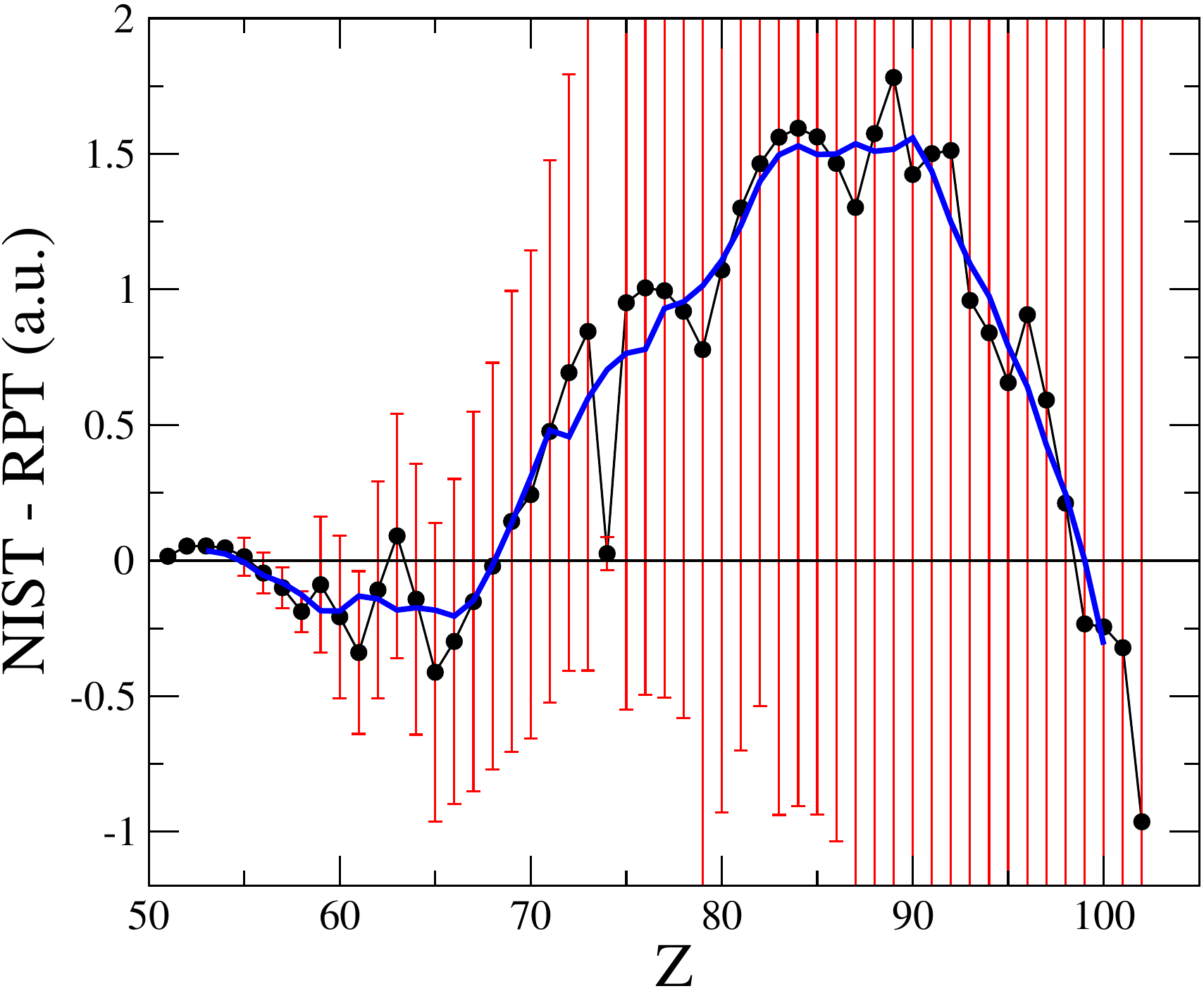}
\caption{\label{fig51} (Color online) The Sb-like sequence ($N=51$). The $Z=74~$\cite{W-ions1} point is distinguished.}
\end{figure}
\end{center}

RPT coefficients:
\begin{eqnarray}
\nonumber a_2 &=& 0.03125 + 0.00113932~ (Z/137.036)^2,\\
\nonumber a_1 &=& -2.61122 -0.172008~ (Z/137.036)^2,\\
\nonumber a_0 &=& 40.6766,\\
\nonumber a_{-1} &=& 628.32.
\label{sb_coeff}
\end{eqnarray}

Conditions at $Z=N-1$:
\begin{eqnarray}
\nonumber E_a({\rm Sn}) &=& 0.0408532, \\
s &=& 0.254716.
\label{rest_sb}
\end{eqnarray}

The slope was computed from $E_a({\rm Sn})$ and $R_{cov}({\rm Sn})=2.645617$.

\textbf{Discussion:}

$I_p$ at $Z=74$ is underestimated in -0.678 a.u. We have stressed previously the problem of $Z=61-63$ ions for Sn-like case. A similar comment applies here.

\subsection{The Te-like sequence ($N=52$)}
\label{Te}

\begin{center}
\begin{figure}[!ht]
\includegraphics[width=0.9\linewidth,angle=0]{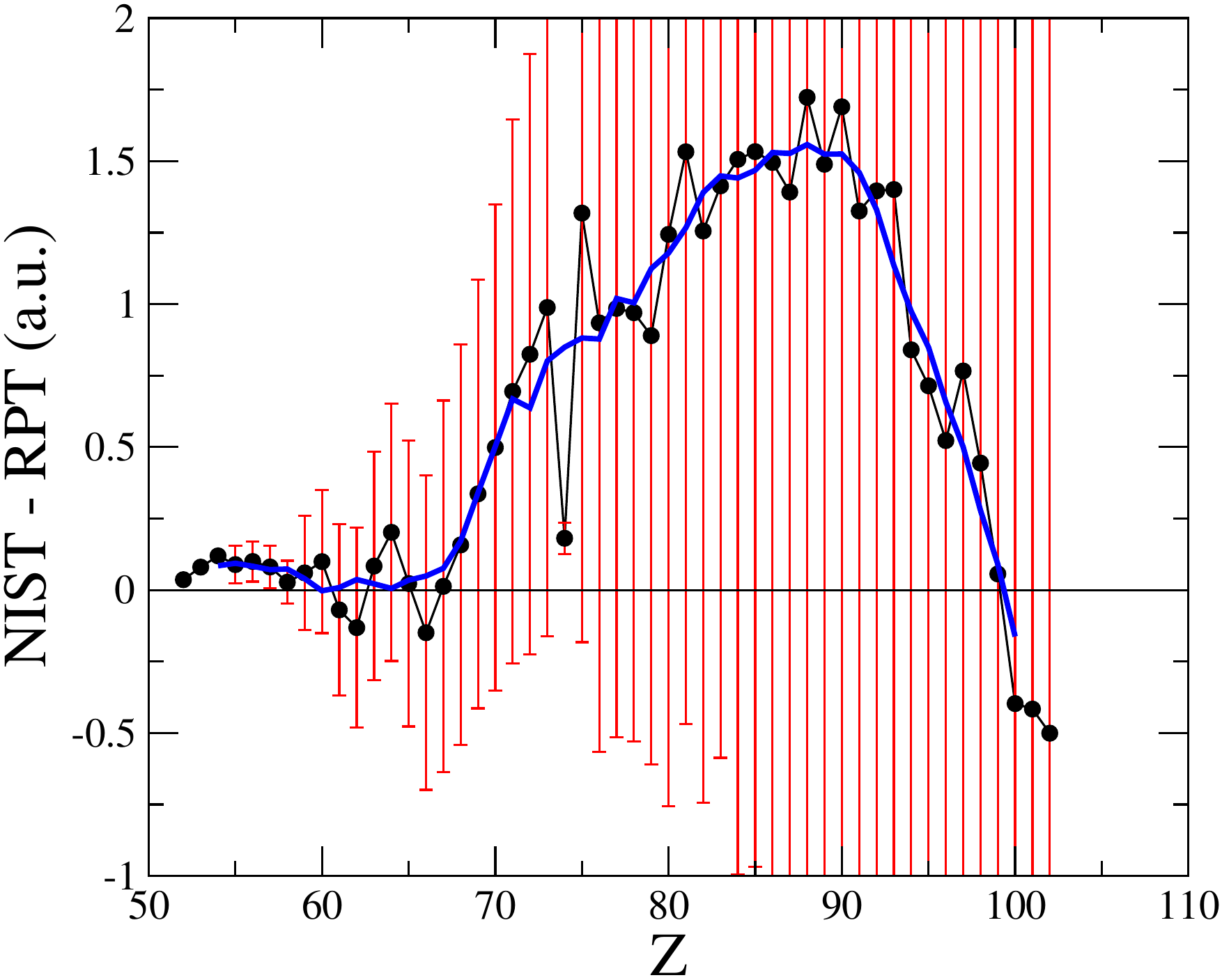}
\caption{\label{fig52} (Color online) The Te-like systems ($N=52$). A marked deviation at $Z=74~$\cite{W-ions1} is noticed.}
\end{figure}
\end{center}

RPT coefficients:
\begin{eqnarray}
\nonumber a_2 &=& 0.03125 + 0.00113932~ (Z/137.036)^2,\\
\nonumber a_1 &=& -2.65793 -0.174483~ (Z/137.036)^2,\\
\nonumber a_0 &=& 41.4216,\\
\nonumber a_{-1} &=& 699.342.
\label{te_coeff}
\end{eqnarray}

Conditions at $Z=N-1$:
\begin{eqnarray}
\nonumber E_a({\rm Sb}) &=& 0.0384262, \\
s &=& 0.252618.
\label{rest_te}
\end{eqnarray}

The slope was computed from $E_a({\rm Sb})$ and $R_{cov}({\rm Sb})=2.645617$.

\textbf{Discussion:}

$I_p$ at $Z=74$ is 0.668 a.u. below the average curve.

\subsection{The I-like sequence ($N=53$)}
\label{I}

RPT coefficients:
\begin{eqnarray}
\nonumber a_2 &=& 0.03125 + 0.000488281 ~(Z/137.036)^2,\\
\nonumber a_1 &=& -2.71192 -0.133452 ~(Z/137.036)^2,\\
\nonumber a_0 &=& 44.5066,\\
\nonumber a_{-1} &=& 670.563.
\label{i_coeff}
\end{eqnarray}

\begin{center}
\begin{figure}[!ht]
\includegraphics[width=0.9\linewidth,angle=0]{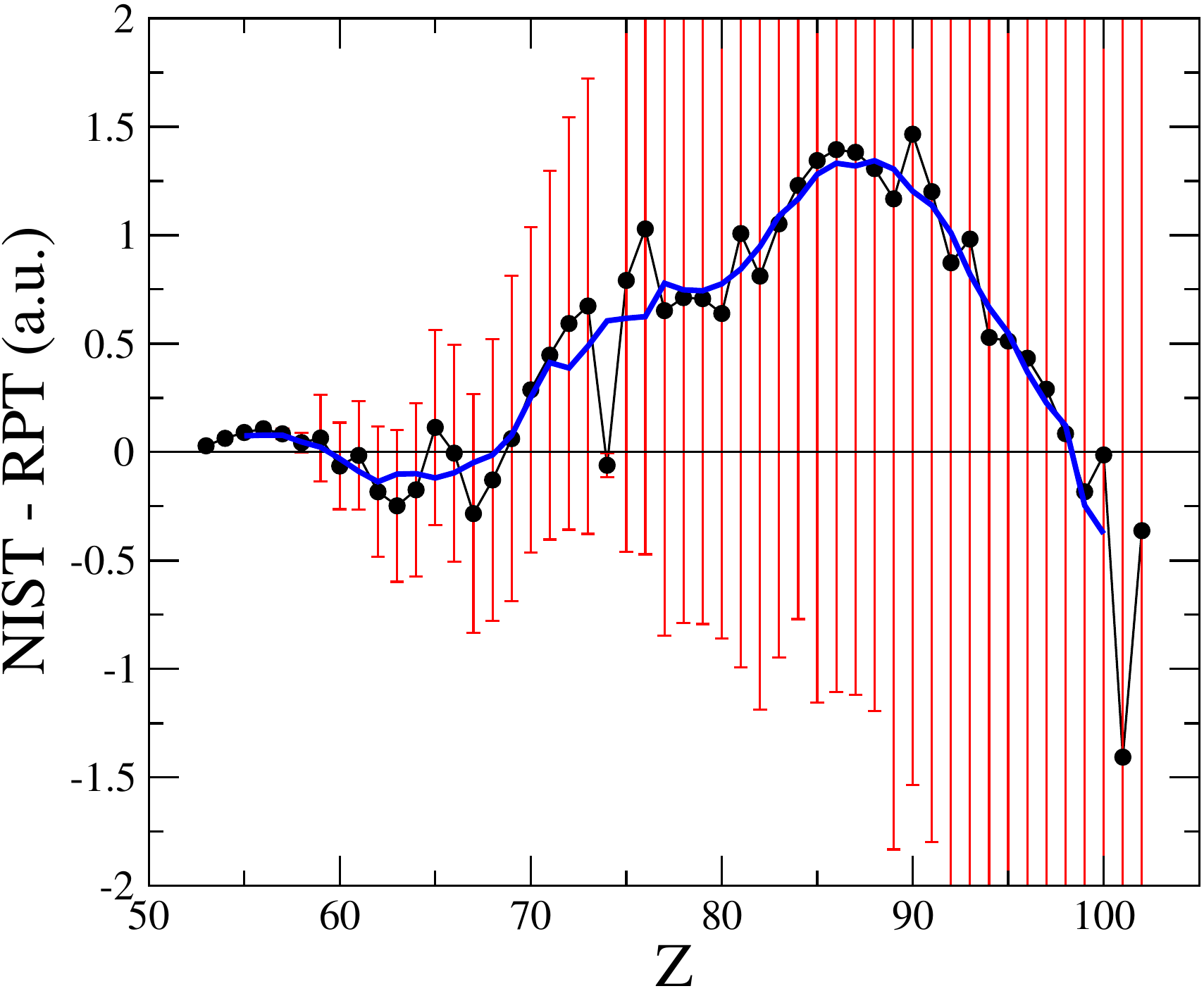}
\caption{\label{fig53} (Color online) The case of I-like systems ($N=53$). An inconsistency is detected at $Z=74$.}
\end{figure}
\end{center}

Conditions at $Z=N-1$:
\begin{eqnarray}
\nonumber E_a({\rm Te}) &=& 0.0724331,\\
s &=& 0.278176.
\label{rest_i}
\end{eqnarray}

The slope was computed from $E_a({\rm Te})$ and $R_{cov}({\rm Te})=2.588925$.

\textbf{Discussion:}

$I_p$ at $Z=74$ is 0.668 a.u. below the average curve. Notice that underestimation of this point is systematical since the Ag sequence.

\subsection{The Xe-like sequence ($N=54$)}
\label{Xe}

RPT coefficients:
\begin{eqnarray}
\nonumber a_2 &=& 0.03125 + 0.000488281~ (Z/137.036)^2,\\
\nonumber a_1 &=& -2.75916 -0.134776~ (Z/137.036)^2,\\
\nonumber a_0 &=& 46.3287,\\
\nonumber a_{-1} &=& 694.4.
\label{xe_coeff}
\end{eqnarray}

Conditions at $Z=N-1$:
\begin{eqnarray}
\nonumber E_a({\rm I}) &=& 0.112416, \\
s &=& 0.293706.
\label{rest_xe}
\end{eqnarray}

The slope was computed from $E_a({\rm I})$ and $R_{cov}({\rm I})=2.570028$.

\begin{center}
\begin{figure}[!ht]
\includegraphics[width=0.9\linewidth,angle=0]{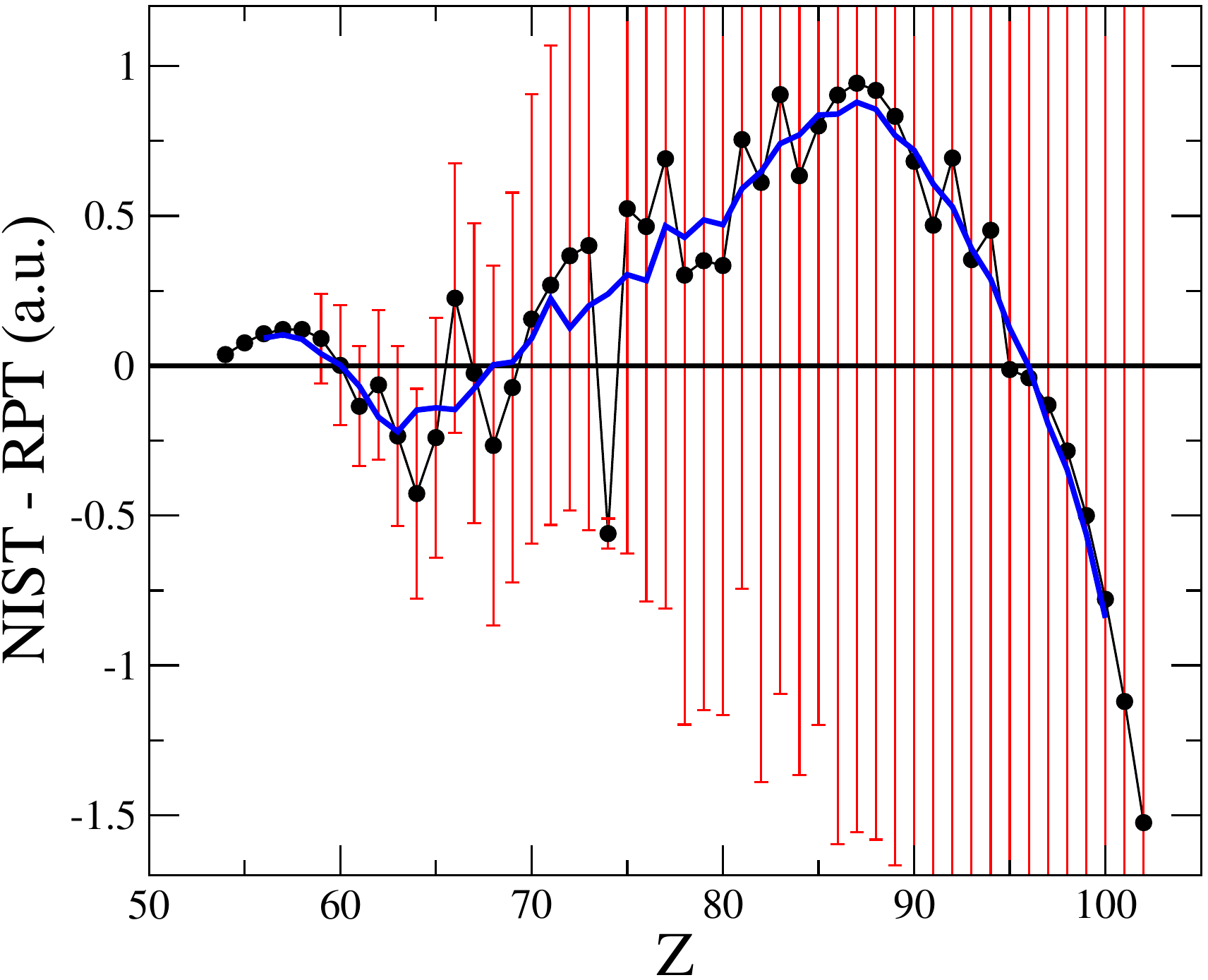}
\caption{\label{fig54} (Color online) The Xe-like systems ($N=54$). Only the $Z=74$ point shows a clear inconsistency.}
\end{figure}
\end{center}

\textbf{Discussion:}

$I_p$ at $Z=74$ is underestimated in 0.799 a.u.

\end{document}